\documentclass[10pt,letterpaper]{article}
\usepackage[top=0.85in,left=2.75in,footskip=0.75in]{geometry}

\usepackage{amsmath,amssymb}

\usepackage{changepage}

\usepackage[utf8x]{inputenc}

\usepackage{textcomp,marvosym}

\usepackage{cite}

\usepackage{nameref,hyperref}

\usepackage{microtype}
\DisableLigatures[f]{encoding = *, family = * }

\usepackage[table]{xcolor}

\usepackage{array}

\newcolumntype{+}{!{\vrule width 2pt}}

\newlength\savedwidth

\raggedright
\usepackage[aboveskip=1pt,labelfont=bf,labelsep=period,justification=raggedright,singlelinecheck=off]{caption}

\makeatletter
\renewcommand{\@biblabel}[1]{\quad#1.}
\makeatother

\usepackage{lastpage,fancyhdr,graphicx}
\usepackage{epstopdf}

\fancyheadoffset[L]{2.25in}
\fancyfootoffset[L]{2.25in}
\usepackage{xcolor}
\usepackage{amsmath}
\usepackage{lipsum}
\usepackage{tikz}
\usepackage{lipsum}
\usepackage{pgfplots}
\usepackage{pgfkeys}
\pgfplotsset{compat=1.14}
\usepackage{pgfkeys}
\usepackage{appendix}
\newenvironment{customlegend}[1][]{%
        \begingroup
        \csname pgfplots@init@cleared@structures\endcsname
        \pgfplotsset{#1}%
    }{%
        \csname pgfplots@createlegend\endcsname
        \endgroup
    }%

    \def\addlegendimage{\csname pgfplots@addlegendimage\endcsname}
\pgfplotsset{
cycle list={%
{draw=black,mark=star,solid},
{draw=black, mark=square,solid},
{draw=black,mark=+,solid},
{black,mark=o},}}

\begin{document}
\vspace*{0.2in}
\date{}
\begin{flushleft}
{\Large
\textbf\newline{Vaccination strategies on dynamic networks with indirect transmission links and limited contact information} 
}
\newline
\\
Md Shahzamal\textsuperscript{1,2*},
Raja Jurdak\textsuperscript{13},
Bernard Mans\textsuperscript{2},
Frank de Hoog\textsuperscript{4}
and Dean Paini\textsuperscript{5}
\\
\bigskip
\textbf{1} Data61, CSIRO, Brisbane, Australia
\\
\textbf{2} Macquarie University, Sydney, Australia
\\
\textbf{3} Queensland University of Technology, Brisbane, Australia
\\
\textbf{4} Data61, CSIRO, Canberra, Australia
\\
\textbf{5} Health and Biosecurity, CSIRO, Canberra, Australia
\\
\bigskip

%
%





* md.shahzamal@mq.edu.au

\end{flushleft}
\section*{Abstract}
Infectious diseases are still a major global burden for modern society causing 13 million deaths annually. One way to reduce the morbidity and mortality rates from infectious diseases is through preventative or targeted vaccinations. Current vaccination strategies, however, rely on the highly specific individual contact information that is difficult and costly to obtain, in order to identify influential spreading individuals. Current approaches also focus only on direct contacts between individuals for spreading, and disregard indirect transmission where a pathogen can spread between one infected individual and one susceptible individual that visit the same location within a short time-frame without meeting. This paper presents a novel vaccination strategy that relies on coarse-grained contact information, both direct and indirect, that can be easily and efficiently collected. Rather than tracking exact contact degrees of individuals, our strategy uses the types of places people visit to estimate a range of contact degrees for individuals, considering both direct and indirect contacts. We conduct extensive simulations to evaluate the performance of our strategy in comparison to the state of the art's vaccination strategies. Results show that our strategy achieves comparable performance to the oracle approach and outperforms all existing strategies when considering indirect links.



\section*{Introduction}
In addition to basic health and survival concerns, infectious diseases also represent a global burden for modern society due to their substantial economic and social impacts. To reduce these impacts various strategies have been devised to reduce the number of casualties. One strategy is to vaccinate the population before an outbreak occurs. This vaccination is called preventive vaccination as it is applied to prevent future outbreaks of the known infectious diseases. However, if there is a new strain of a virus, post-outbreak vaccinations (once generated) can be implemented to hinder the spread of a new disease for an ongoing outbreak. There have been several occurrences of well-documented preventive and post-outbreak vaccinations strategies to prevent spreading of infectious diseases~\cite{yang2016optimal,gong2013efficient, cohen2003efficient}. The development of current vaccination strategies can be divided into two steps: i) studying various vaccination strategies in a virtual environment by simulating disease spread on social contact networks with vaccination strategies; and ii) using the results of the simulations to inform the actual strategy~\cite{patel2005finding, burke2006individual,penny2008should}. 

Existing vaccination strategies aim to identify a set of individuals that are influential disease spreaders for vaccination to reduce the rate at which the infectious disease can spread~\cite{al2018analysis,scholtes2016higher,kas2013incremental}. An essential task for implementing a vaccination strategy, therefore, is to identify an appropriate set of individuals and vaccinate them to modify the spreading dynamics. The size of the selected set of individuals should be minimal and should only require realistic resources (taking into account vaccination cost, information collection cost and manpower) to achieve the required prevention.  

Previous research has found that infection transmission from an infected individual to a susceptible individual depends on the movement and interaction patterns of both susceptible and infected individuals along with other biological factors~\cite{mao2009efficient}. Thus, the interaction patterns are often analysed to find appropriate vaccination strategies~\cite{mao2009efficient,miller2007effective}. Applying detailed information about individual interactions to develop a strategy will often result in an effective vaccination. However, the higher-order interaction information is quite difficult to obtain for real-world social contact networks because of the collection complexity and privacy issues. Computing diffusion control parameters such as betweenness centrality, eigenvalue and closeness~\cite{scholtes2016higher,kas2013incremental,taylor2017eigenvector} are often impractical for vaccination purposes and thus vaccination strategies are often solely developed based on locally obtainable contact information. This paper, therefore, focuses on studying vaccination strategies using the local contact information.

There has been a wide range of methods to reduce the spread of disease on contact networks using the local contact information. The simplest one is the random vaccination (RV) where a proportion of a population is randomly chosen to be vaccinated~\cite{madar2004immunization,cohen2003efficient,lelarge2009efficient}. Unfortunately, this strategy requires a large number of individuals to be vaccinated. Another simple vaccination approach is the acquaintance vaccination (AV) where a random individual is approached and asked to name a friend~\cite{deijfen2011epidemics}. A name recommended by multiple individuals increases the preference to be vaccinated~\cite{cohen2003efficient}. This strategy avoids the random selection of RV strategy and provides an opportunity to select individuals who have contact with many other individuals. It is a targeted vaccination where the most influential individuals are removed from the disease transmission path. However, this AV strategy also requires a large number of individuals to be vaccinated. There have been several modifications to the AV strategy in order to make it more efficient~\cite{holme2017three}. However, these strategies are only studied on static contact networks. The work of~\cite{lee2012exploiting} has upgraded the AV strategy by prioritising the most recently contacted neighbours and assigning weights to their links to capture contact frequency. These vaccination strategies use the recommendation of a neighbouring individual from another individual. However, such information is often inaccurate and hence an optimal set of individuals may not be selected for vaccination.

Another limitation of current contact-based vaccination strategies is that they only focus on direct contacts between individuals. Our recent work~\cite{shahzamal2019indirect,shahzamal2017airborne,shahzamal2018impact,shahzamal2018graph} introduced the concept of indirect transmission, where disease can transmit through indirect interaction (in addition to direct interactions), which is representative of many infectious diseases. For example, a susceptible individual may get infected by visiting the locations where an airborne disease infected individual has been even after that individual has left the location. This is because infectious particles generated by the infected individual persist in the environment, which can transfer to the visiting susceptible individuals. Current vaccination strategies are not designed to capture indirect transmissions, thus potentially missing highly influential individuals with many indirect links. This work investigates the development of appropriate vaccination strategies on dynamic contact networks capturing indirect transmission links as well as direct transmission links.

In particular, this paper proposes a local contact information based strategy, called the individual's movement based vaccination (IMV) strategy, where individuals are vaccinated based on their movement behaviours. It is common for individuals to not provide accurate contact information when asked about their past interactions. Instead, they give a rough estimation of their contact information. In the proposed strategy, individuals are ranked based on their movements relative to public places. Individuals are asked about the frequencies of visits to the different classes of locations (classified based on the intensity of individual visits to those locations, such as the temporal popularity of the places). The locations are labelled with a range of values indicating the number of individuals that could be met when the location is visited by an individual. Based on this coarse-grained information, ranking scores are estimated for the sample individuals (picked up as candidates for the vaccine) from a population, and the highest ranked individuals are vaccinated. The performance of the proposed vaccination strategy is analysed for the spread of airborne infectious disease. The susceptible-infectious-recovered (SIR) epidemic model is applied to simulate the disease spreading on dynamic contact networks. First, the final outbreak sizes are obtained without vaccination, then the effectiveness of the proposed strategy is analysed by estimating the reduction in the outbreak sizes compared to those without vaccination. The efficacy of the proposed strategy is compared with three other strategies: random vaccination (RV), acquaintance vaccination (AV), and degree-based vaccination (DV) where higher degree individuals are vaccinated~\cite{pastor2002immunization}.

Experiments were conducted to study vaccination strategies for both preventive and post-outbreak vaccination scenarios. As individuals can only mention their direct interactions, vaccination strategies are first investigated by ranking nodes based on direct interactions. Then, it is determined how the vaccination strategies are affected if both direct and indirect interactions are accounted for selecting the nodes to be vaccinated. As the proposed strategy depends on coarse-grained information, the vaccination performance can deviate from that of the exact contact information based vaccination. Thus, the vaccination strategies are analysed by using the exact contact information. The proposed strategy is designed for dynamic contact networks even though the temporal information is not integrated with the node selection process. An investigation is done to understand the impacts of integrating temporal information with the proposed strategy. The other important focus of this paper is to examine the effectiveness of vaccination strategies with respect to the scale of information collection on the node's contact~\cite{vidondo2012finding}. The aims of this paper are as follows:

\begin{itemize}
\item Investigating the impact of indirect interactions on the performance of the current vaccination strategies,

\item Developing a new vaccination strategy using the local approximate contact information and movement behaviours,

\item Analysing the performance of the developed strategy in both preventive and post-outbreak scenarios and exploring the significance of the other contact information,

\item Investigating how the scale of information availability influences the performance of the vaccination strategies and understanding what is the minimum required threshold to make a vaccination strategy efficient.
\end{itemize}

\section*{Method and materials}
This section describes the proposed vaccination strategy and the methods for analysing its effectiveness to reduce disease spreading.

\subsection*{Proposed vaccination strategy}
The proposed vaccination strategy, called the individual's movement based vaccination (IMV) strategy, is based on the individual's movement behaviours and propensity to interact with each other. The importance of an individual's past movement behaviours and interaction propensity is reflected by an individual ranking score. The ranking scores are used as indicators of individual's influence in spreading the disease in the near future. In the IMV strategy, locations where individuals visit daily such as offices, restaurants, shopping malls and schools etc. are classified based on the number of neighbour individuals an individual can meet if they visit these locations. The locations are intuitively grouped into six classes as in Table~\ref{tab:cls}. Note that this classification is approximate and its accuracy is correlated to the effectiveness of the proposed vaccination strategy. In this strategy, individuals are asked the number of times they have visited the different classes of locations within a previous time period. The class range is used as individuals may not remember nor notice the exact number of the individuals they have contacted in their past visits to different locations. Instead, they give a rough estimate of their contacts. Individuals also forget to mention their visits for a short duration. Thirdly, it is easier to gather movement information of individual instead of collecting information for every single visit. 

\begin{table}[h!]
\centering
\begin{tabular}{|c|c|c|}
\hline
class & contact sizes & locations\\
\hline
class-1 & 1-5 & home, store\\ \hline
class-2 & 6-15 & coffee shop, bus stop \\ \hline
class-3 & 16-25 & shopping mall, office, local train station, small park \\ \hline
class-4 & 26-50 & central train station, large park\\ \hline
class-5 & 51-100 & university, college, central train station\\ \hline
class-6 & 101- & university, college, airport\\
\hline
\end{tabular}
\vspace{2ex}
\caption{Classification of visits nodes do during their daily activities}
\label{tab:cls}
\end{table}

A generic method is now developed to find the rank of an individual based on the given movement information. Assuming that a susceptible individual $v$ who has been at a location where an infected individual $u$ has visited gets infected with probability $\beta$. The probability of $v$ for not being infected due to this visit is $1-\beta$. If an infected individual $u$ meets $d$ individuals during this visit, the probability of transmitting disease to the neighbours through this visit is given by 
\begin{equation*}
w=1-\left(1-\beta\right)^{d}
\end{equation*}
where $\left(1-\beta\right)^{d}$ is the probability that no neighbour individual is infected from this visit~\cite{alvarez2019dynamic}. Here, the assumption is that the individual $u$ is only the source of infection. All the neighbour individuals are susceptible and contact with $u$ independently. Under these assumptions, $w$ is the spreading potential for a visit of an infected individual based on the number of individuals they meet at the visited location. Now, the spreading potential for visiting a location belonging to a class $i$ can be approximated as 
\begin{equation*}
w_i=\frac{1}{2}\left(2-\left(1-\beta\right)^{d_{i}^{1}}-\left(1-\beta\right)^{d_{i}^{2}}\right)
\end{equation*}
where $d_{i}^{1}$ is the lower limit of class $i$ and $d_{i}^{2}$ is the upper limit of class $i$. In fact, this is the average spreading potential for the class $i$ locations. Then, the ranking score of an individual for visits to different classes of locations is given as
\begin{equation} \label{rank}
W=\sum_{1}^{6} f_i\times w_i 
\end{equation}
where $f_i$ is the frequency of visit to a location belong to class $i$.

In this method, $W$ can be interpreted as the maximum number of disease transmission events during the observation period. As this score is a relative value, it will carry significant information even if the same neighbours are met repeatedly. This is because repeated interaction increases the disease transmission opportunity. In addition, $W$ indicates how easily a susceptible individual $v$ get infected due to his movement behaviours and the propensity of interactions.
The IMV strategy can account for super-spreaders defined by the degree-based strategy and the intensity of interactions among individuals through $d_i$ and $f_i$. It also accounts for the importance of places that 
DV strategy does not consider. It is also easier to remember the visited locations than how many people one has met.

\subsection*{Experimental setup}
The proposed vaccination strategy is analysed through simulating infectious disease spreading on the dynamic contact networks. We have used two types of contact networks: 1) a real contact network constructed from the GPS locations of Momo users; and 2) a synthetic contact network generated by the SPDT graph model~\cite{shahzamal2019indirect}. Both networks consider indirect transmission links along with direct links. Simulations on both networks allow us to understand the consistency of the obtained results. We have also run simulations for other popular vaccination strategies to compare the performance. Here, we first describe the construction of the contact networks followed by disease propagation model, selected baseline vaccination strategies and performance metrics.

\subsubsection*{Data set and contact networks}
This study applies movement information collected from users of a location based social discovery network \emph{Momo}\footnote{https://www.immomo.com}. The Momo App enables users to interact with nearby users by sharing their current locations. Whenever a user launches the Momo app, the current location is forwarded to the Momo server. The server sends back the latest location updates of all users nearby. These location updates have been previously collected by the authors of~\cite{thilakarathna2017deep} using a set of network API communicating with the Momo server. The API retrieved location updates every 15 minutes over a period of 71 days (from May to October 2012). The data set contains 356 million location updates from about 6 million Momo users around the world, but primarily in China. Each database entry includes GPS coordinates of the location, time of update and user ID. For this study, the updates from Beijing are separated as it is the city with the highest number of updates for the period of 32 days from 17 September, 2012 to 19 October, 2012. This data contains almost 56 million location updates from 0.6 million users. 

All possible disease transmission links are extracted from this data set according to the SPDT diffusion model. As the first step, it is identified that an host user (assumed infected with disease) $v$ is staying at a location. Consecutive updates, $X=\{(x_{1},t_{1}),(x_{2},t_{2}),\ldots (x_{k},t_{k})\}$ where $x_{i}$ are the co-ordinate values and $t_{i}$ are the update times, from the user $v$ within a radius of 20m (travel distance of airborne infection particles~\cite{han2014risk}) of the initial update's location $x_{1}$ are indicative of the user staying within the same proximity of $x_1$. For the host user $v$, its visit to the proximity of $x_{1}$ will represent an active visit if a neighbour user $u$ has location updates starting at $t^{'}_{1}$ while $v$ is present, or within $\delta$ seconds after $v$ leaves the location. The user $u$ should have at least two updates within 20m of $x_{1}$ to ensure that it is in fact staying at the same proximity, and therefore can be exposed to the infected particles, rather than simply passing by. The stay period of host user $v$ at the proximity of $x_1$ is ($t_{s}=t_{1}, t_l=t_{k}$), where $t_k$ represents the end of the current stay period. If $u$'s last update within 20m around $x_{1}$ is $(x^{'}_{j},t^{'}_{j})$, the created SPDT link has a link duration ($t_s^{\prime}=t^{\prime}_{1},t_l^{\prime}=t^{'}_{j}$) due to active visit ($t_{s}=t_{1}, t_l=t_{k}$). All links to other users for this active visit ($t_{s}=t_{1}, t_l=t_{k}$) are computed. Similarly, all visits made by $v$ are searched over the updates of 32 days and SPDT links are extracted. Each link between the two same users are distinguished by the time intervals $(t_s,t_l)$ and $(t_s^{\prime},t_l^{\prime}
)$.

The above process is executed for all host users present in the data set to find all possible SPDT disease transmission links and provide a SPDT contact network of 338K users. This network includes possible direct and indirect transmission links due to direct and indirect co-location interactions among users. However, users appear in the system for on average 3-4 days and then disappear for the remainder of the simulation period. Thus, the link density in the network is sparse. In this sparse SPDT network, infected individuals cannot apply their full potential to infect other individuals due to absence from the networks and thus underestimate diffusion dynamics. Thus, we reconstruct a dense SPDT network (DDT) from this network repeating the links from the available days of a user to the missing days for that user~\cite{stehle2011simulation,toth2015role}. If a user has links for multiple days, a day will be randomly chosen and will be copied to a randomly chosen day where that user has no links. Then, this 32 days contact network is extended to 42 days (6 weeks). In this extension, all links of a randomly selected day are copied to a day within 32 to 42 days. Thus, a DDT network for 42 days with 338K users is obtained.

A synthetic SPDT contact network (GDT) is generated for 42 days with 368K nodes using the SPDT graph model introduced in our paper. In the GDT network, a node activates for a period of time which mimic a visit to a location where other nodes are present and disease can transmit. The length of active periods is drawn from a geometric distribution. During the active period, a node contact with a number of neighbouring nodes drawn from a power law distribution. The parameters of power law distribution are heterogeneous as individual have heterogeneous propensity to engage in contact with others. The delay a node make before join a host node and the duration the node stay with the host node are also drawn from the geometry distributions. Thus, the GDT network can simulate disease spreading with the proposed vaccination strategy. The GDT network can verify the experimental results obtained from the DDT contact network. If indirect links from both DDT and GDT networks are excluded, they provide the contact networks with direct links only. Networks with the direct links are called DST and GST respectively. From these networks, the contact information of the first 7 days is applied for ranking the nodes to vaccinate and the rest of the contact information is applied to simulate disease spread. 

\subsubsection*{Disease propagation}
For propagating disease on the selected contact networks, we consider a generic Susceptible-Infected-Recovered (SIR) epidemic model. In this model, nodes remain in one of the three compartments, namely, Susceptible (S), Infectious (I) and Recovered (R). If a node in the susceptible compartment receives a SPDT link from a node in the infectious compartment, the former is subject to exposure $E_l$ of infectious pathogens for both direct and indirect transmission links according to the following equation
\begin{equation}
E_l =\frac{gp}{Vr^2}\left[r\left(t_i-t_s^{\prime}\right)+ e^{rt_{l}}\left(e^{-rt_i}-e^{-rt_l^{\prime}} \right)\right]
+\frac{gp}{Vr^2}\left(e^{-rt_l^{\prime}}-e^{-rt_s^{\prime}} \right)e^{rt_{s}}
\end{equation}
where g is the particle generation rate of infected individual, p is the pulmonary rate of susceptible individual, V is the volume of the interaction area, r is the particles removal rates from the interaction area, $t_s$ is the arrival time of infected individual, $t_l$ is the leaving time of infected individual, $t_s^{\prime}$ is the arrival time of susceptible individuals and $t_l^{\prime}$ is the leaving time of susceptible from the interaction location and $t_i$ is given as follows: $t_i=t_l^{\prime}$ when SPDT link has only direct component, $t_i=t_l$ if SPDT link has both direct and indirect components, and otherwise $t_i=t_s^{\prime}$. If $t_s<t_s^{\prime}$, $t_s$ is set to $t_s^{\prime}$ for calculating appropriate exposure~\cite{shahzamal2019indirect}. If a susceptible individual receives $m$ SPDT links from infected individuals during an observation period, the total exposure $E$ is 
\begin{equation}\label{eq:expo}
E=\sum_{k=0}^{m}E_{l}^{k}
\end{equation}
where $E_{l}^{k}$ is the received exposure for k$^{th}$ link. The probability of infection for causing disease can be determined by the dose-response relationship defined as 
\begin{equation}\label{eq:prob}
P_I=1-e^{-\sigma E}
\end{equation}
where $\sigma$ is the infectiousness of the virus to cause infection~\cite{fernstrom2013aerobiology}. This value depends on the disease types and even virus types. If the susceptible individual move to the infected compartment with the probability $P_I$, it continues to produce infectious particles over its infectious period $\tau$ days until they enter the recovered state, where $1/\tau$ is the rate of recovering from the disease. In this model, no event of birth, death or entry of a new individual is considered.

The simulations are forwarded in our experiments with one day interval~\cite{stehle2011simulation,toth2015role}. We chose an initial set of seed nodes, based on the requirements of the experiment, to start simulations assuming that it will be capable of showing the full epidemic curve in the studied simulation duration of 35 days. During each day of disease simulation, the received SPDT links for each susceptible individual from infected individuals are separated and infection probabilities are calculated by Eqn.~\ref{eq:prob}. The volume $V$ of proximity in Eqn.~\ref{eq:expo} is fixed to 2512 m$^{3}$ assuming that the distance, within which a susceptible individual can inhale the infectious particles from an infected individual, is 20m and the particles will be available up to the height of 2m~\cite{han2014risk,fernstrom2013aerobiology}. The other parameters are assigned as follows: particle generation rate $g=0.304$ PFU (plaque-forming unit)/s and pulmonary rate $q=7.5$ liter/min ~\cite{yan2018infectious,lindsley2015viable,han2014risk}. Infectious particles may require 7.5 min to 300 min to be removed from interaction areas after their generation. We assign $r=\frac{1}{60 b}$ to Eqn~\ref{eq:expo} where $b$ is the particle removal time randomly chosen from [7.5-300] min given a median particle removal time $r_t$. The parameter $\sigma$ is set to 0.33 as the median value of required exposures for influenza to induce disease in 50$\%$ susceptible individuals is 2.1 PFU~\cite{alford1966human}. Susceptible individuals stochastically switch to the infected states in the next day of simulation according to the Bernoulli process with the infection probability $P_I$ (Eqn~\ref{eq:prob}). Individual stays infected up to $\tau$ days randomly picked up from 3-5 days maintaining $\bar{\tau}=3$ days (except when other ranges are mentioned explicitly)~\cite{huang2016insights}.

\subsubsection*{Baseline vaccination strategies}
The performance of the proposed vaccination strategy is compared to three vaccination strategies:

Random vaccination (RV): this is a simple way of vaccination where nodes are chosen randomly to be vaccinated~\cite{rushmore2014network,pastor2002immunization}. To implement this process in preventive vaccination scenarios, a percentage $P$ of nodes are chosen randomly without knowing their contact behaviours and are vaccinated. The RV strategy is also applied in post-outbreak scenarios where a percentage $P$ of neighbouring nodes (whom the host node has contacted) of an infected node are chosen for vaccination. This approach for post-outbreak scenarios is called ring vaccination with random acquaintance selection and is widely used. 

Acquaintance vaccination (AV): in this strategy, a node is randomly approached and asked to name a neighbouring node to be vaccinated~\cite{madar2004immunization,takeuchi2006effectiveness}. To rank the nodes in this strategy, each node present in the network during the first seven days is asked to name a neighbour node and then a list is prepared with each recommended name (the same name can be listed multiple times as they can be recommended by multiple neighbouring nodes). Then, the number of repetitions of each name is counted and ranking scores are obtained for all nodes in the network. The nodes who have contact with a large number of nodes may be named more frequently. Then, $PN/100$ nodes are chosen from the top-ranked nodes based on the naming score to vaccinate percentage $P$ of nodes, where $N$ is the number of total nodes in the selected networks. AV fails to capture indirect transmission events as individuals do not have visibility into indirect contacts they have had, these contacts are due to visits to the same place at different times. 

Degree-based vaccination (DV):
this vaccination strategy vaccinates the nodes that have the largest number of contacts (high degree nodes) as they are more prone to get infected and spread disease~\cite{pastor2002immunization}. It limits the problem of the above strategies that require substantial resources to prevent the spread of a disease. The contact set sizes for the first seven days of nodes in both networks are separated and the nodes are ranked based on the contact set sizes during this time. Then, $PN$ nodes are chosen from the top-ranked node based on the contact set sizes to vaccinate percentage $P$ of nodes. However, this strategy requires the exact information of all contacts a node has. It is often hard in practice to count explicitly all other individuals an individual has met, and may be infeasible in real-world implementations. However, it is of interest to understand how effective the proposed strategy is compared to DV strategy, as an oracle approach. As for AV, the DV approach does not consider indirect transmission links and suffers from the same limitations. 

\subsubsection*{Characterising metric}
For analysing the performance of a vaccination strategy, we first simulate disease spreading on the selected contact networks without vaccination and obtain the outbreak sizes after 35 days. These simulations are run from multiple seed nodes and the average outbreak size $z_{r}$ is computed. This indicates the propensity of the disease to spread in a network without vaccination and is used as the reference for comparing the effectiveness of the applied vaccination strategies. The performance of a vaccination strategy is quantified by how much reduction it can achieve in terms of average outbreak sizes comparing to $z_{r}$. Thus, the effectiveness of a vaccination strategy with a vaccination rate $P$ is given by
\[\eta=\frac{z_{r}- z_{c|P}}{z_r}\times 100\]
where $z_{c|P}$ is the average outbreak sizes of the candidate vaccination strategy with the vaccination rate $P$. In the simulation, the outbreak size indicates the number of new infections caused after running simulation over 35 days.
\section*{Results and discussion}
\subsection*{Preventive vaccination}
The preventive vaccination is implemented to prevent possible future outbreaks of disease. In this section, the effectiveness of the selected vaccination strategies to prevent future outbreaks in the empirical SPDT contact network (DDT network) and the generated synthetic SPDT contact network (GDT network) are investigated. First, simulations are conducted to understand the upper bound of the efficiency of the applied strategies where it is assumed that contact information of all nodes are available and can be collected for vaccinations. Then, the performance is investigated with the simulation scenarios where contact information of only a proportion of nodes are available for ranking and only these nodes are chosen to be vaccinated. However, all nodes participate in propagating disease on the networks.

\subsubsection*{Performance analysis}
\begin{figure}[h!]
\begin{tikzpicture}
    \begin{customlegend}[legend columns=4,legend style={at={(0.12,1.02)},draw=none,column sep=3ex ,line width=2pt,font=\small}, legend entries={RV, AV, IMV, DV, direct, indirect}]
    \addlegendimage{solid, color=blue}
    \addlegendimage{color=red}
    \addlegendimage{color=olive}
    \addlegendimage{color=violet}
    \addlegendimage{color=black}
    \addlegendimage{dashdotted, color=black}
    \end{customlegend}
 \end{tikzpicture}
 \centering
\includegraphics[width=0.495\linewidth, height=5.0 cm]{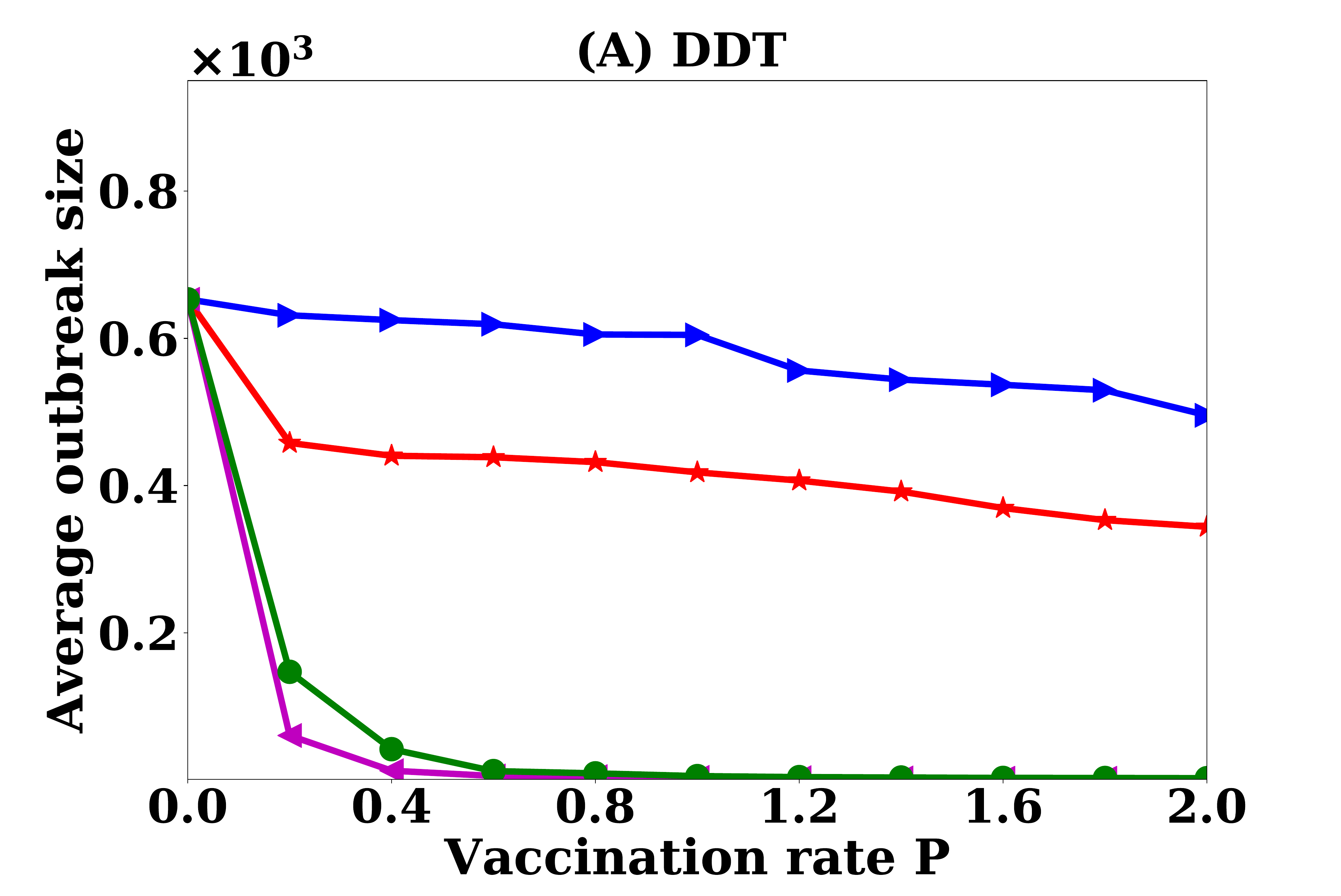}
\includegraphics[width=0.495\linewidth, height=5.0 cm]{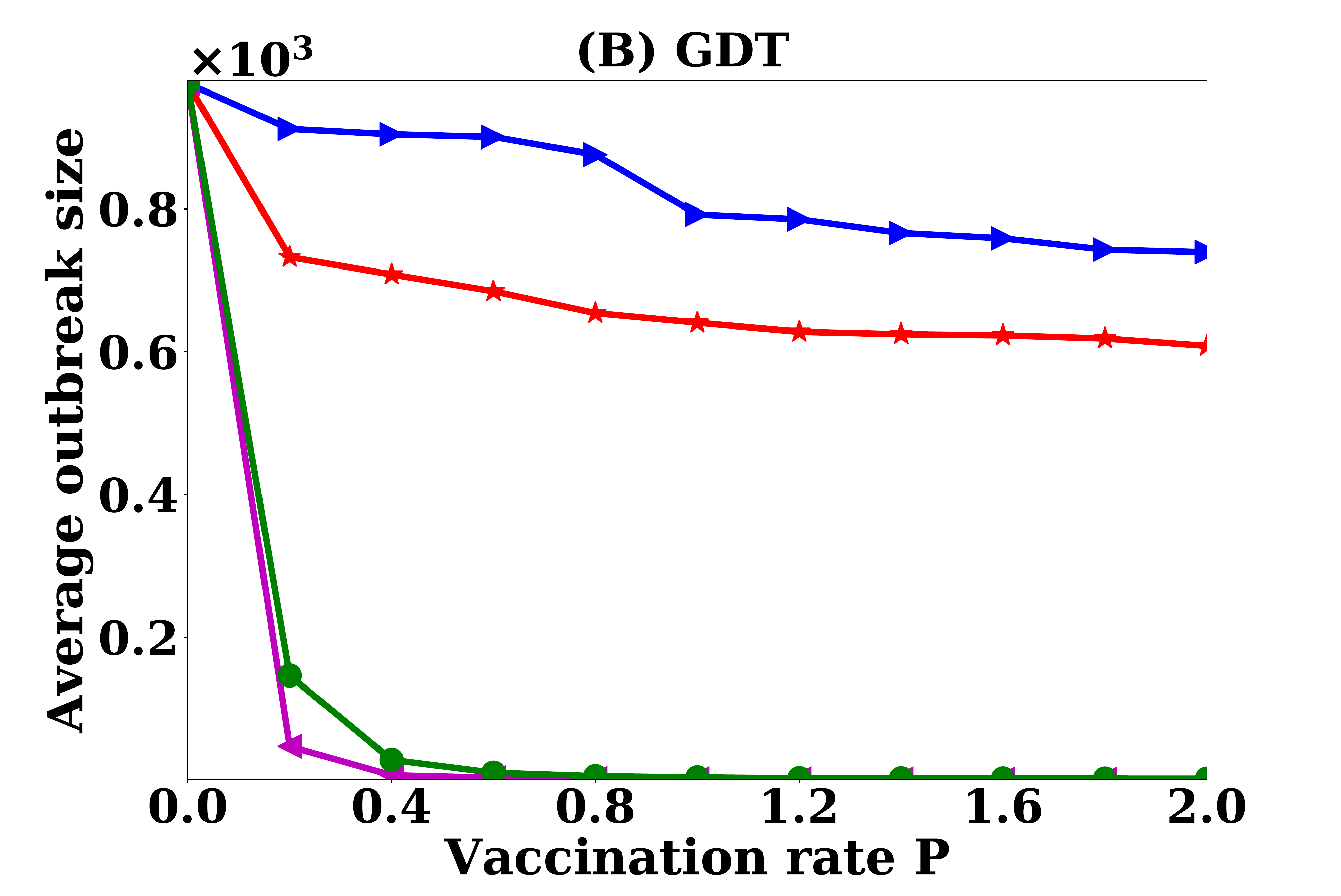}

\includegraphics[width=0.495\linewidth, height=5.0 cm]{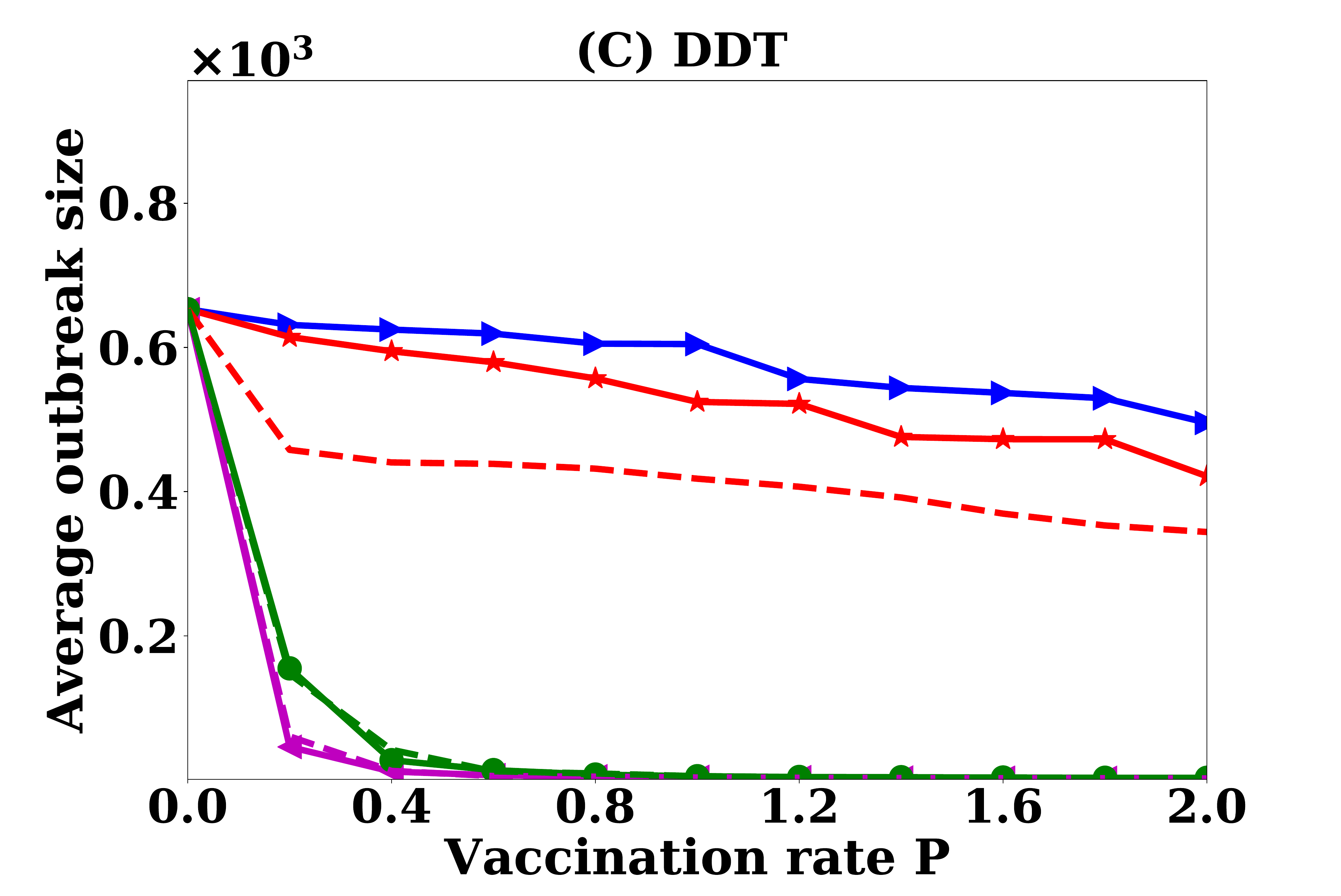}
\includegraphics[width=0.495\linewidth, height=5.0 cm]{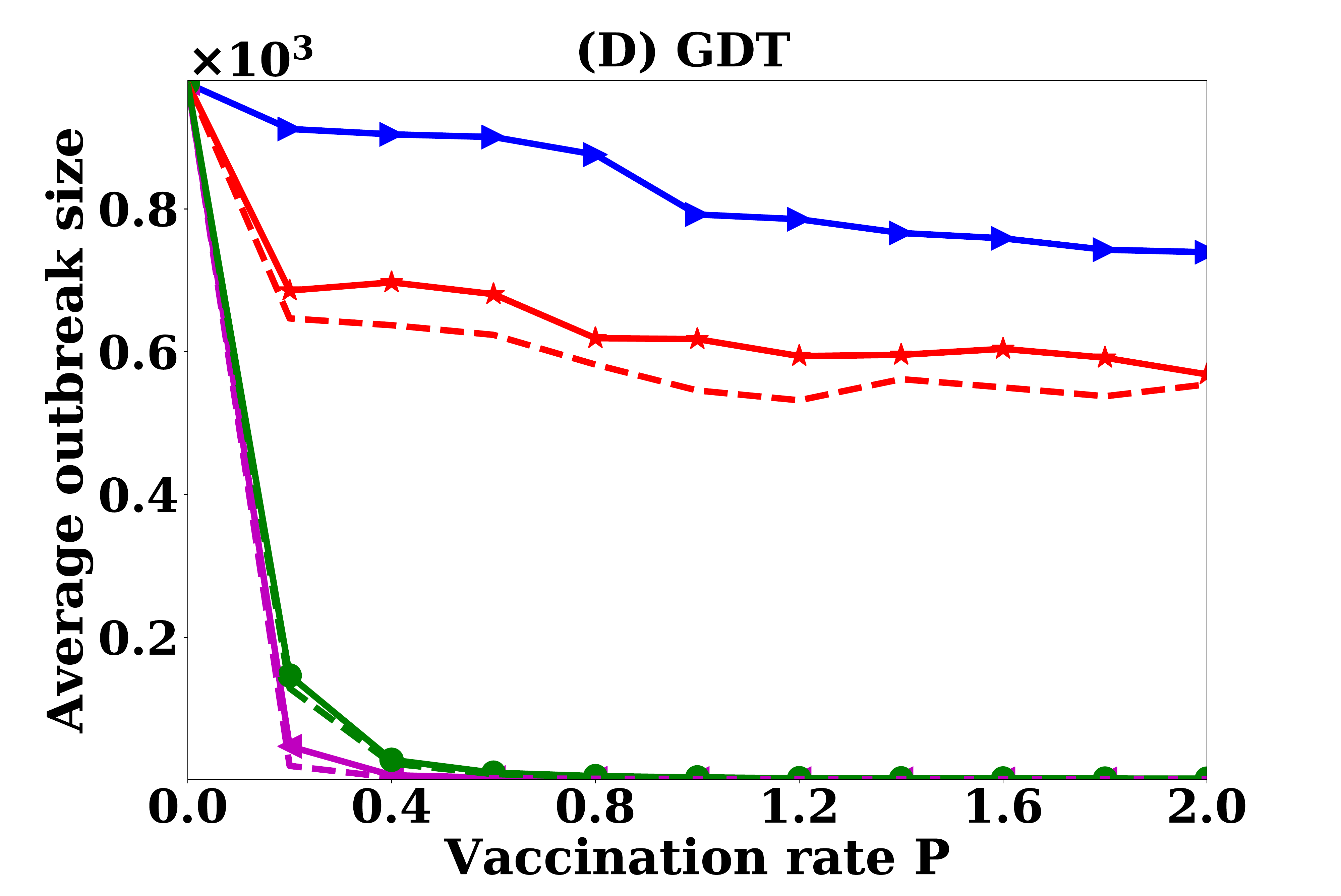}
\caption{Average outbreak sizes at various vaccination rates $P$ (percentage of total nodes) of different strategies: (A, B) nodes are vaccinated with contacts created for direct interactions and (C, D) comparison of outbreak sizes for vaccinating nodes with contacts based on the direct interactions (solid lines) and contact based on any direct or indirect interactions (dashed lines)}
\vspace{-1.5em}
\label{fig:avac}
\end{figure}

\begin{figure}[h!]
\begin{tikzpicture}
    \begin{customlegend}[legend columns=4,legend style={at={(0.12,1.02)},draw=none,column sep=3ex ,line width=2pt,font=\small}, legend entries={RV, AV, IMV, DV}]
    \addlegendimage{solid, color=blue}
    \addlegendimage{color=red}
    \addlegendimage{color=olive}
    \addlegendimage{color=violet}
    \end{customlegend}
 \end{tikzpicture}
 \centering
\includegraphics[width=0.495\linewidth, height=5.0 cm]{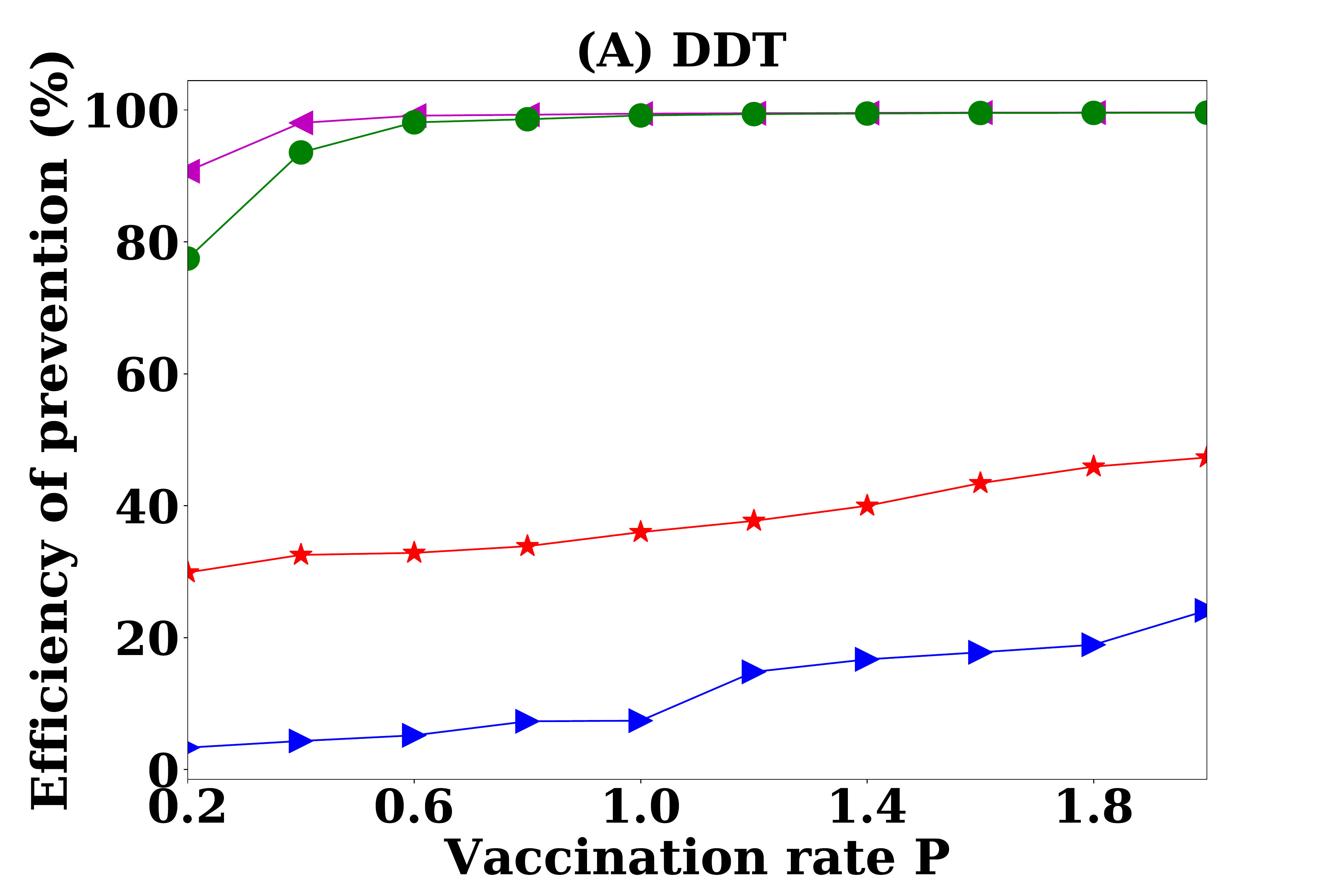}
\includegraphics[width=0.495\linewidth, height=5.0 cm]{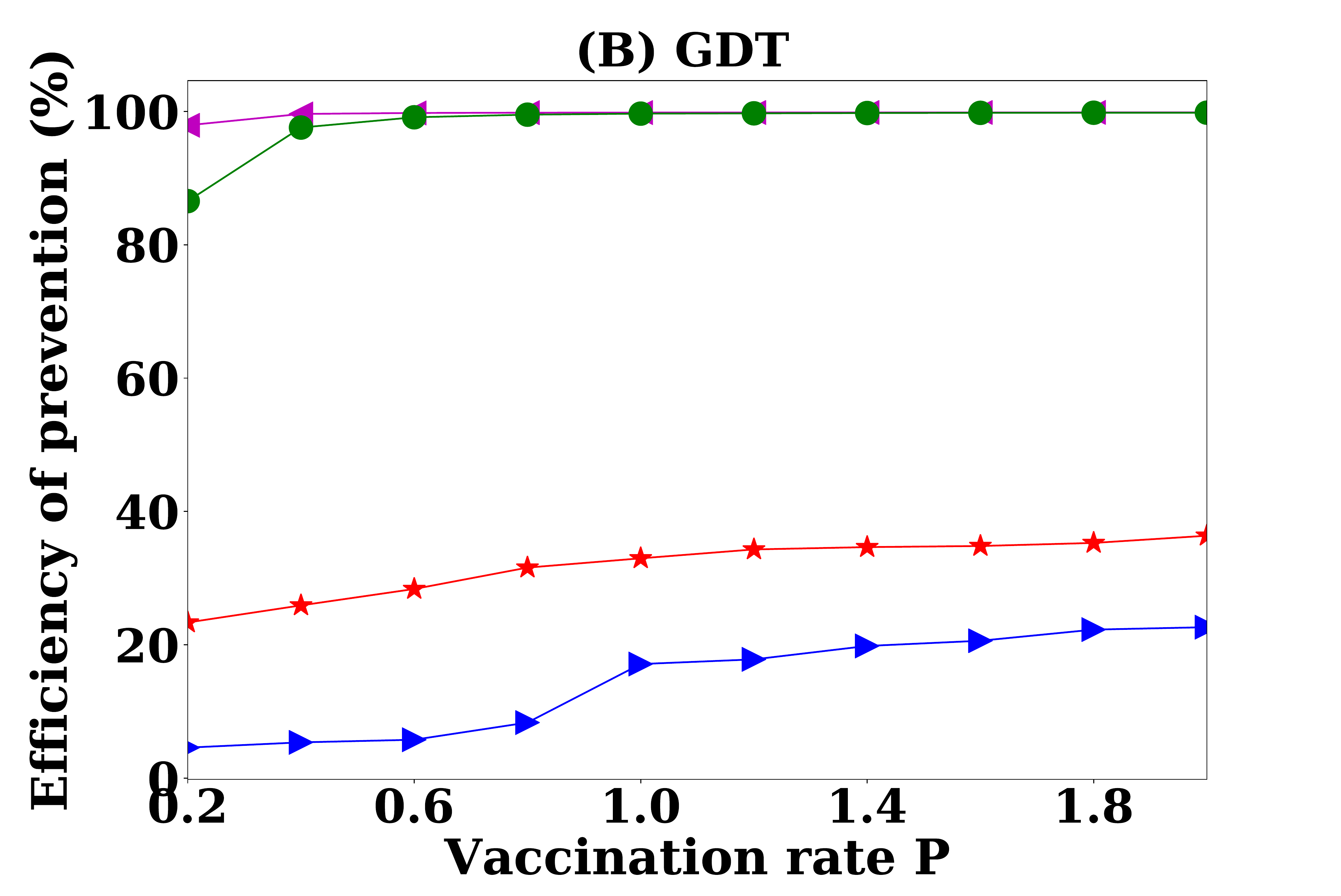}\\

\includegraphics[width=0.495\linewidth, height=5.0 cm]{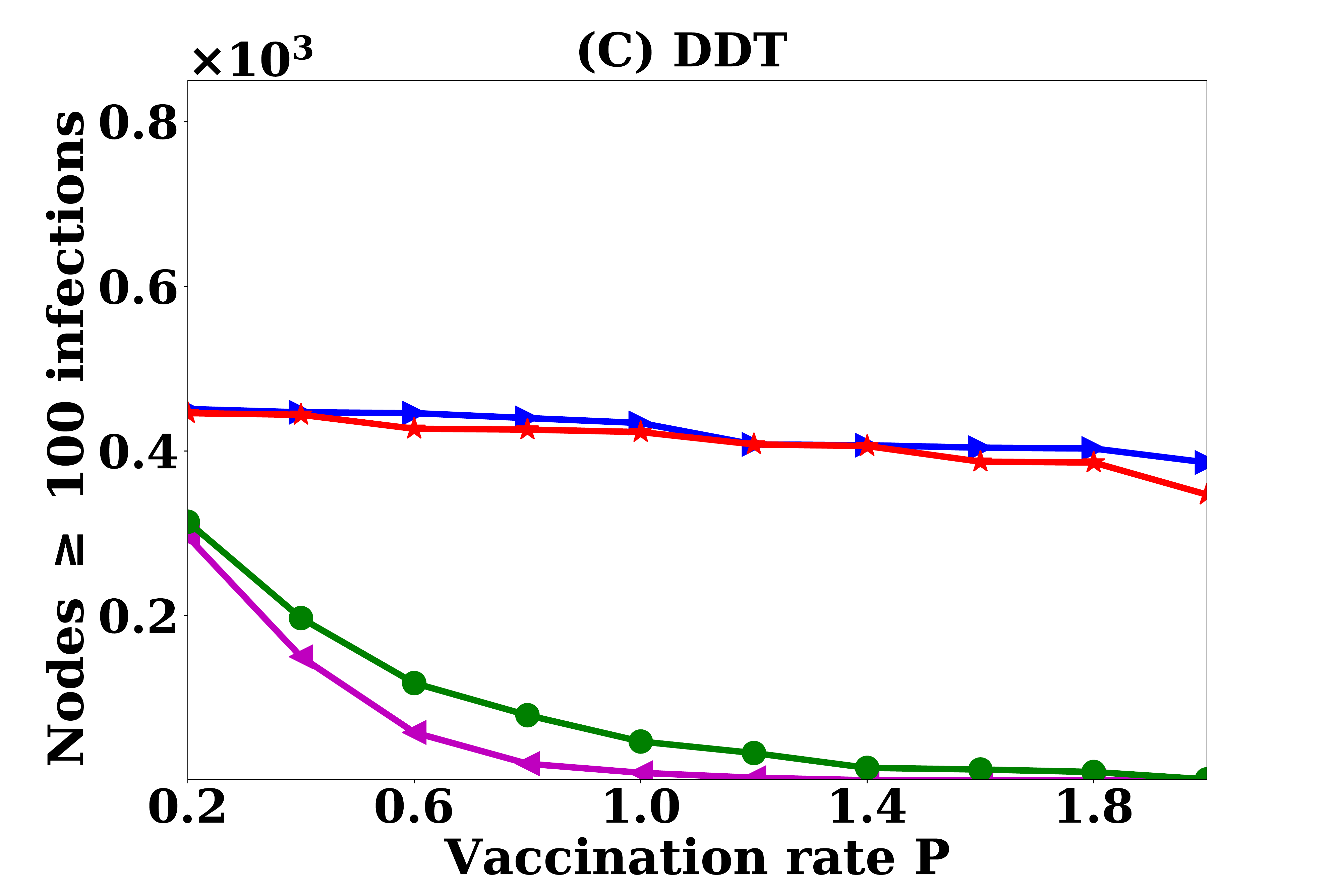}
\includegraphics[width=0.495\linewidth, height=5.0 cm]{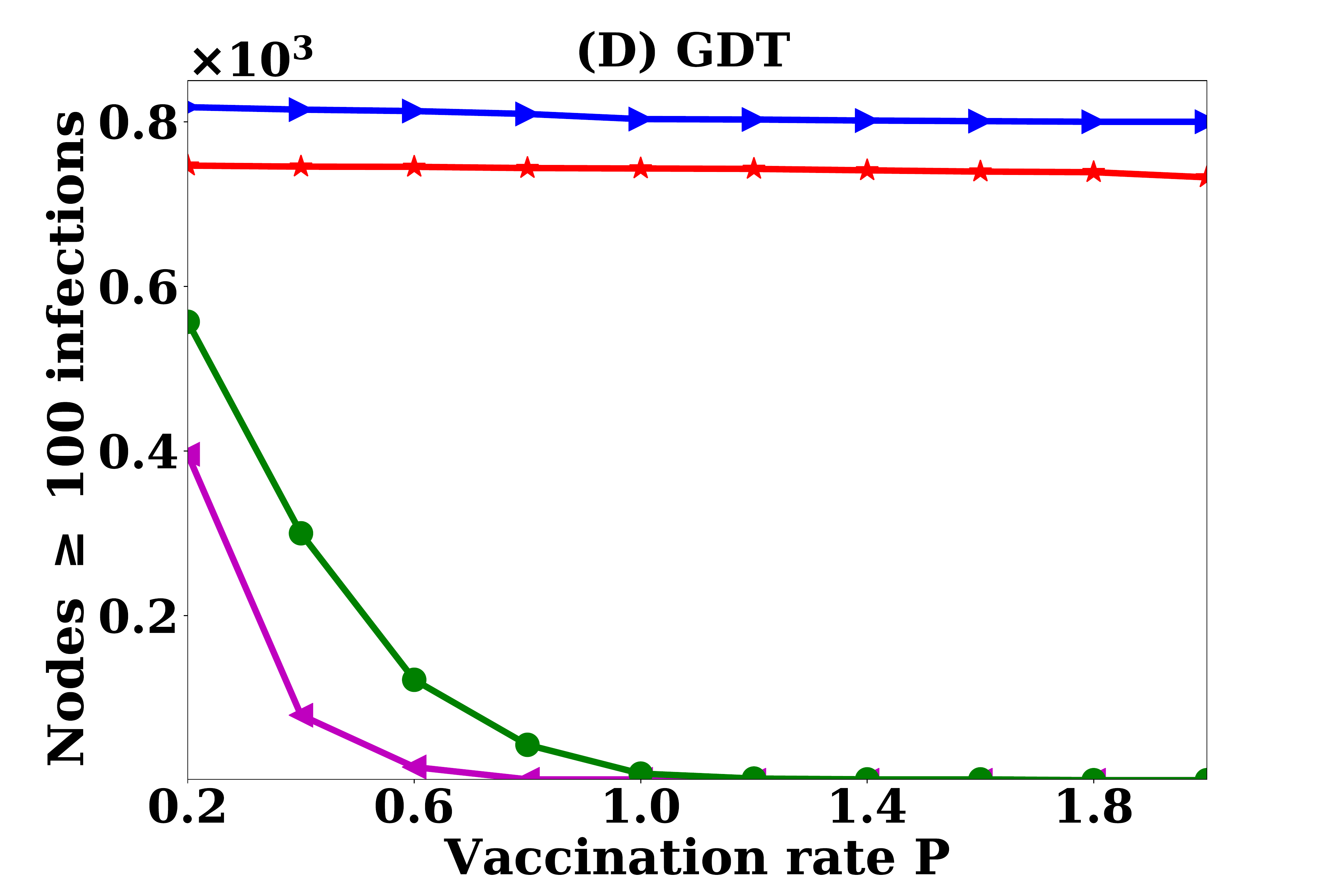}\\

\includegraphics[width=0.495\linewidth, height=5.0 cm]{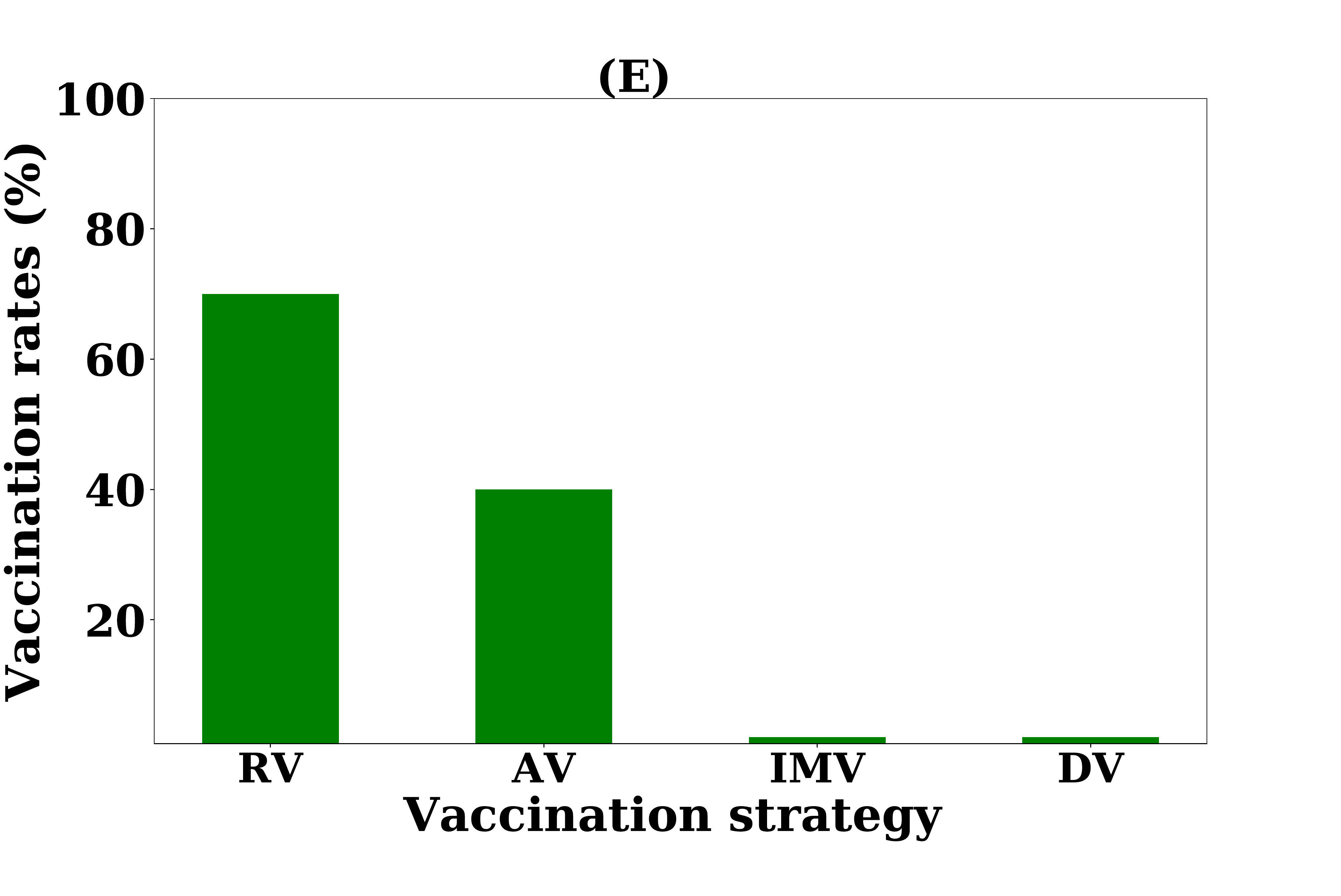}
\includegraphics[width=0.495\linewidth, height=5.0 cm]{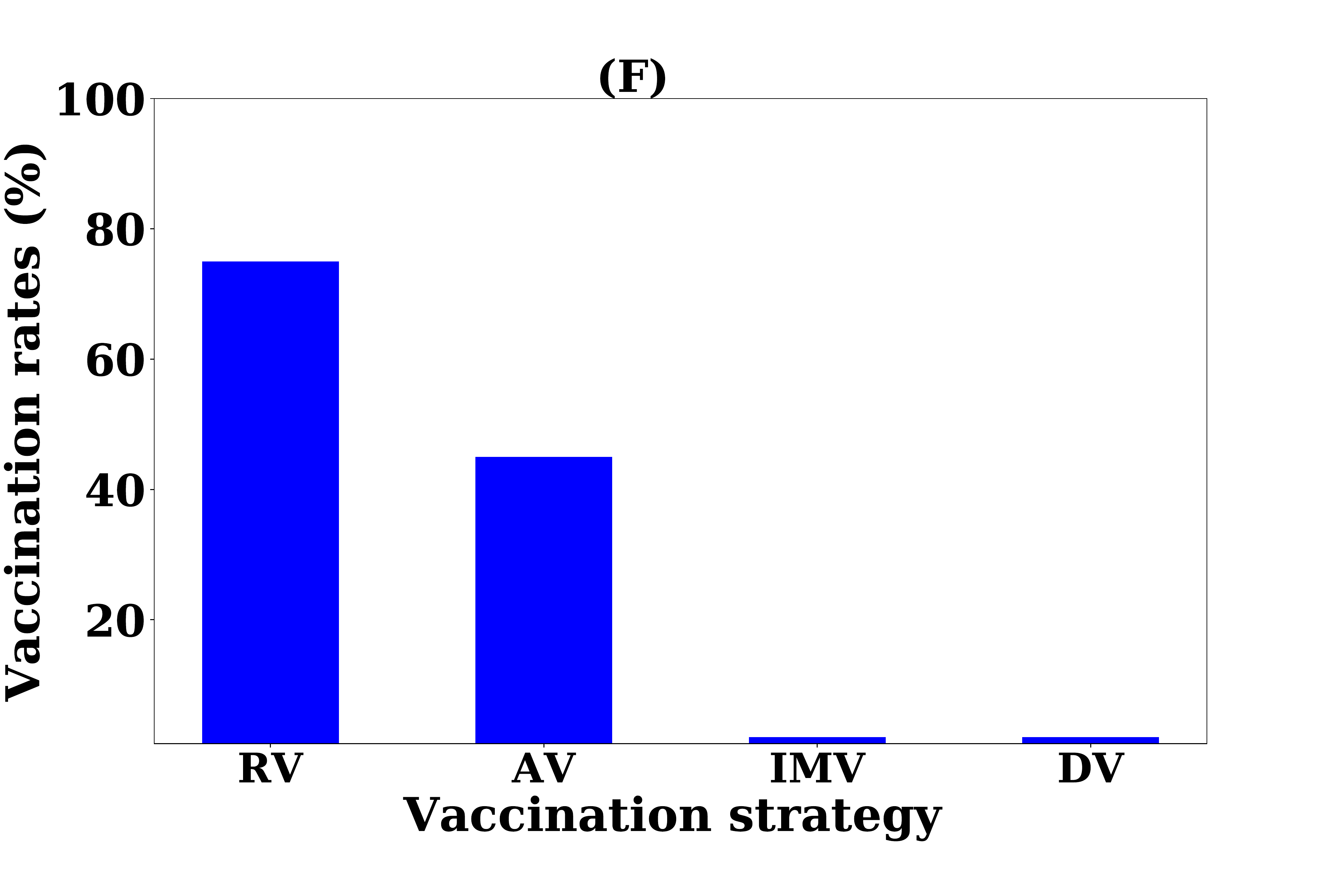}
\vspace{-1.0em}
\caption{Efficiency of vaccination strategies to prevent disease spreading: (A, B) preventive efficiency as a measure of reduction in average outbreak sizes due to vaccination, (C, D) number of seed nodes out of 5000 having outbreak sizes greater than 100 infections, and (E, F) required vaccination rates for strategies to keep outbreak sizes below 100 infections from all seed nodes}
\label{fig:vacef}
\vspace{-1.0em}
\end{figure}

The performance of vaccination strategies is studied with varying vaccination rates $P$ (percentage of nodes) in the range [0.2-2]\% with the step of 0.2\%. Thus, $PN/100$ nodes are chosen based on the vaccination strategies and are vaccinated by assigning their status as recovered. Then, a random node is chosen as a seed node and the outbreak size is obtained by running the disease spreading simulation over 35 days from the seed node. This process is iterated through 5,000 different seed nodes. Seed nodes are infectious for 5 days and then recover. When individuals are asked about their movements, they usually provide information based on the number of individuals they have seen in particular locations. They will not be aware of the number of individuals who have SPDT links through indirect interactions. Thus, the nodes are ranked based on the contacts they have seen, i.e. considering only the direct interactions where both the host and neighbour nodes stay together in a location. Then, the performance is also examined accounting for the indirect links along with the direct links. The average of outbreak sizes for vaccination strategies with various $P$ are presented in Figure~\ref{fig:avac}. The average outbreak sizes without vaccination are obtained by simulating disease from 5000 single seed nodes in both networks which are 653 infections in the DDT network and 976 infections in the GDT network. Based on these values, the preventive efficiency is measured at each $P$ for all vaccination strategies and shown in Figure~\ref{fig:vacef} (a,b). To clearly understand the efficiency of a strategy, the number of seed nodes having outbreak sizes of more than 100 infections are also counted and presented in Figure~\ref{fig:vacef}(c,d). This indicates the efficiency of a strategy to hinder super-spreader nodes from spreading disease. If there are no seed nodes with outbreaks of 100 infections in a network, it is assumed that the protection from infection is significant for the applied strategy.

The results show that the average outbreak sizes for random vaccination (RV) are higher in both DDT and GDT networks at all values of $P$. Furthermore, increasing $P$ does not reduce the outbreak sizes significantly for RV (see Fig.~\ref{fig:avac}). For changing $P$ from 0.2\% to 2\%, the preventive efficiency of RV strategy is reached to 18\% in the DDT network and 16\% in the GDT network (see Fig.~\ref{fig:vacef}). On average 9\% of seed nodes have outbreak sizes of more than 100 infections in the DDT network and 16\% nodes in the GDT network at $P=0.2$\%. These numbers slightly reduce for changing $P$ from 0.2\% to 2\%. Therefore, the preventive performance of RV strategy in this range of vaccination rate is very poor. The proposed IMV strategy, however, shows significantly small average outbreak sizes even at $P=0.2$\% with the preventive efficiency of 78\% in the DDT network and 82\% in the GDT network. Unlike RV strategy, the average outbreak sizes quickly decrease with increasing $P$ in IMV strategy. If the vaccination rate is $P=0.6$\% (with vaccinating 2880 nodes), the average outbreak size becomes below 10 infections in both networks and a preventive efficiency of 98\% is achieved. However, 1.5\% seed nodes have outbreaks of greater than 100 infections and it becomes zero at $P=0.2$\% vaccination. Simulations are also run by increasing $P$ for RV strategy until the preventive efficiency reaches the stage where no seed node has outbreak greater than 100 infections. It is found that the RV strategy requires 70\% of nodes to be vaccinated to achieve such preventive efficiency in both the DDT and GDT networks (see Fig.~\ref{fig:vacef}). Thus, the proposed IMV strategy achieves significantly higher efficiency than the random vaccination strategy.

The AV strategy improves the preventive efficiency over the RV strategy. However, the results show that the proposed IMV strategy is still much better than the AV strategy. At $P=0.2$\%, the efficiency is one third of the efficiency of IMV. Moreover, the IMV strategy shows nearly 0\% seed nodes with outbreak size of greater than 100 infections at vaccination rate $P=1.4$\% in GDT network while the AV strategy still has 14\% seed nodes with outbreak sizes of greater than 100 infections. The AV strategy requires about 40\% of nodes to be vaccinated to achieve preventive efficiency where no seed node has outbreaks greater than 100 infections (see Fig.~\ref{fig:vacef}). Thus, the IMV strategy achieves much better efficiency than the AV strategy. The IMV strategy also achieves the efficiency that is close to that of the DV strategy (degree based vaccination). The DV strategy has the preventive efficiency of 92\% in the DDT network and 98\% in the GDT network. This is a bit higher than the IMV strategy. However, the efficiency of IMV strategy becomes closer to that of DV strategy as $P$ increases. To achieve strong preventive efficiency, both strategies require almost the same rates of vaccination. Thus, the coarse-grained information based IMV strategy achieves the performance of degree based vaccination strategy. 

The above results are obtained by ranking the nodes based on the contacts created for direct interactions, i.e. neighbour nodes who are connected with direct links are only considered in the ranking process. Figure~\ref{fig:avac} shows (dashed lines) the results for vaccination strategies with indirect interactions, i.e. neighbour nodes connected with indirect links are also considered in calculating the ranking scores for a strategy. It is observed that the DV strategy and proposed IMV strategy do not vary the average outbreak sizes largely for any vaccination rates due to including indirect links. As the movement information is the same for creating direct or indirect interactions, the performance of IMV strategy does not decrease. The performance of the other neighbour based strategy AV increases slightly when considering indirect interactions as some nodes may become important with indirect links and this is not captured by the direct interaction based implementation. In addition, some new neighbour nodes appear when indirect links are counted and this changes the score of RV strategy. Thus, the nodes are ranked more precisely in the AV strategy when indirect links are counted and the effectiveness of vaccination increases.

\subsubsection*{Sensitivity analysis}
The performance of the strategies is now studied for the various scale of information availability regarding contact of nodes. For IMV strategy, the coarse-grained information of an individual's movement is applied through classifying the locations into groups. It is interesting to know how much improvement can be achieved if exact contact information of nodes is applied by ranking process. Accordingly, a new ranking score is defined as
\begin{equation}\label{eq:vtrans}
    w_i=1-(1-\beta)^{d_i}
\end{equation}
where $w_i$ is the probability of transmitting disease to neighbouring nodes for a visit $i$ at a location where infected nodes meet $d_i$ number of other nodes. Thus, the node's rank can be given by
\begin{equation}
    W=\sum_{i=1}^{n}w_i
\end{equation}
This equation now finds the score considering the exact number of neighbouring nodes for each visit during the observation period.
The other important factor for disease spread through visiting the location is the duration of stay. If an infected individual stays longer at a location, they may transmit disease to more susceptible individuals. Thus, the temporal information is integrated with the ranking process as follows. It is assumed that the probability of transmitting disease for visiting a location increases exponentially. Therefore, the transmission probability $\beta$ is defined as 
\begin{equation}
    \beta=1.6 \beta_{0} \left(1-e^{-\frac{t_{i}}{t_0}}\right)
\end{equation}
where $t_i$ is the stay time of the node for a visit $i$ to a location, $t_0$ is the average stay time of users of Momo App, and $\beta_{0}$ is the disease transmission probability if a susceptible node stays with an infected node for $t_0$ time. Integrating this time dependency on transmission probability of Equation~\ref{eq:vtrans} can capture the impact of the duration of stay at the visited locations. Therefore, the IMV strategy is also studied based the node's rank with temporal information. Finally, the performance of all vaccination strategies is studied for the scenarios where $F$ proportion of nodes provides contact information for the ranking process and nodes to be vaccinated are chosen from them. The experiment is conducted for $F=\{0.25, 0.5, 0.75\}$. 

\begin{figure}[h!]
\includegraphics[width=0.5\linewidth, height=5.0 cm]{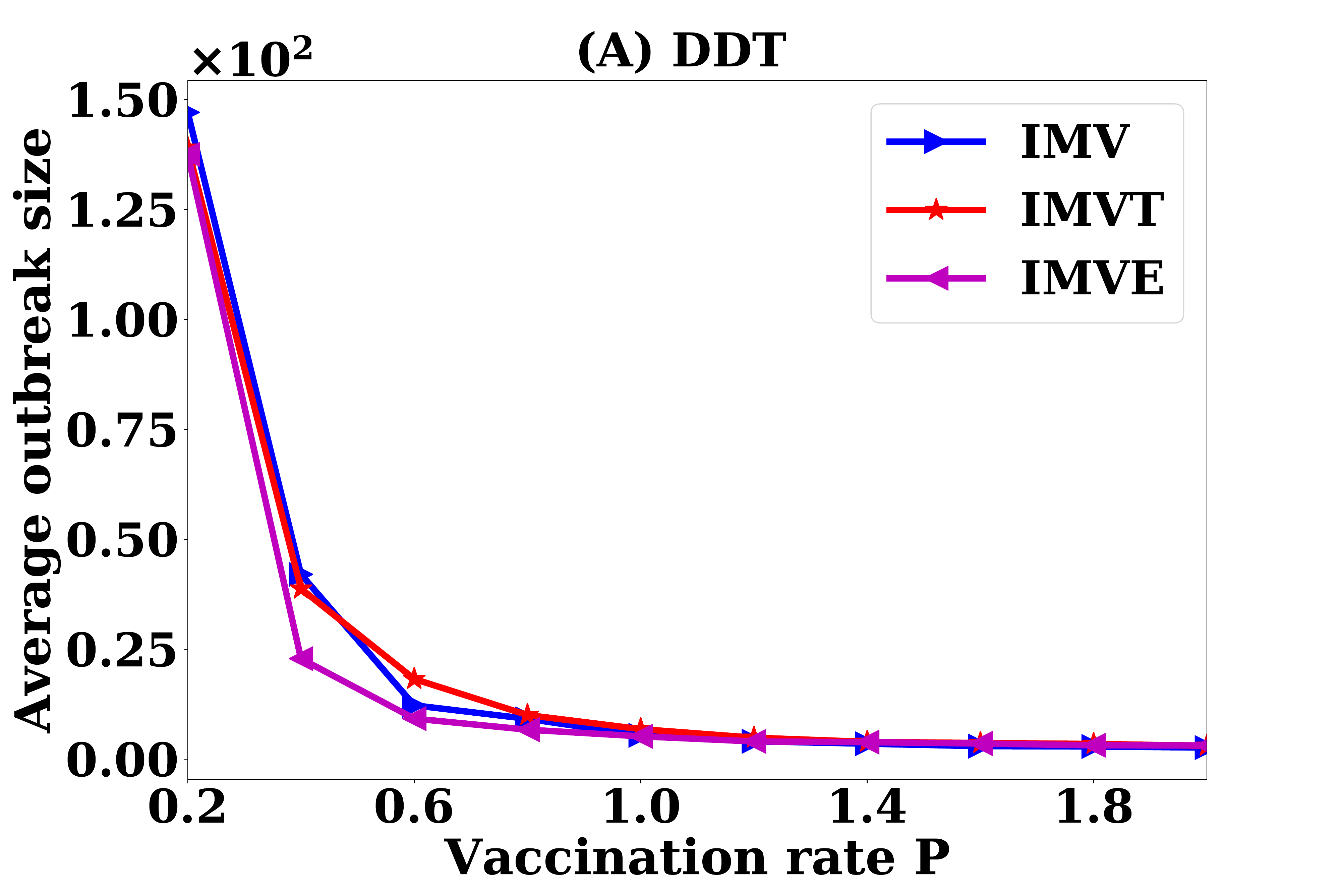}
\includegraphics[width=0.5\linewidth, height=5.0 cm]{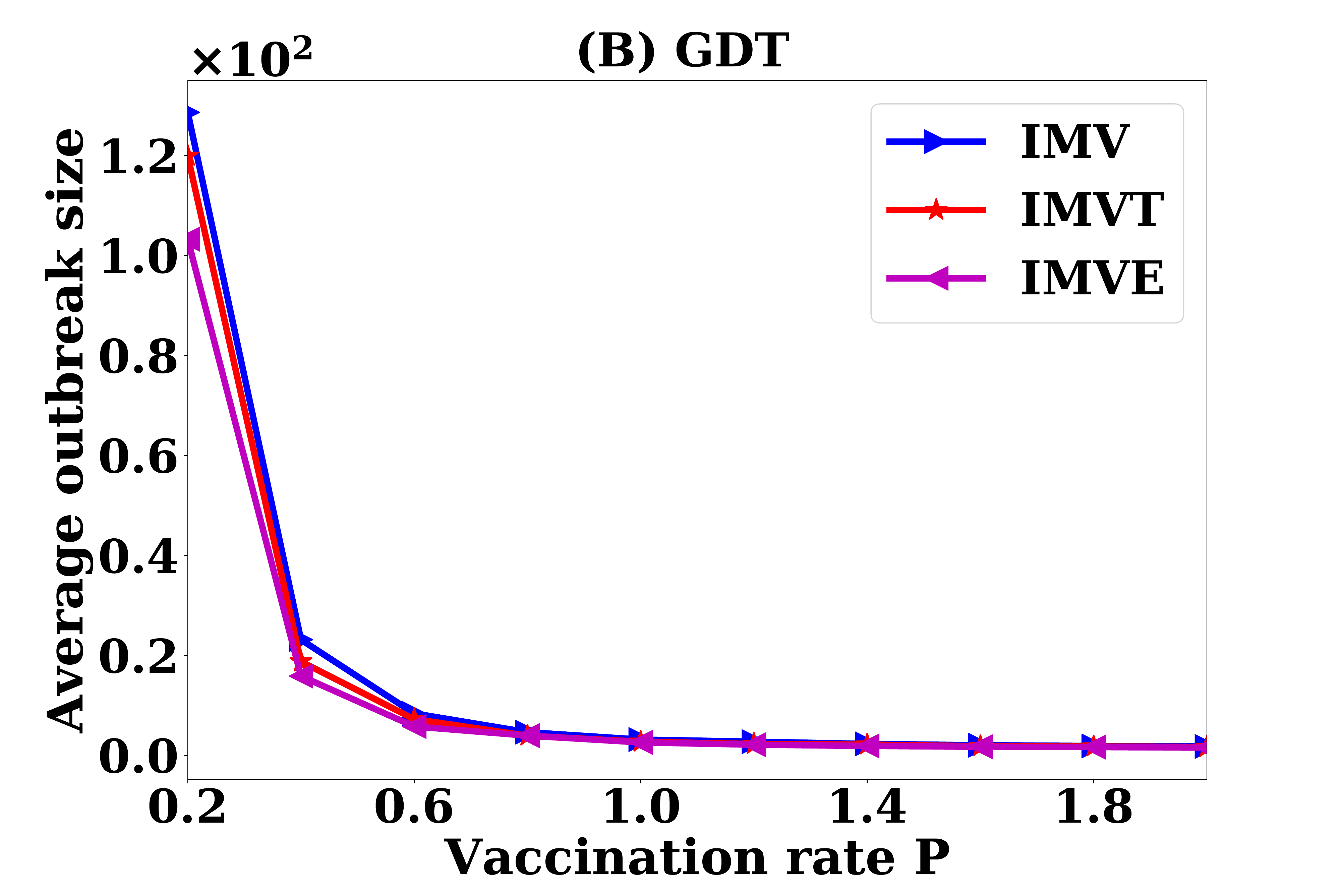}\\
\caption{Variations in average outbreak sizes of IMV strategy for accounting the temporal information and exact contact information}
\label{fig:vacd}
\end{figure} 

Nodes are now ranked with the exact contact information and temporal information for vaccinating with the IMV strategy. Theses vaccination strategies are named as IMVE (exact information based) and IMVT (with temporal information based). Similar to the previous experiments, simulations are conducted from 5,000 different single seed nodes and average outbreak sizes are computed for each $P$ in the range [0.2,2]\% with a step of 0.2\%. The results are presented in Figure~\ref{fig:vacd}. Applying exact information, the IMVE strategy shows a slightly lower average outbreak sizes with the preventive efficiency of 81\% in the DDT network and 85\% in the GDT network at $P=0.2$\% while they are 78\% and 84\% in the IMV strategy. 
However, if $P$ is increased, these variations compared to the IMV strategy is minimised. At $P=0.8$\%, the average outbreak sizes close to one for all versions of IMV strategy. Thus, it is concluded that the inclusion of temporal information (stay duration) does not have a substantial impact when the requirement of preventive efficiency is high and a high vaccination rate is applied. The coarse-grained contact information is sufficient to capture the individual movement information. 

\begin{figure}[h!]
\includegraphics[width=0.5\linewidth, height=5.0 cm]{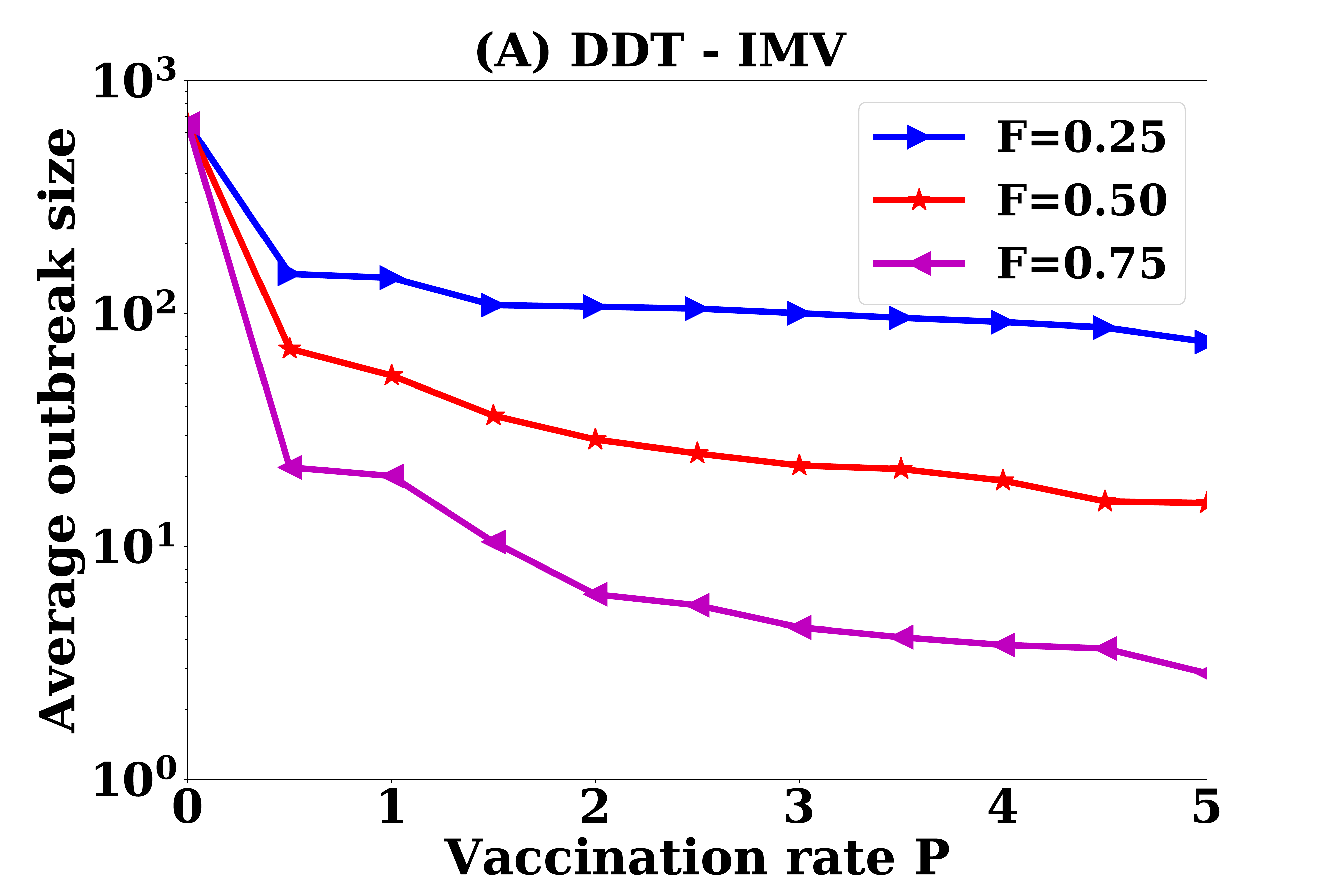}
\includegraphics[width=0.5\linewidth, height=5.0 cm]{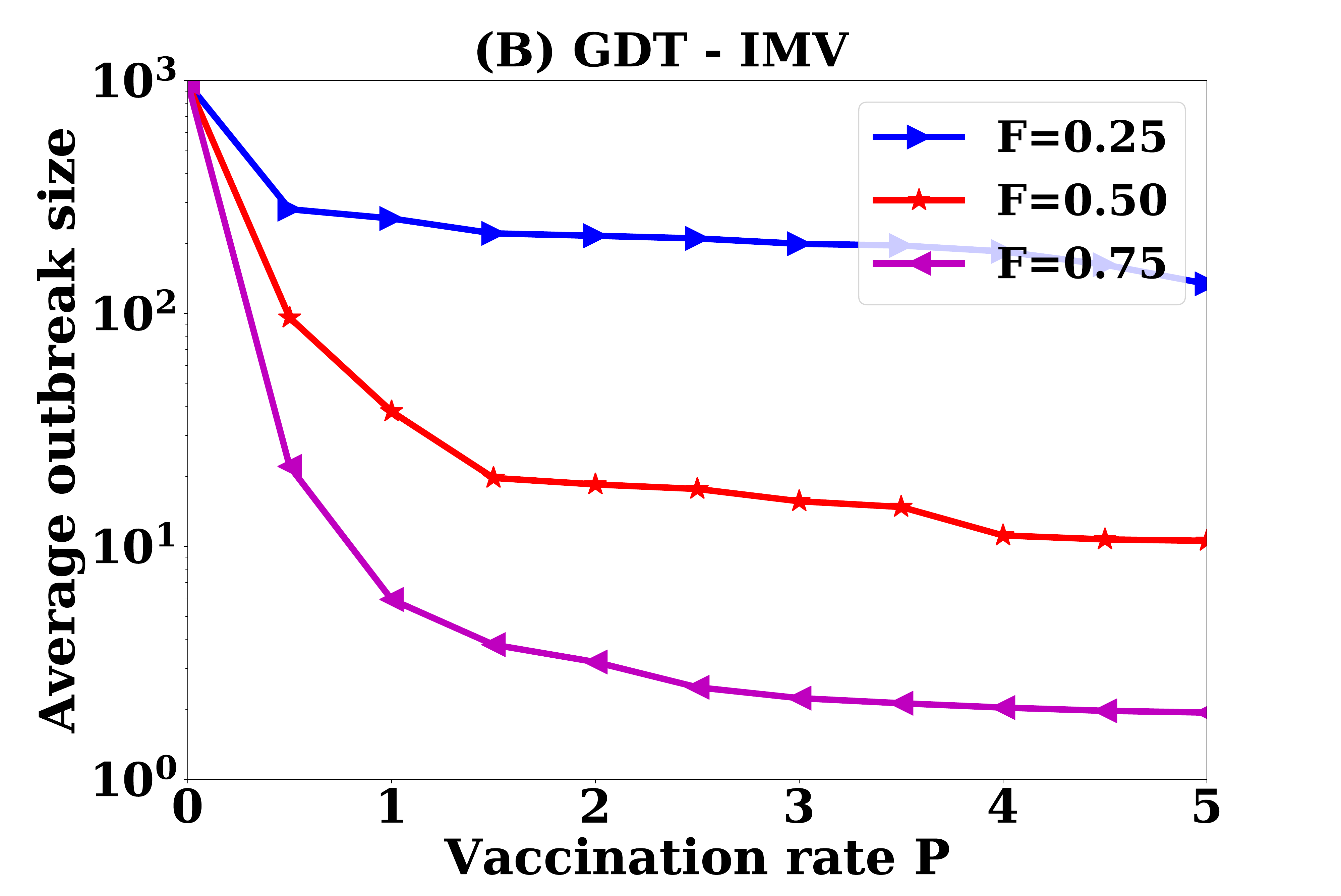}\\
\includegraphics[width=0.5\linewidth, height=5.0 cm]{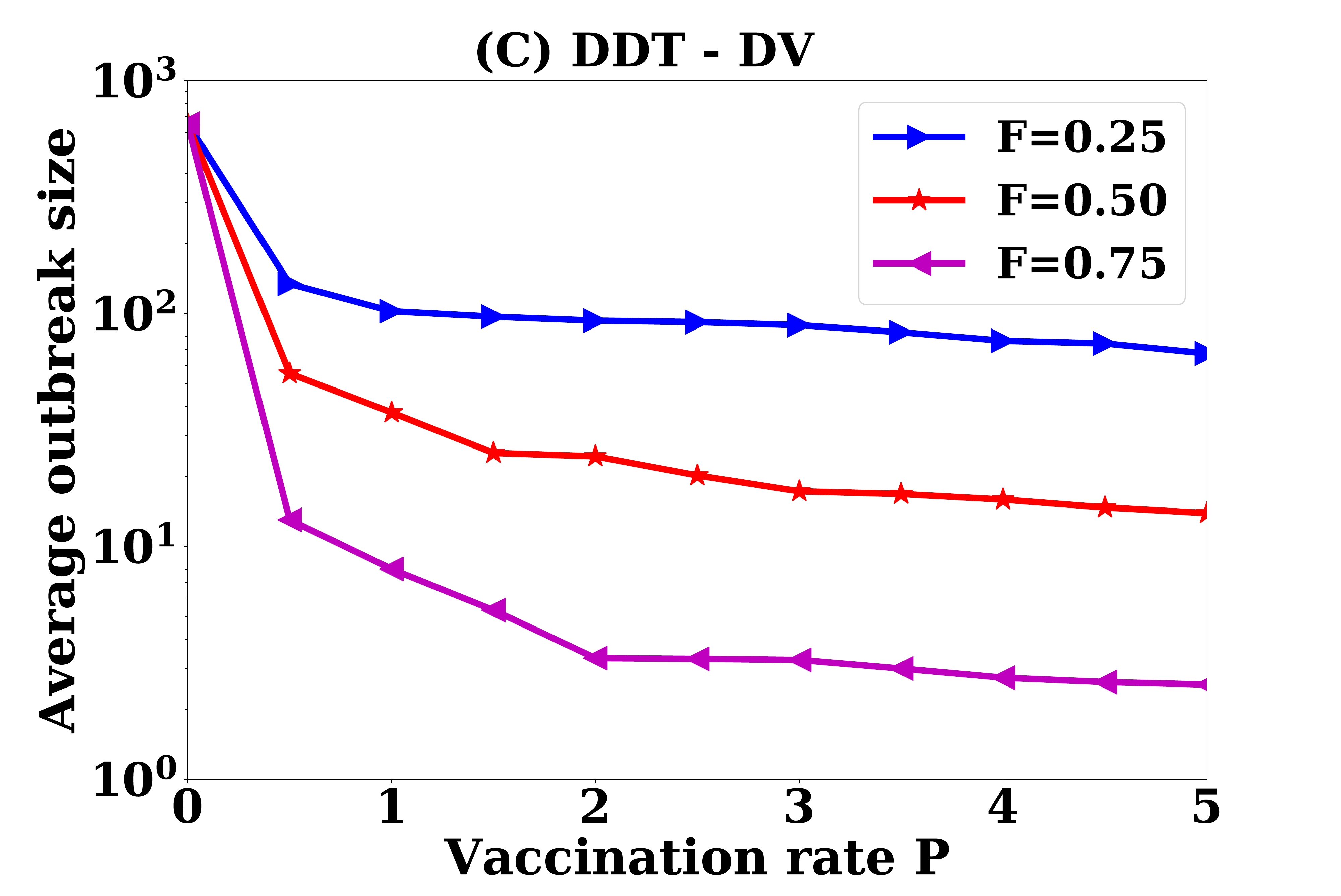}
\includegraphics[width=0.5\linewidth, height=5.0 cm]{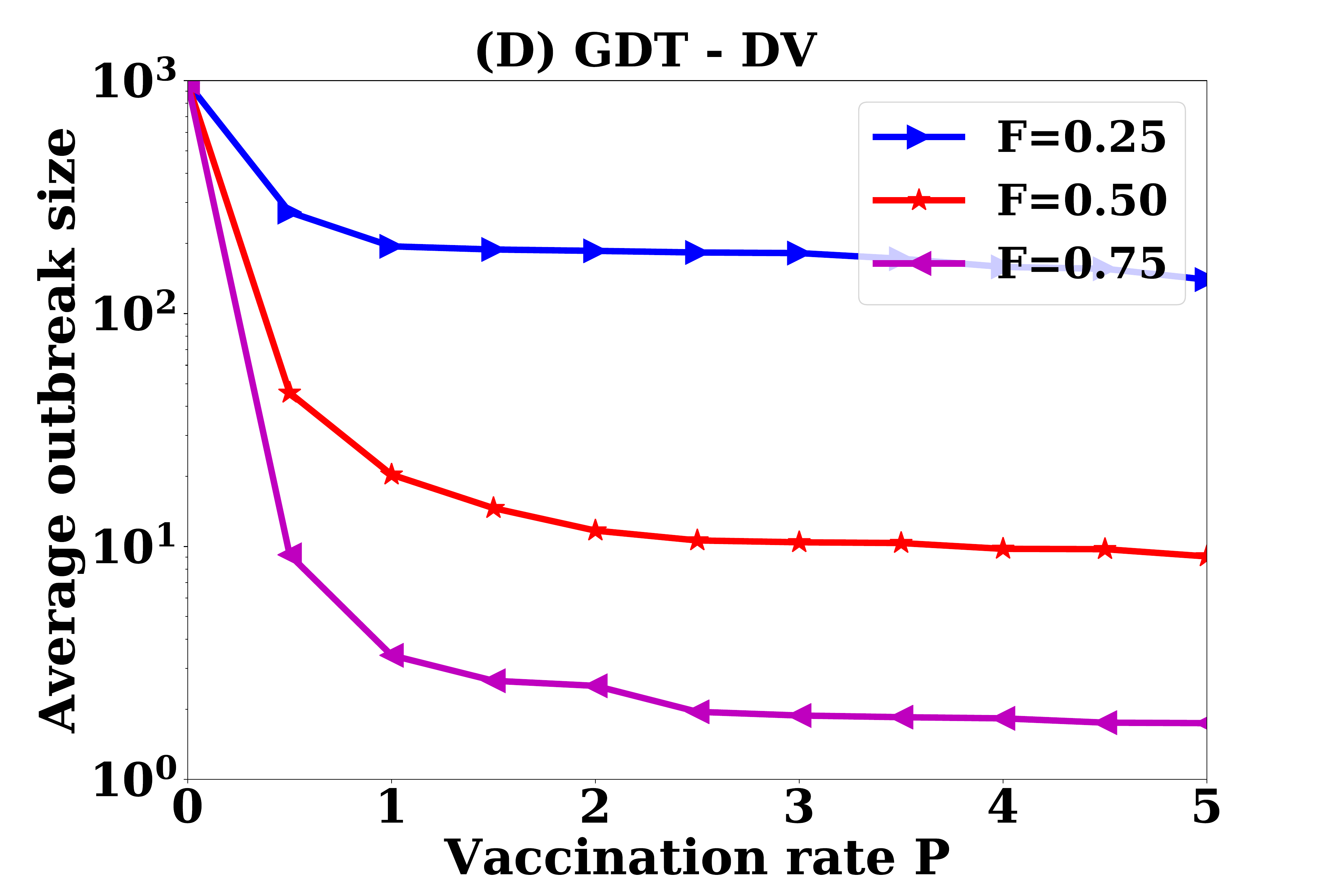}\\
\includegraphics[width=0.5\linewidth, height=5.0 cm]{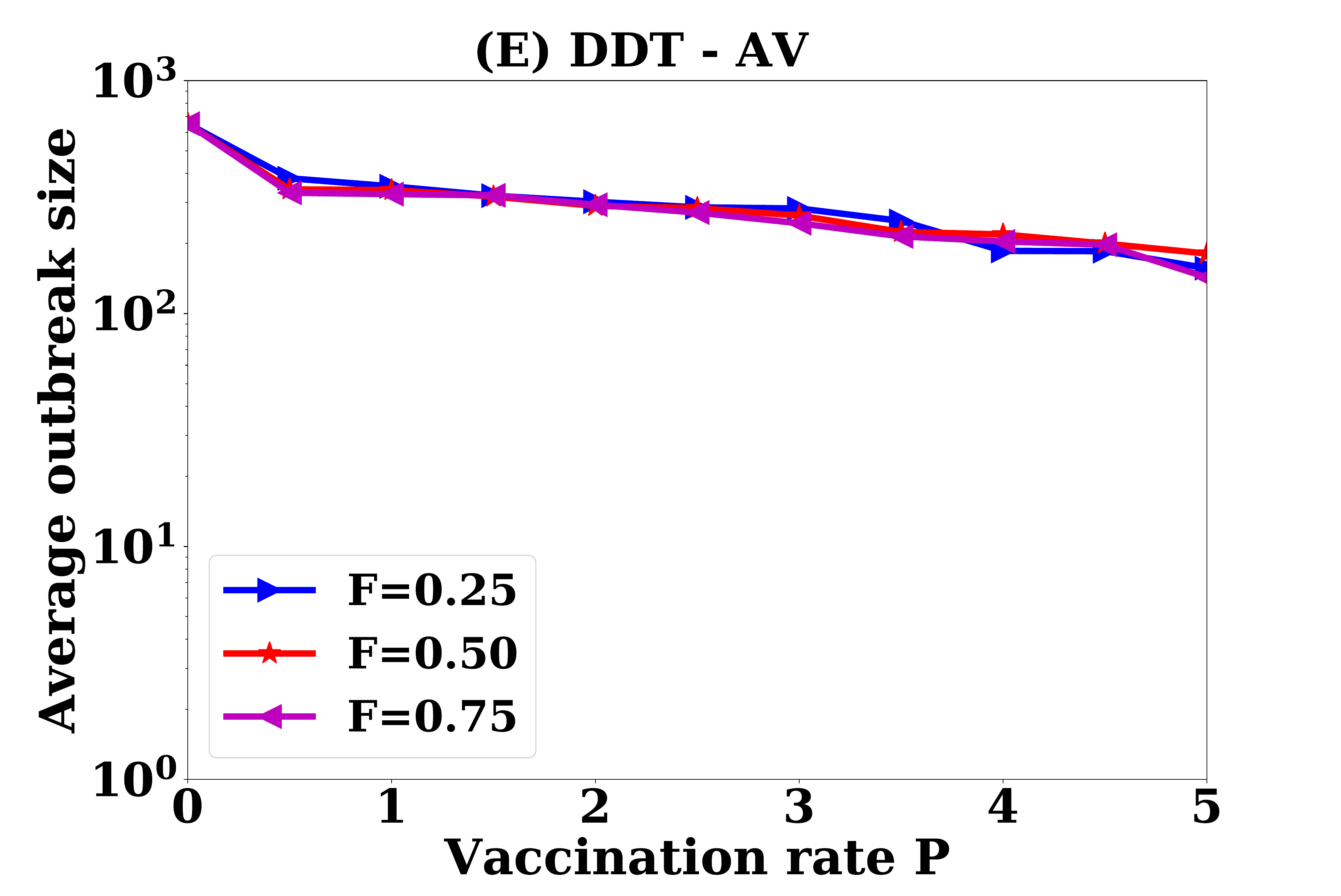}
\includegraphics[width=0.5\linewidth, height=5.0 cm]{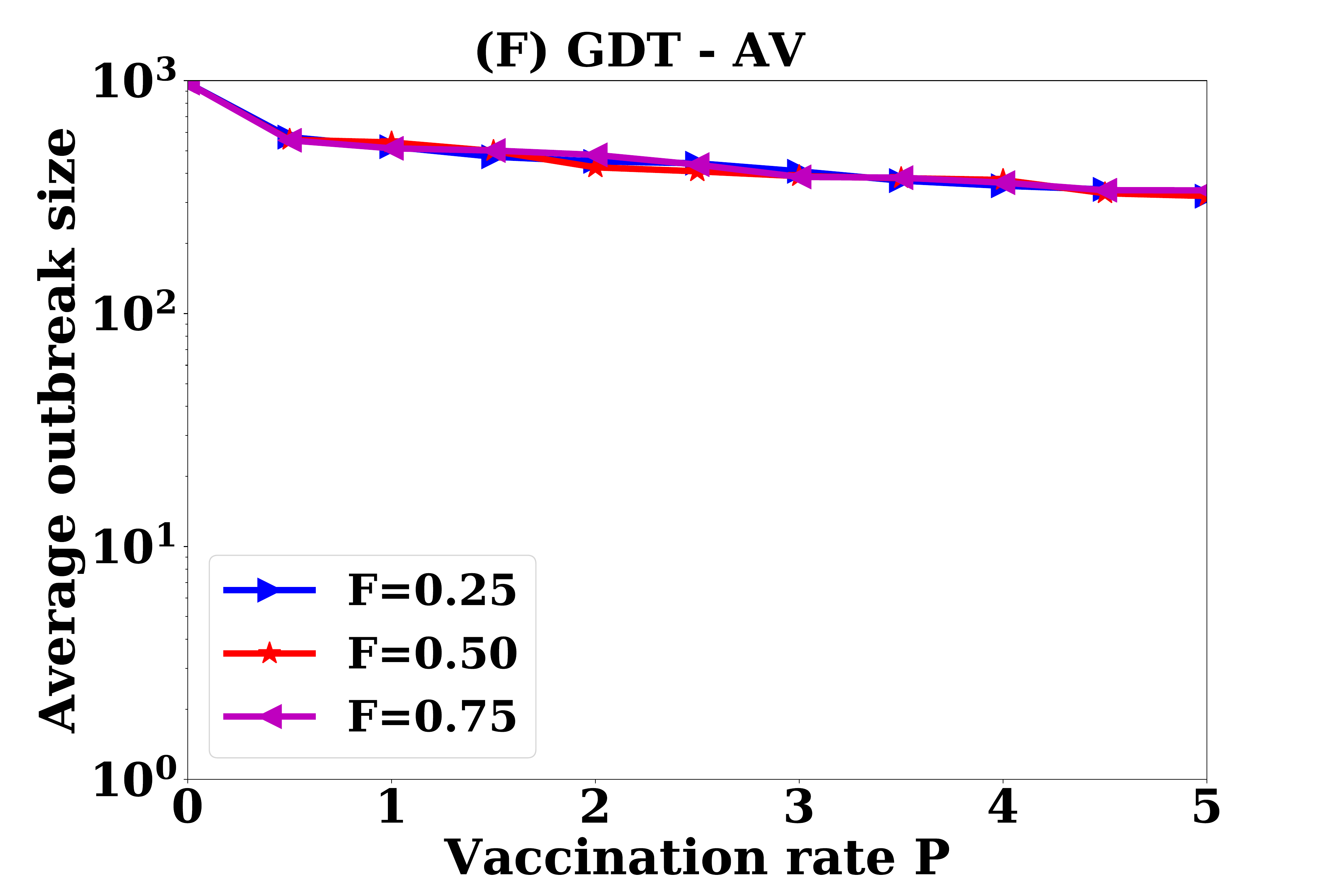}

\caption{Performance of vaccination strategies at various scale of information availability $F$: (A,B) proposed vaccination strategy, (C,D) degree based vaccination strategy and (E,F) acquaintance vaccination strategy}
\label{fig:sens}
\end{figure} 

\begin{figure}[h!]
\includegraphics[width=0.49\linewidth, height=5.5 cm]{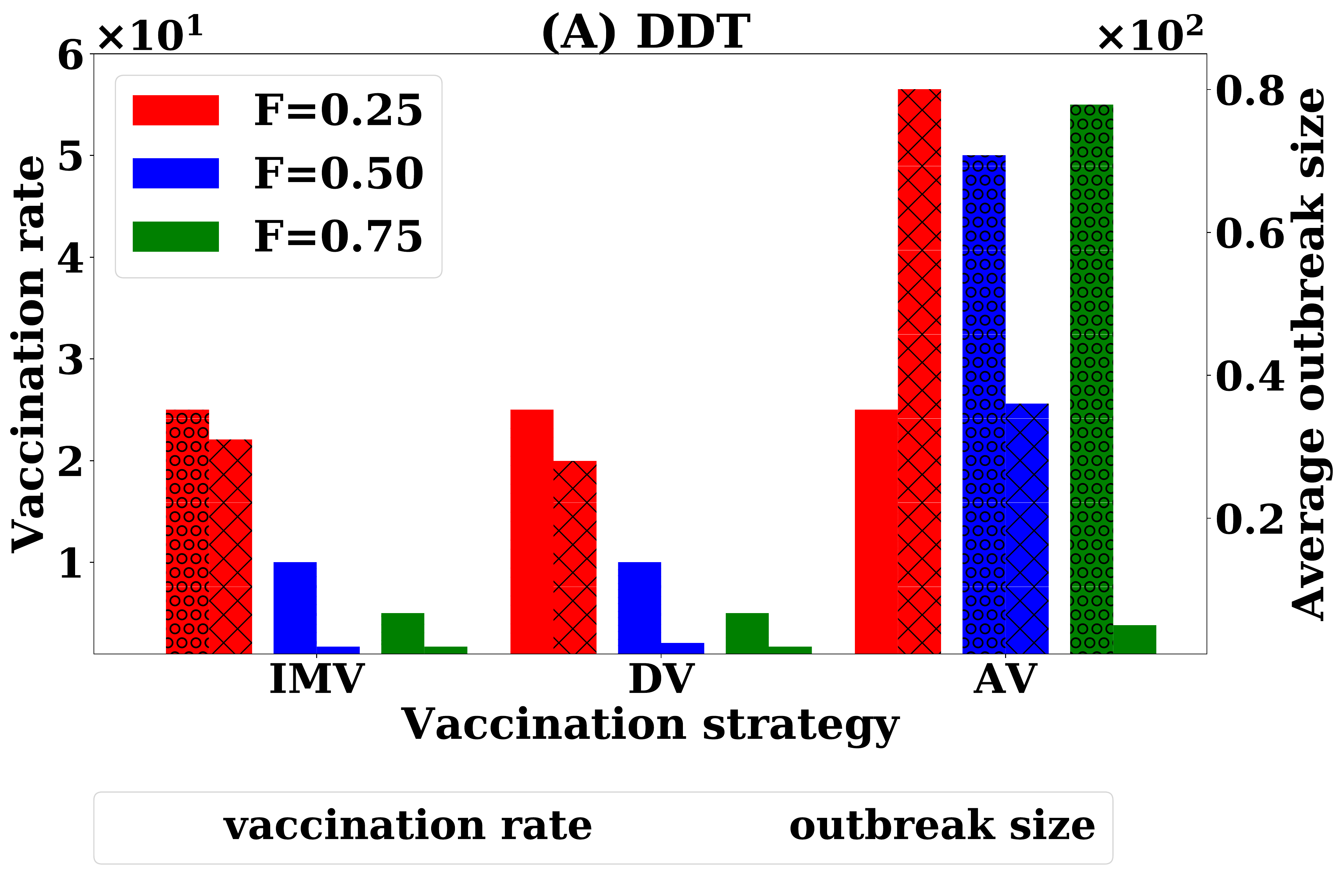} \quad
\includegraphics[width=0.49\linewidth, height=5.5 cm]{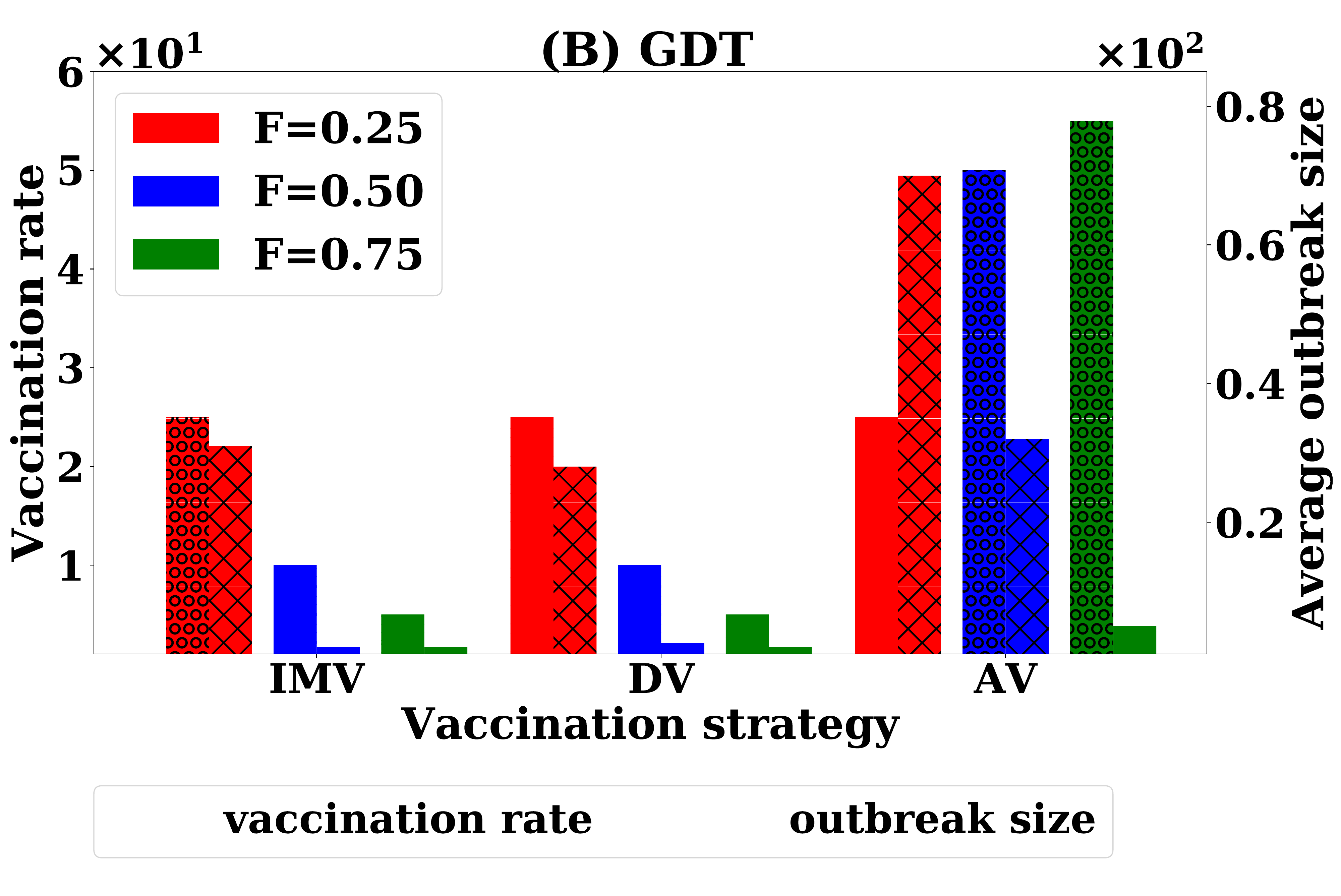}\\ [3ex]
\includegraphics[width=0.49\linewidth, height=5.5 cm]{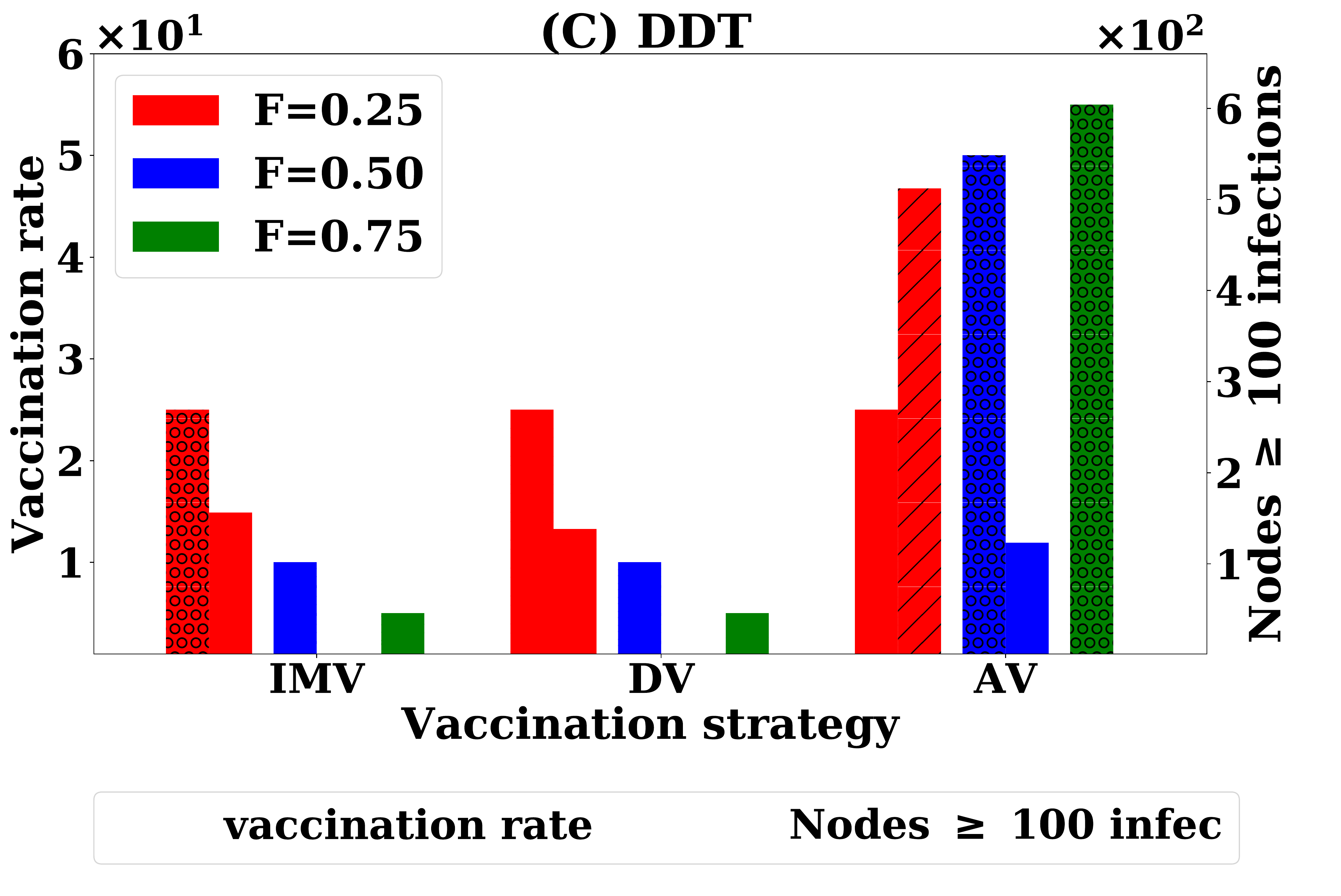}\quad
\includegraphics[width=0.49\linewidth, height=5.5 cm]{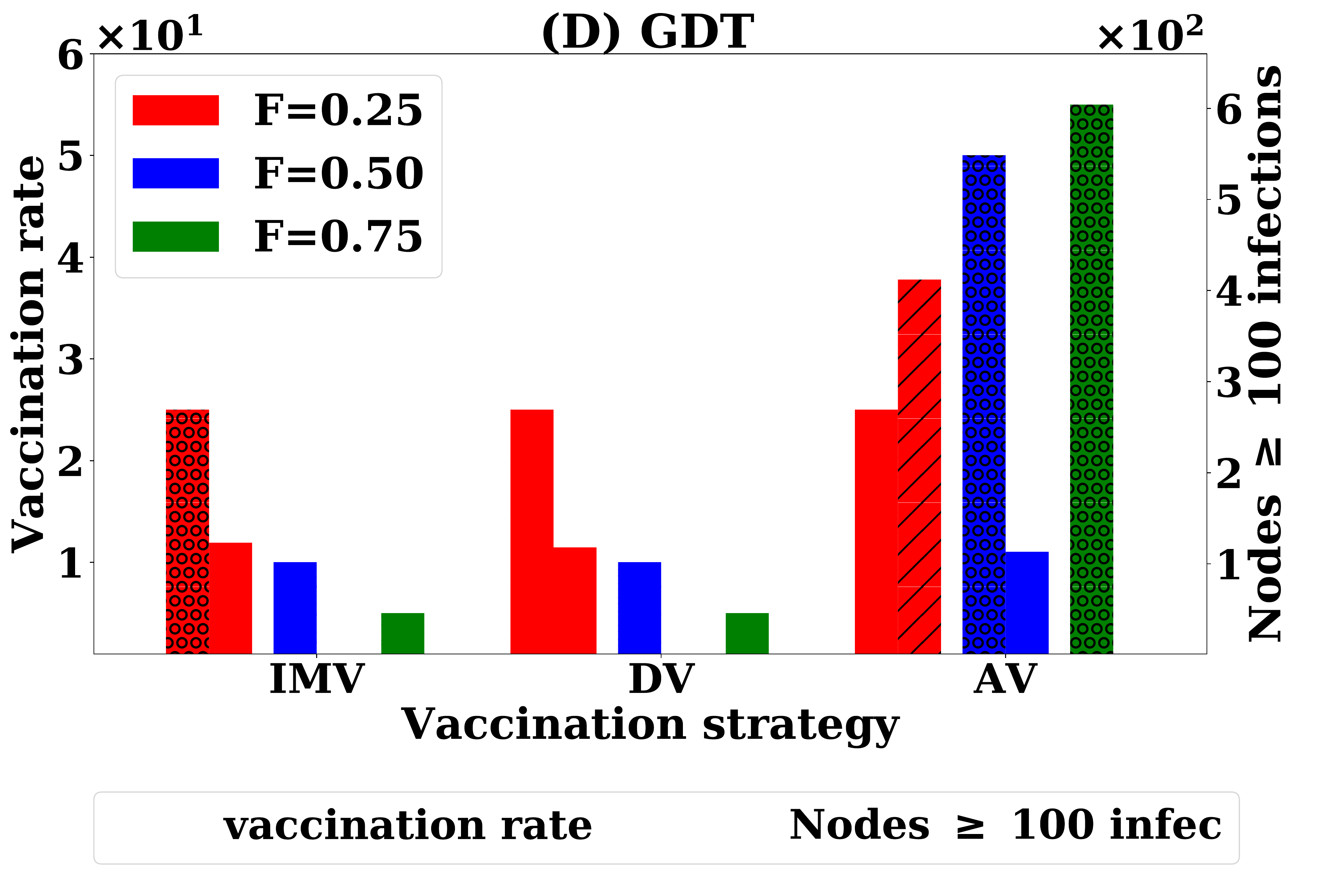}

\caption{Trade-off among information collection cost, vaccination cost and infection cost for various strategy - vaccination rate is on the left y-axis while others are on the right y-axis: (A, B) average outbreak sizes with vaccination rates for different $F$, and (C, D) number of seed nodes having outbreak greater than 100 infections with $P$ and $F$}
\label{fig:sens3}
\end{figure}

In the previous experiments, the upper bound of the efficiency for vaccination strategies was studied assuming that contact information of all nodes can be collected. However, it is not possible to collect movement information of all individuals in real world scenarios. Thus, it is required to understand the effectiveness of the strategies with the scale of information collected regarding nodes contact. Therefore, this experiment analyses the efficiency of the strategies varying the proportion $F$ of nodes that are picked for collecting their movement information. The required vaccination rates $P$ for a strategy also depend on the value of $F$. In the simulations, a proportion $F$ of nodes are picked up randomly from the first seven days of networks and their ranks are calculated based on the applied vaccination strategy. Then, a vaccination rate $P$ is implemented with the sampled nodes. The impact of scale of information availability is studied for varying $F$ to 0.25, 0.50 and 0.75. The efficiency of strategies is first analysed varying $P$ from 0.5 to 5\% with the step of 0.5\%. The simulations are conducted for IMV, DV and AV strategies on both the DDT and GDT networks. The average outbreak sizes are presented in Figure~\ref{fig:sens}. Then, the simulations are also run for larger values of $P\geq 5$\% until the preventive efficiency is achieved where no seed node has the outbreak of more than 100 infections. The results are presented in Figure~\ref{fig:sens3} showing the trade-off between the performance and vaccination rates with information availability scale $F$.

The results show that there is a trade-off among vaccination rate, infection cost and information collection cost for implementing a vaccination strategy. If the contact information of $F=0.25$ proportion nodes is applied to select the nodes to be vaccinated, the average preventive efficiency is about 85\% for IMV and DV strategy in the DDT network and 87\% in the GDT network even at the vaccination rate $P=5$\% (Figure~\ref{fig:sens}). Then, the vaccination rate is increased to $P=25$\% for understanding maximum efficiency at $F=0.25$. There is still substantially high average outbreak sizes in both DDT and GDT networks. Besides, 3.3\% of seed nodes have outbreak sizes more than 100 infections and 3\% in GDT network. With the information collection cost $F=0.25$, therefore, vaccination cost is high along with less protection of infection. If $F$ is increased to $F=0.50$, the preventive efficiency increases to 97\% in GDT network for IMV strategy while it is 95\% in DDT network. However, the degree based vaccination (DV) shows high preventive efficiency compared to that of IMV strategy with lower $P$. But, the average outbreak sizes are not below 20 infections within the vaccination rate of $P=5$\%. However, the preventive efficiency with no seed nodes having outbreak sizes greater than 100 infections is achieved vaccinating 10\% of nodes in both networks. At $F=0.5$, the strong protection from infection can be achieved with the vaccination cost of 10\%. Further increase of $F$ to 0.75 makes possible to reduce outbreak sizes below 10 infections in both networks for DV strategy at $P=2$\%. The number of nodes having outbreak sizes greater than 100 infections become zero for DV strategy. On the other hand, the proposed IMV model reduces outbreak sizes below 10 infections at $P=2$\% in GDT network and at $P=3$\% in DDT network. The number of nodes having outbreak sizes greater than 100 infections become 3 nodes at $P=3$\%. With $F=0.75$, the vaccination cost is about 3\% in IMV strategy while it was 1\% for $F=1$. For AV strategy, average outbreak sizes is very high up to $F=0.5$ and there is a large number of seed nodes having outbreak sizes greater than 100 infections. However, the strong preventive efficiency is achievable at $F=0.75$ with vaccinating of 55\% nodes where no seed node has outbreak greater than 100 infections. With the low scale of information, the IMV strategy performs better than AV strategy and close to DV strategy.

\subsection*{Post-outbreak vaccination}
The proposed vaccination strategy is now studied for the post-outbreak scenarios which can be implemented in two ways. The first way is similar to the mass vaccination and is called population-level vaccination where a proportion $P$ of the population is randomly chosen and is vaccinated. In the simplest version, behaviours of the selected individuals are not considered, randomly picked up to vaccinate, and hence information collection cost is minimal. However, the resource cost is often too high as it requires to vaccinate a large number of individuals. This approach is improved by vaccinating individuals who have specific characteristics relevant to disease spreading. For example, if an individual has interactions with many other individuals, he is chosen to be vaccinated (degree-based vaccination). In this section, the applied three ranking methods (IMV, DV and AV) of nodes from the previous sections are used to select the nodes to be vaccinated. In addition, random vaccination (RV) method is also examined along with these three ranking based methods. 

In the second way of post-outbreak vaccination implementation, the infected individuals are at the focus point where susceptible individuals who have contact with infected individuals are vaccinated to hinder further spreading of disease. Ring vaccination is one of the strategies from this class. This approach of implementing post-outbreak vaccination is named as node level vaccination. In the ring vaccination, a proportion $P$ of neighbour nodes are vaccinated and the neighbour nodes can be chosen randomly or based on a specific criterion. The two ranking methods (IMV and DV) are applied to select neighbours in the ring vaccination and the performance of node level vaccination is studied. The performance is compared with the strategy of random neighbour selection (RV). Both population level and node level vaccination are investigated in this section on both DDT and GDT contact networks. The population level vaccination is also examined with the scale $F$ of information availability on the node's contact. For node level vaccination, all infected node might not be identified. Thus, the performance is analysed if only a proportion $F$ of infected nodes are identified only.

\subsubsection*{Population level vaccination}
\begin{figure}[h!]
\includegraphics[width=0.5\linewidth, height=5.0 cm]{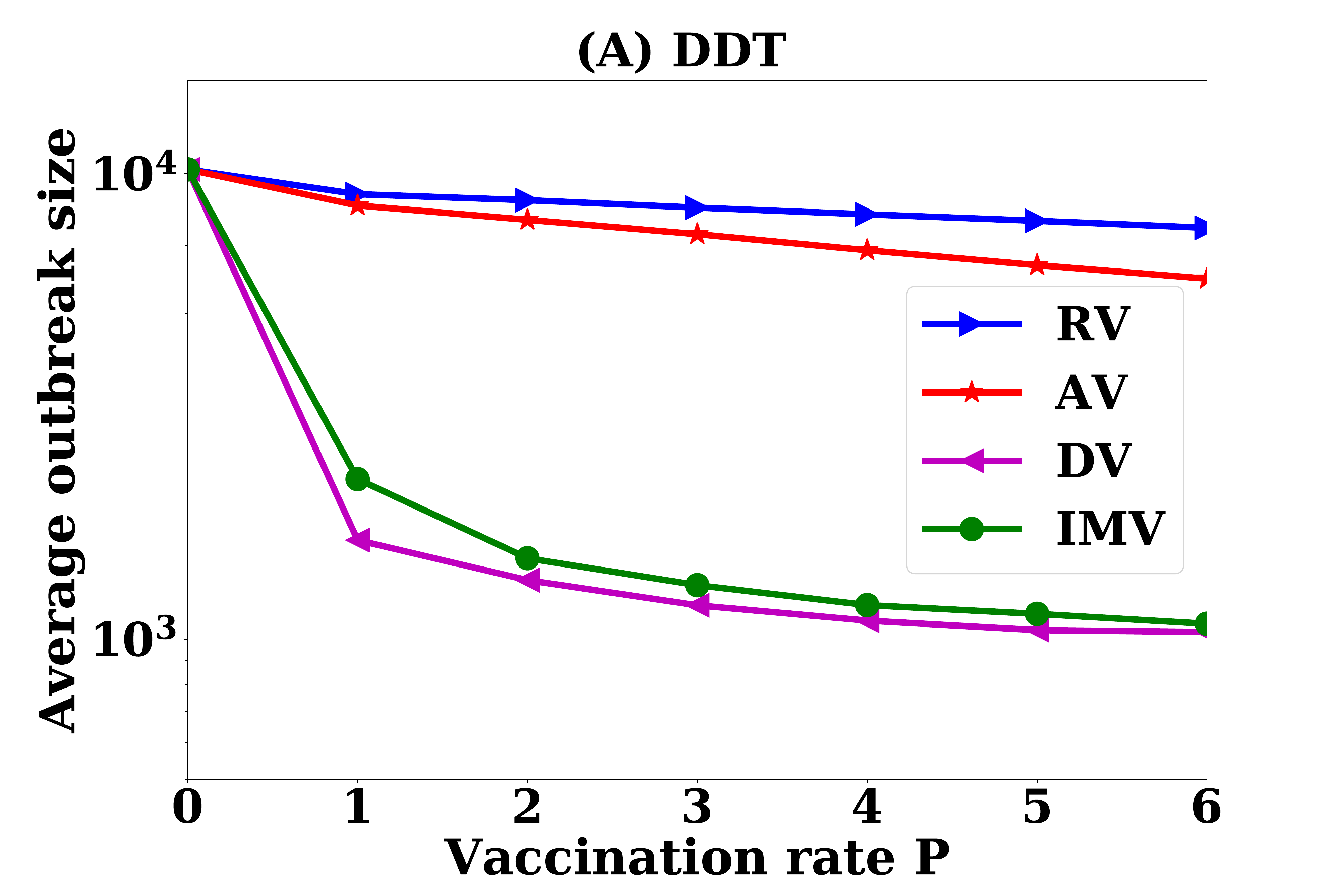}
\includegraphics[width=0.5\linewidth, height=5.0 cm]{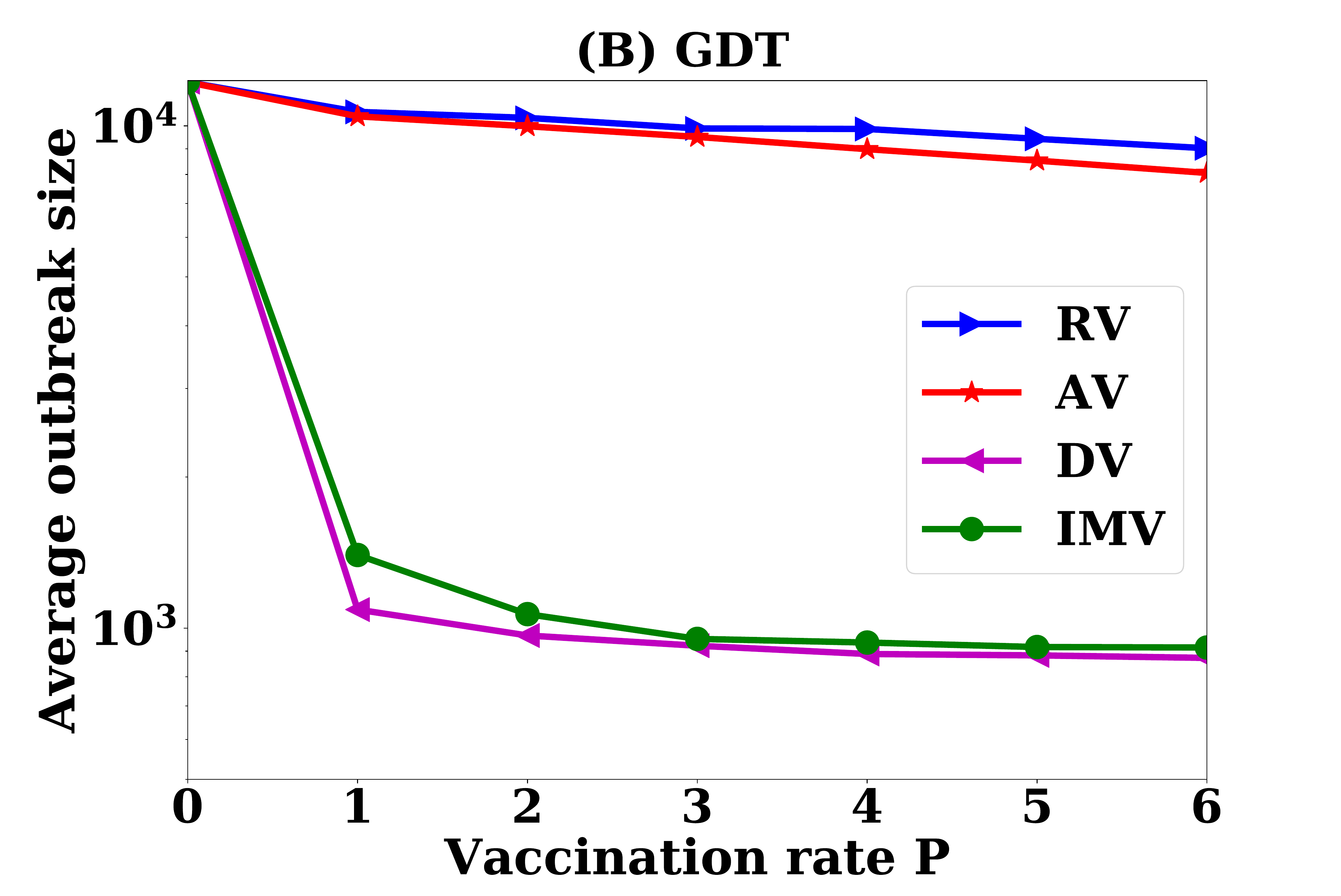}\\
\caption{Average outbreak sizes for different vaccination strategies at various vaccination rates $P$ in post-outbreak vaccination}
\label{fig:vmpf}
\end{figure}

In these simulations, disease starts with 500 seed nodes and continues for 42 days. Nodes are infectious for $\tau$ days randomly chosen in the range [3-5] days. The vaccination is implemented at the 7th day of simulation assuming that these days are required to notice the emergence of disease and to collect the contact information. The node's rank, based on the applied vaccination strategy, is calculated from the contact data of the first seven days. Then, a proportion $P$ of nodes are vaccinated (assigned recovered status) and final outbreak sizes are calculated for 42 days of simulations. For each vaccination strategy, simulations are conducted for different $P$ in the range [1,6]\% with a step of 1\%. At each value of $P$, the simulations are run for 1000 times and the average outbreak sizes are presented in Figure~\ref{fig:vmpf}.  

The simulation of disease spreading without vaccination makes outbreak of on average 10K infections in the DDT network and 12K infections in the GDT network. This average outbreak size is used as the reference to understand the efficiency of the applied vaccination strategy. Random vaccination (RV) strategy does not reduce the average outbreak sizes significantly in both networks (Fig.~\ref{fig:vmpf}). At the vaccination rate $P=1$\% (vaccinating 3600 nodes), the average outbreak size is 9K infections in the DDT network and 10K infections in the GDT network. On changing $P$ from 1\% to 6\% (vaccinating 21600 nodes), there is a 10\% reduction in the average outbreak size for RV strategy. If the ranking score of acquaintance vaccination (AV) is applied to vaccinate nodes, the reduction in outbreak sizes slightly increases. At $P=1$\%, the average outbreak size is similar to that of RV strategy. If $P$ is, however, increased to 6\%, the average outbreak size is reduced by 20\% in the DDT network and 15\% in the GDT network. Within the range of vaccination rates, RV and AV strategies fail to contain the disease spreading. Increasing vaccination rates further for RV and AV strategies show that the RV strategy requires 75\% of nodes to be vaccinated in both networks and the AV strategy requires vaccination of 40\% nodes in DDT network and 35\% in GDT network to contain disease spreading within the outbreaks of 1K infections i.e to reduce outbreak sizes by 90\%.

The proposed vaccination strategy (IMV) shows the significant improvements in reducing the final average outbreak sizes compared to the AV and RV strategies (Fig.~\ref{fig:vmpf}). With vaccinating 1\% of nodes (vaccinating 3600 nodes), the average outbreak sizes are reduced by 80\% in both networks. If the vaccination rate $P$ is increased to 4\% in the DDT network, the outbreak size are reduced by 90\% but a further increase in $P$ does not reduce outbreak sizes substantially. In the GDT network, the average outbreak sizes are reduced by 90\% infections with 3\% (vaccinating 10800 nodes) vaccination. The degree based vaccination shows similar performance to that of IMV strategy. Similar to the preventive vaccination, the DV strategy initially shows better performance and then becomes similar to the IMV strategy at the higher vaccination rates. The coarse-grained information based IMV strategy achieves the same performance of DV strategy for the post-outbreak vaccination scenarios as well.   

\begin{figure}[h!]
\includegraphics[width=0.5\linewidth, height=5.0 cm]{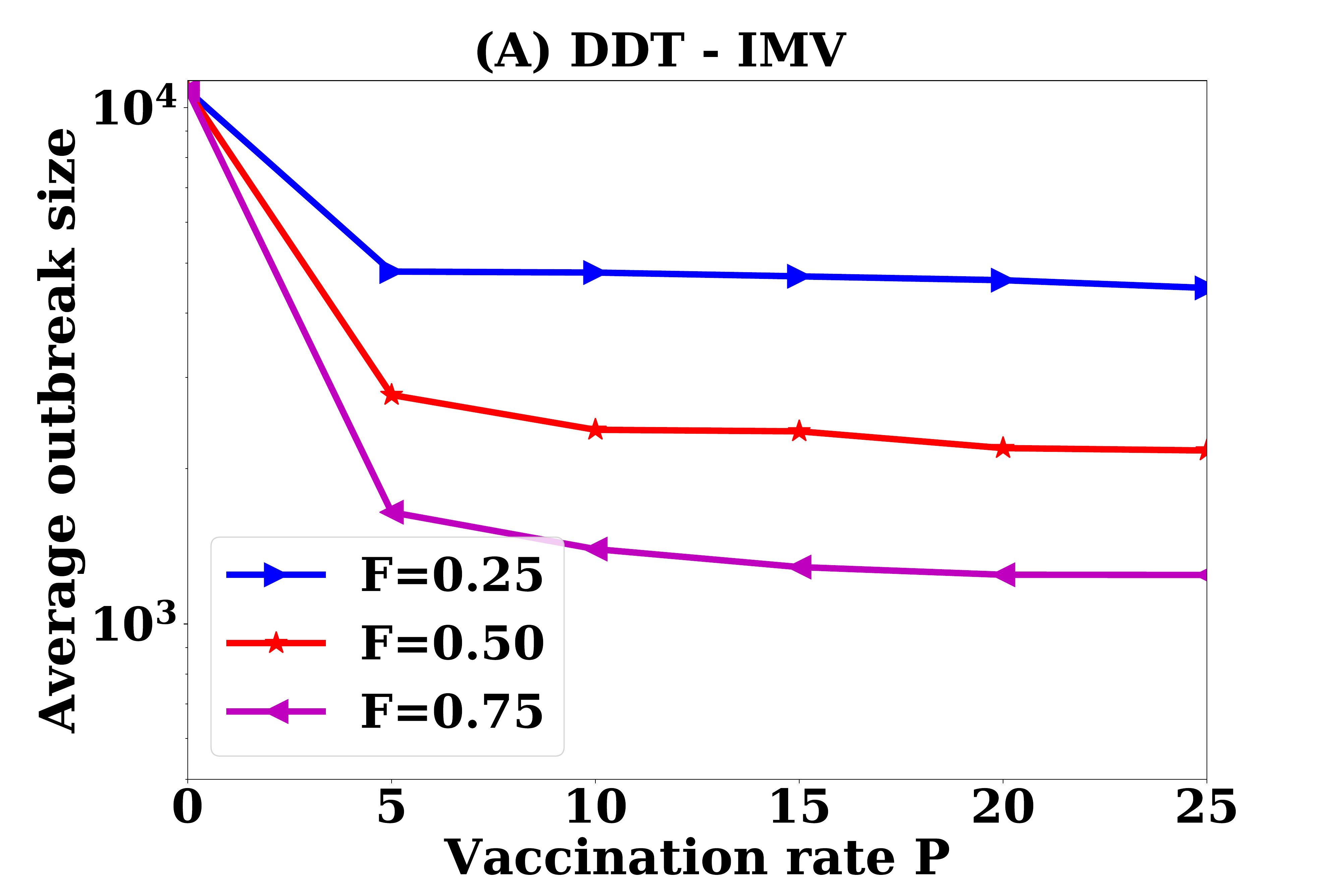}
\includegraphics[width=0.5\linewidth, height=5.0 cm]{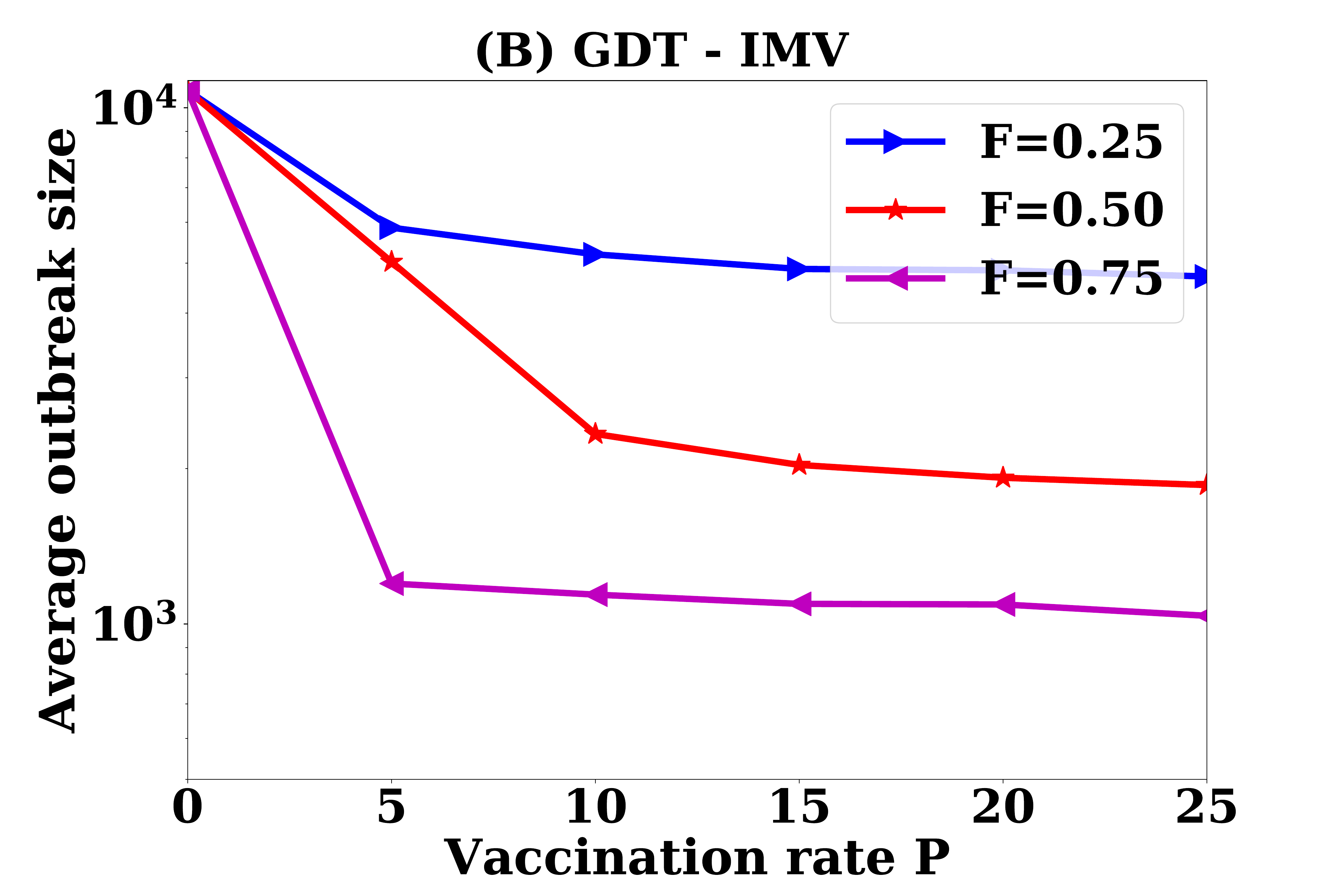}\\
\includegraphics[width=0.5\linewidth, height=5.0 cm]{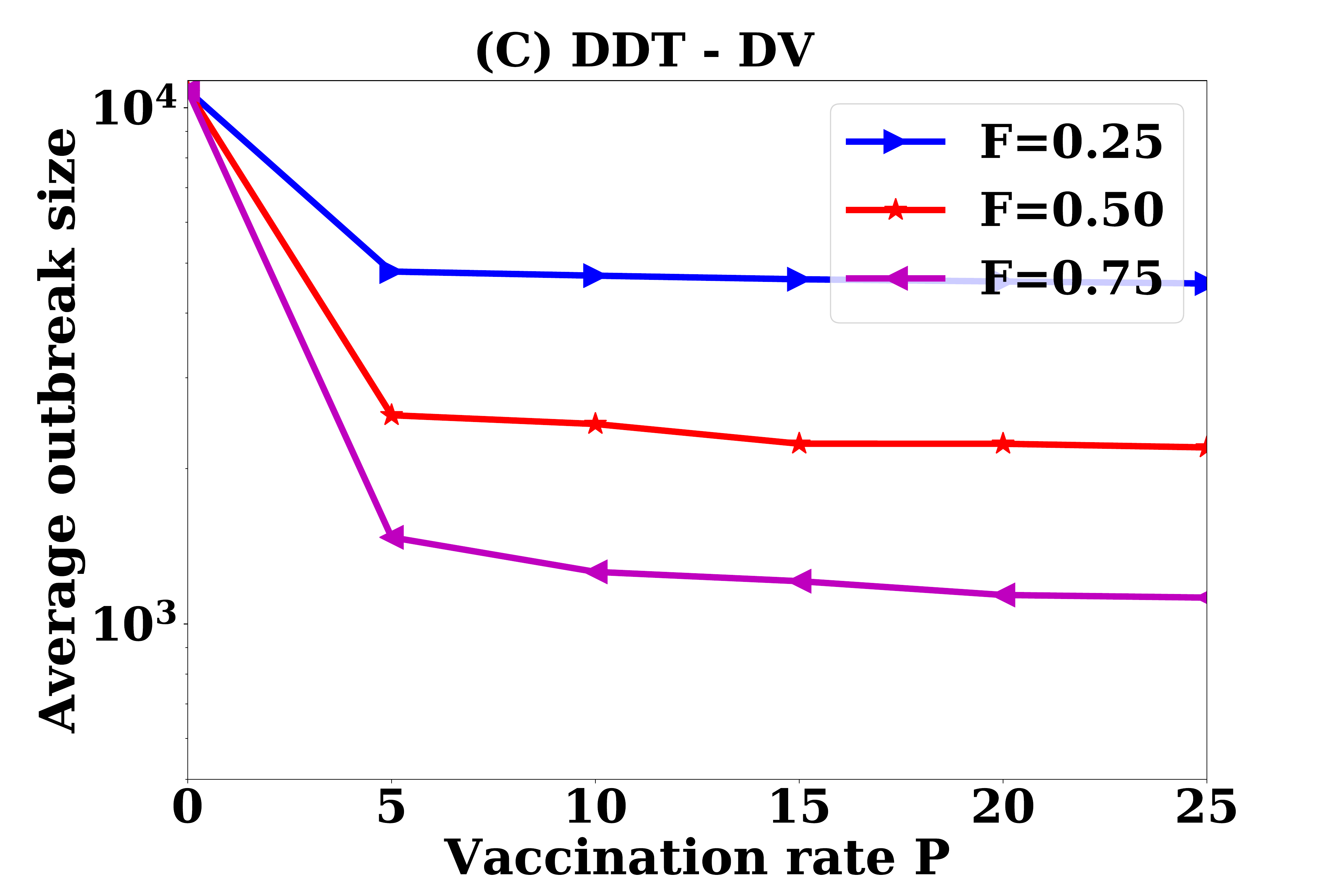}
\includegraphics[width=0.5\linewidth, height=5.0 cm]{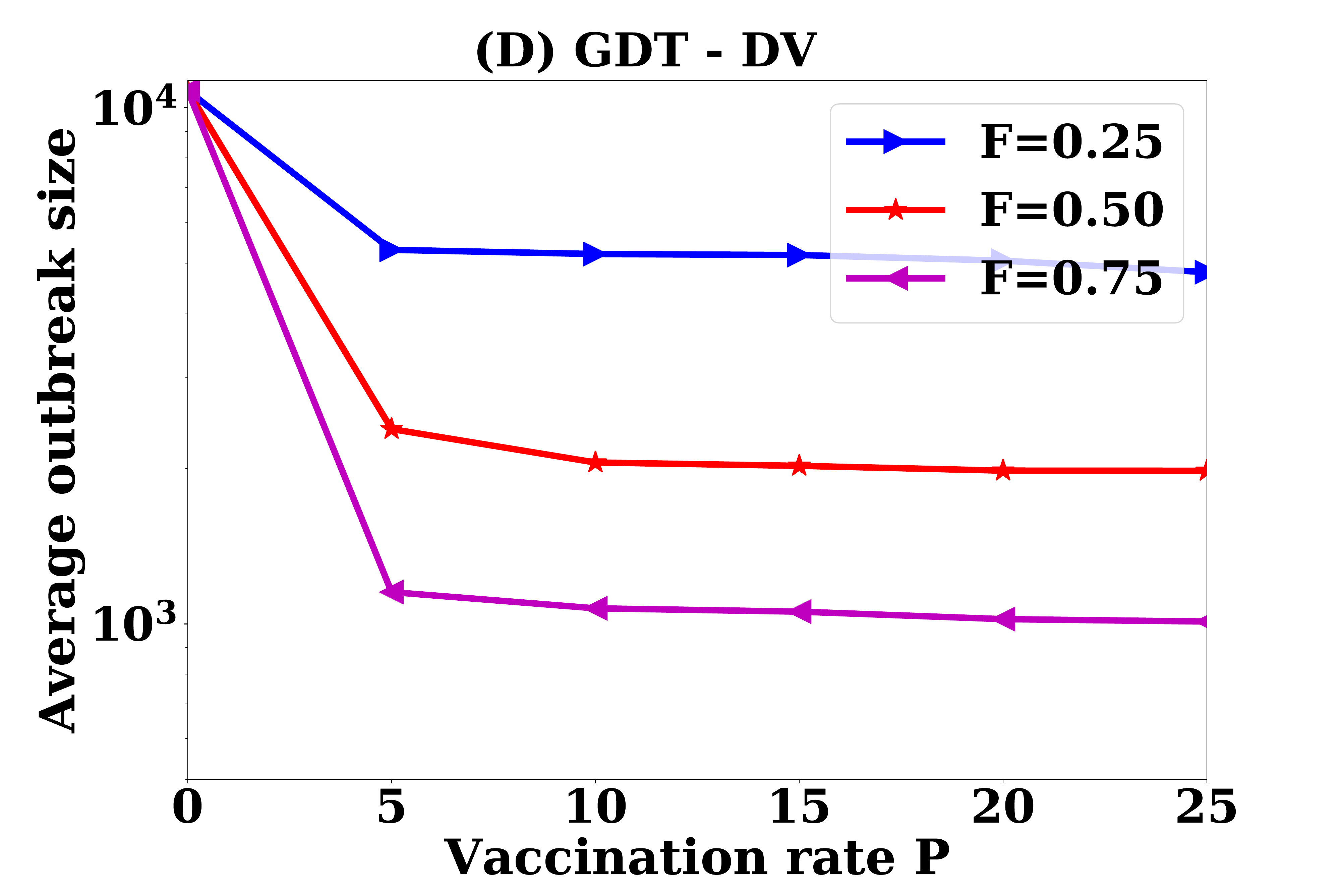}\\
\includegraphics[width=0.5\linewidth, height=5.0 cm]{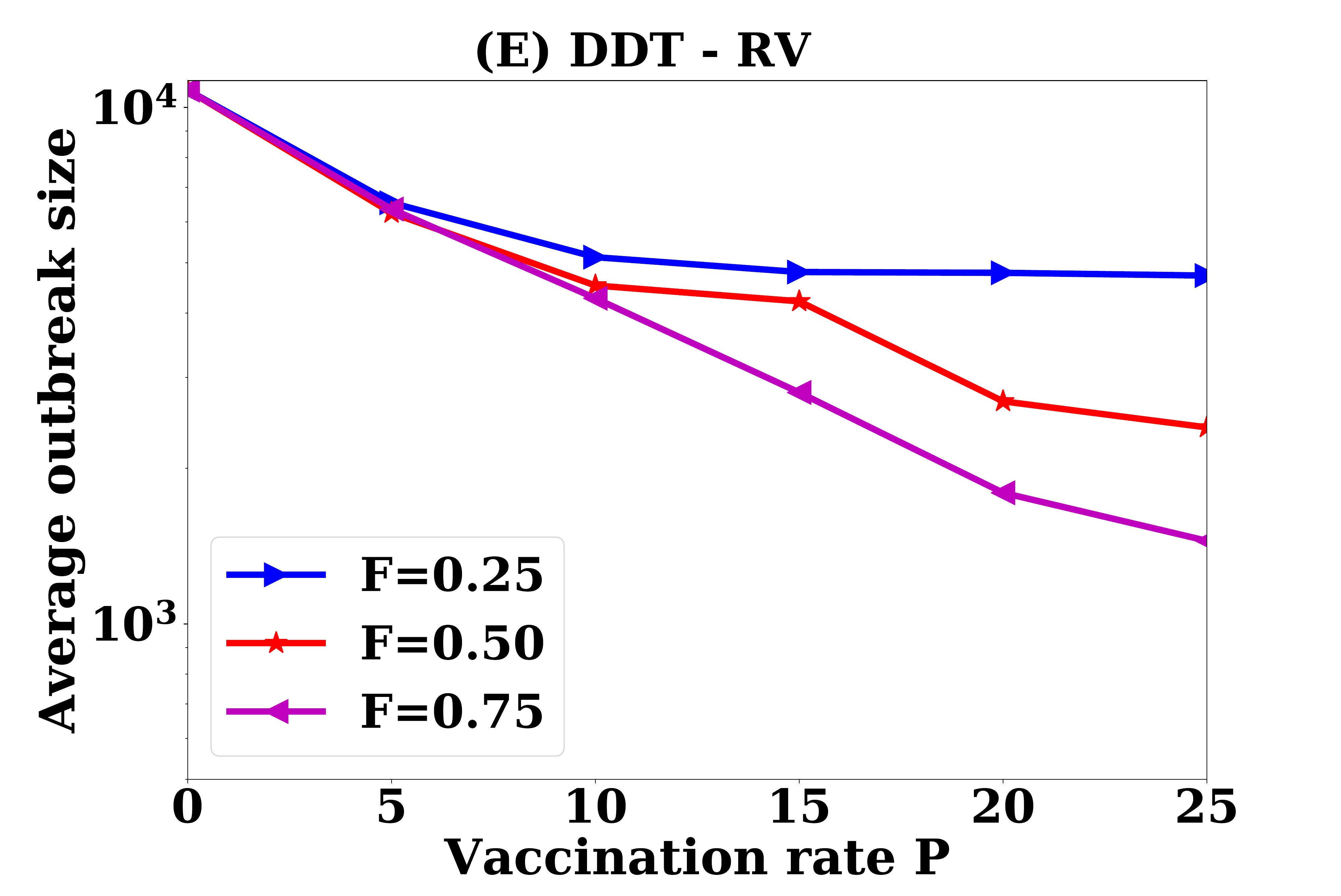}
\includegraphics[width=0.5\linewidth, height=5.0 cm]{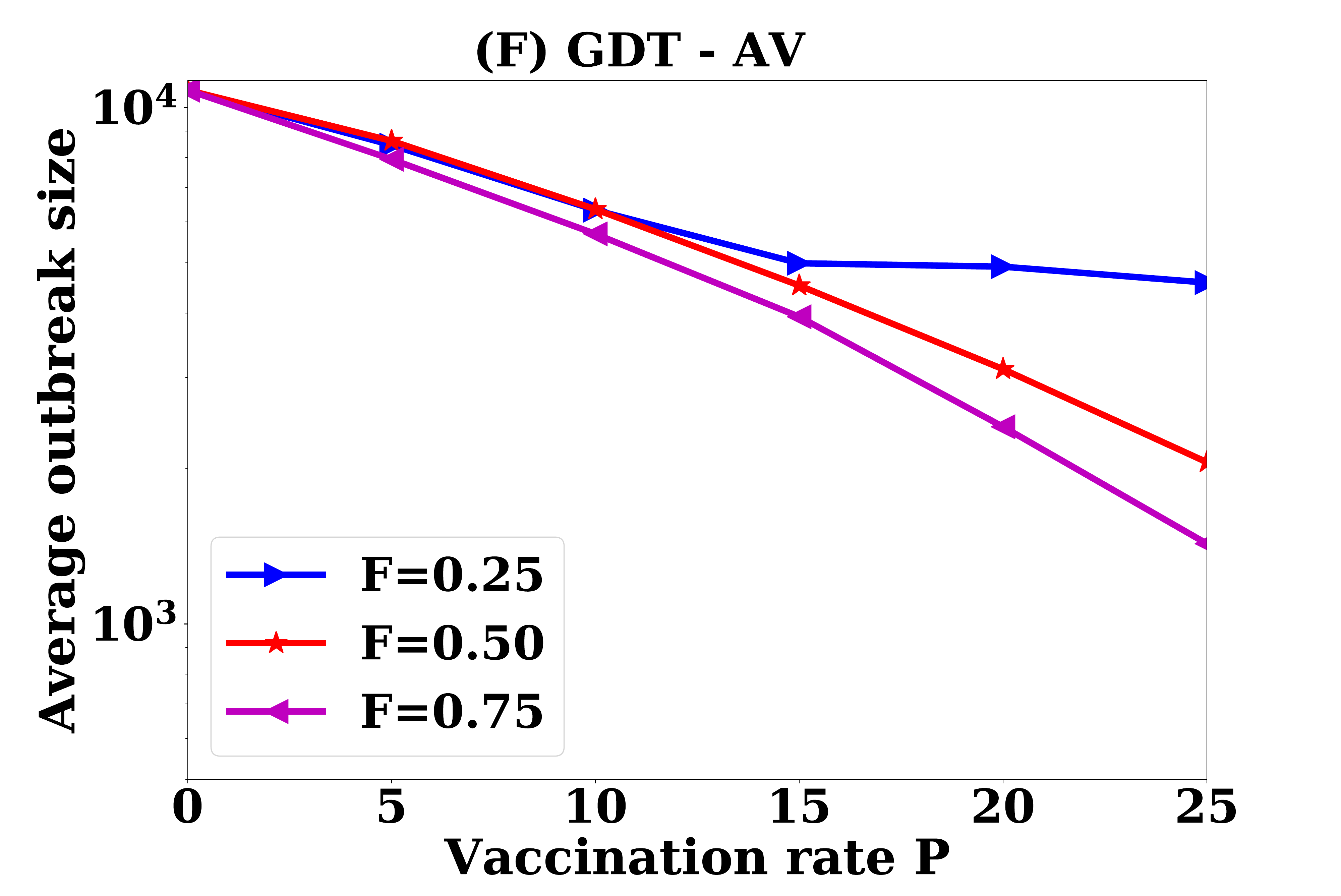}

\caption{Variation in post-outbreak performances with scale of information availability of node contacts: (A,B) proposed vaccination strategy, (C,D) degree based vaccination, and (E,F) acquaintance based vaccination}
\label{fig:fpvac}
\end{figure} 

\begin{figure}[h!]
\begin{tikzpicture}
    \begin{customlegend}[legend columns=2,legend style={at={(0.32,1.00)},draw=none,column sep=3ex ,line width=6pt,font=\small}, legend entries={vaccination rate, outbreak size}]
    \addlegendimage{solid, color=red}
    \addlegendimage{color=blue}
    \end{customlegend}
 \end{tikzpicture}
\centering
\includegraphics[width=0.495\linewidth, height=5.0 cm]{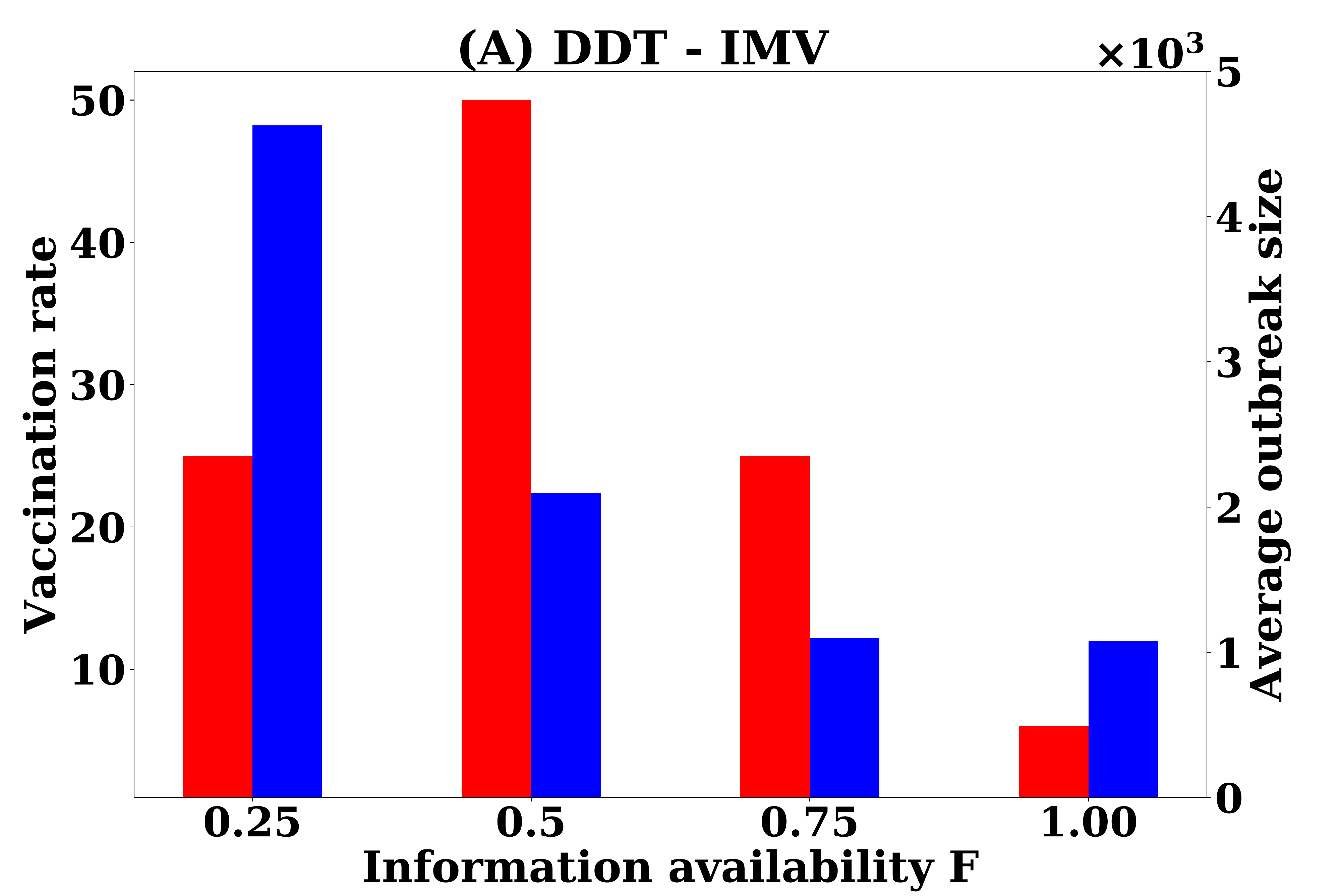}
\includegraphics[width=0.495\linewidth, height=5.0 cm]{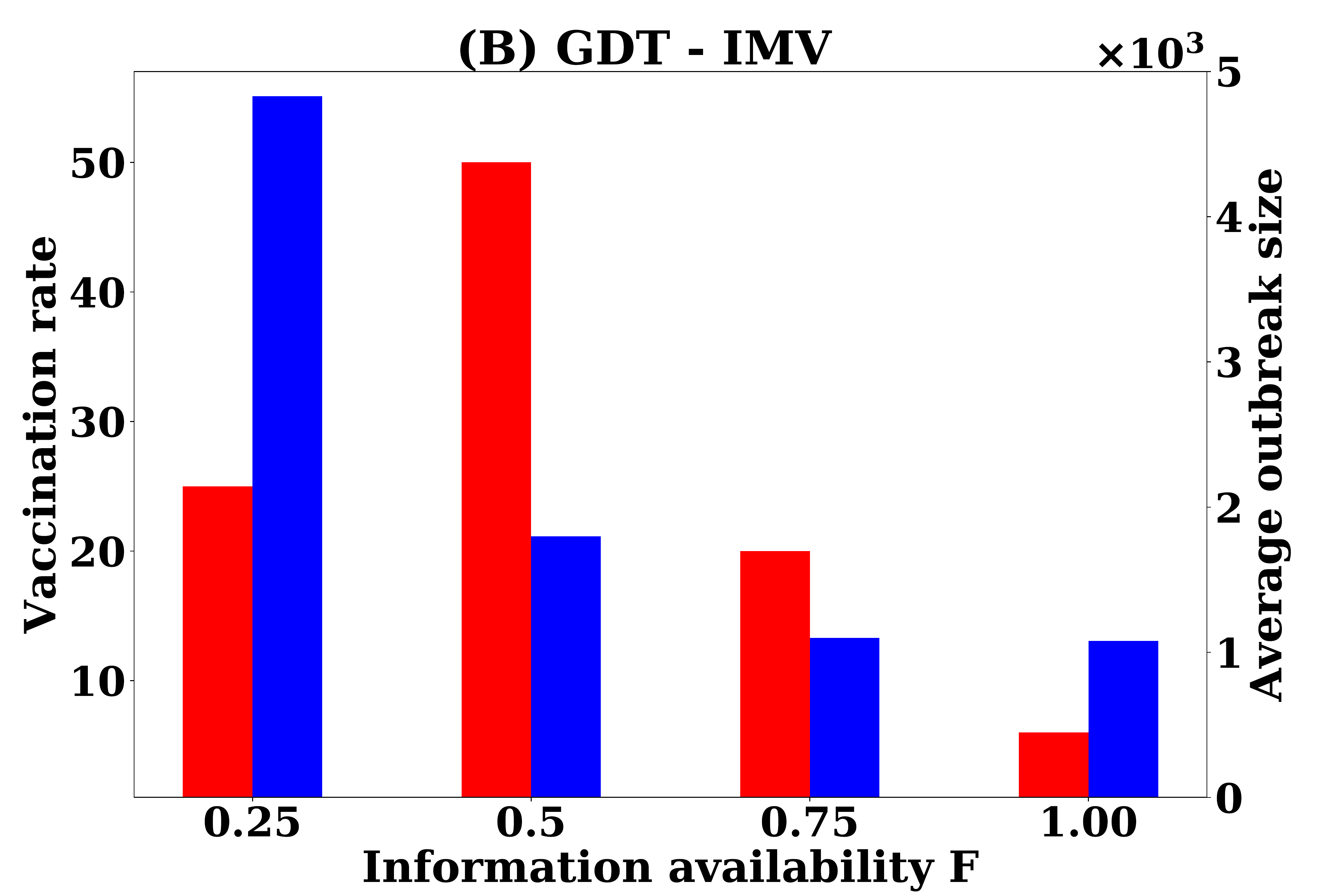}\\ 
\includegraphics[width=0.495\linewidth, height=5.0 cm]{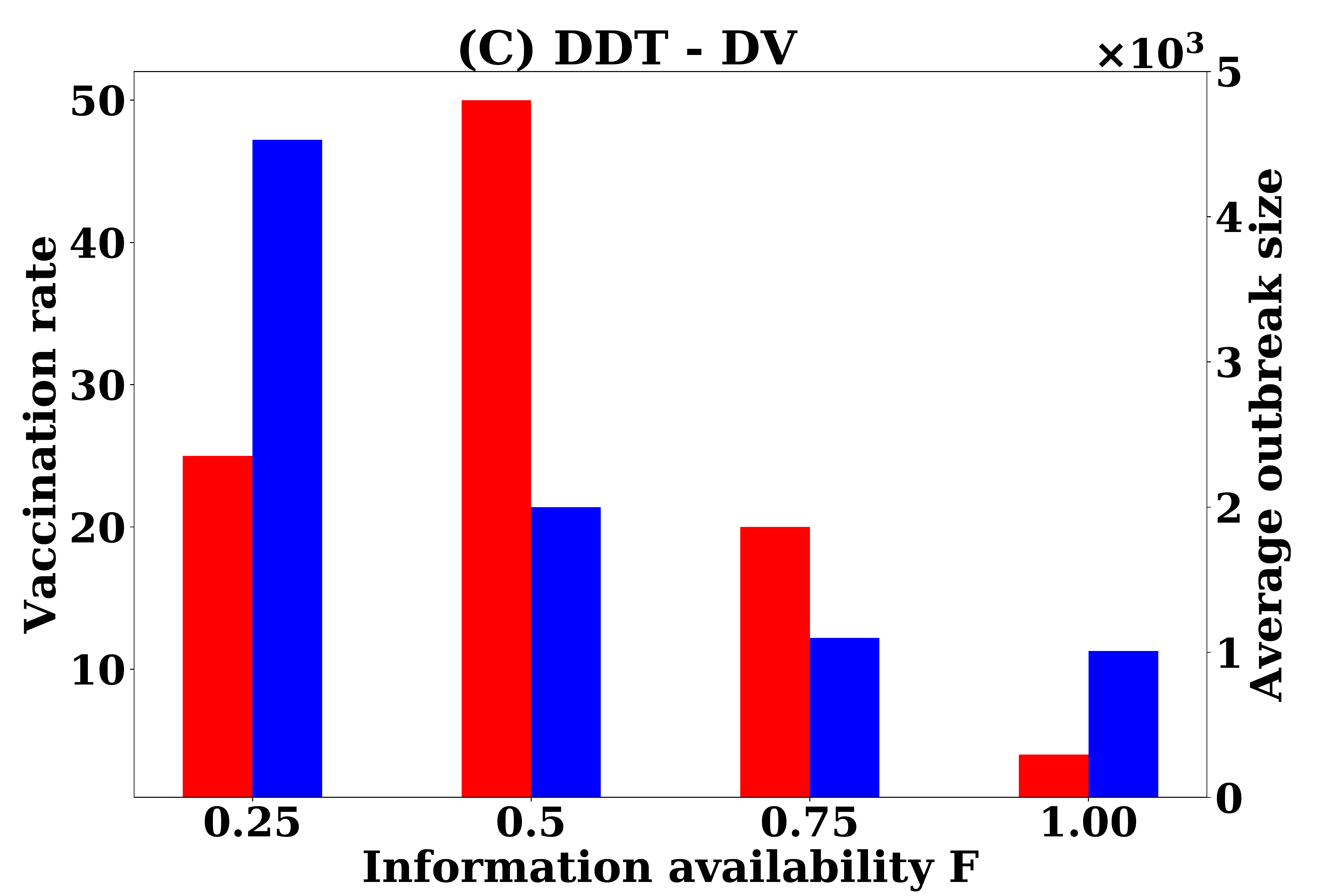}
\includegraphics[width=0.495\linewidth, height=5.0 cm]{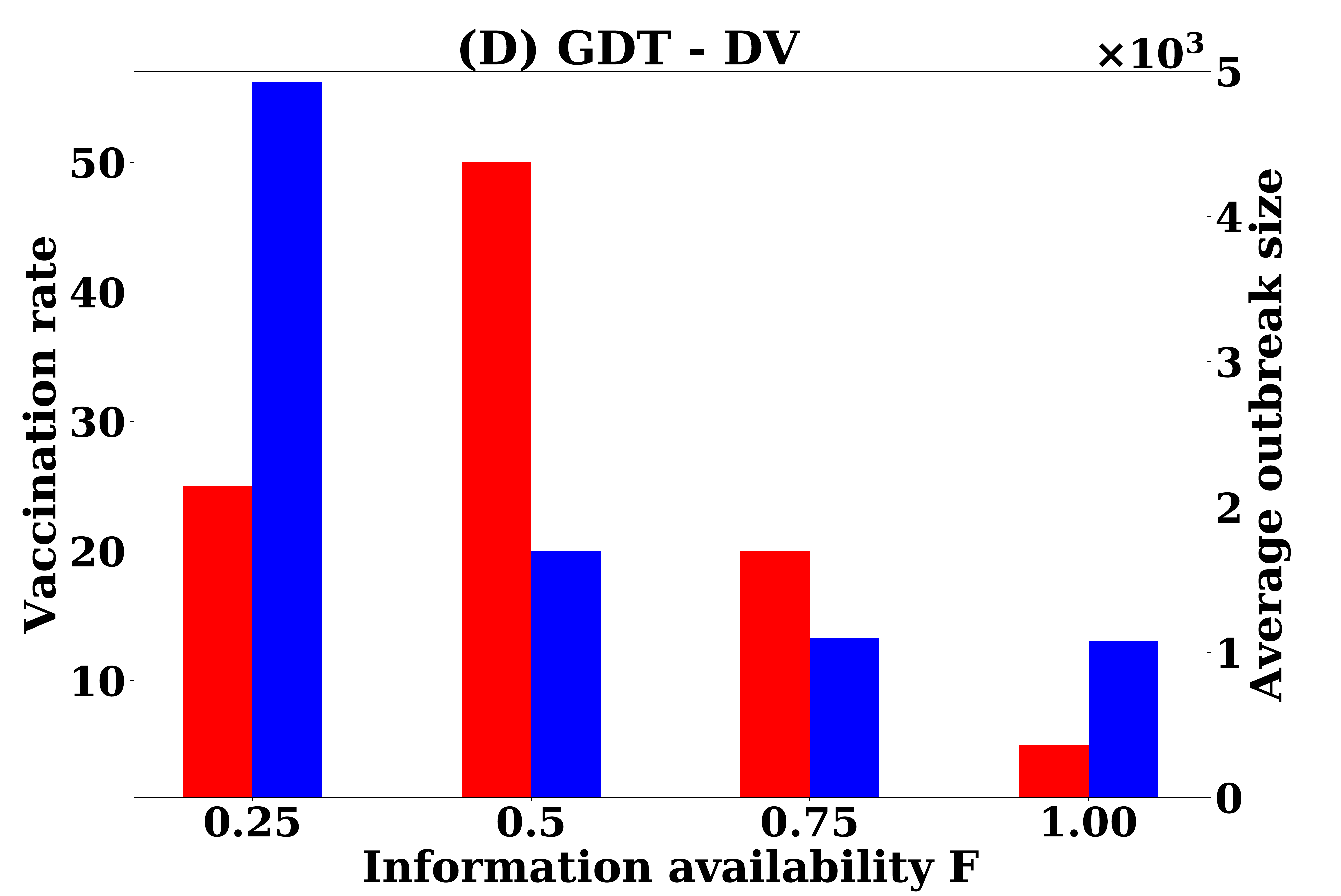}\\ \includegraphics[width=0.495\linewidth, height=5.0 cm]{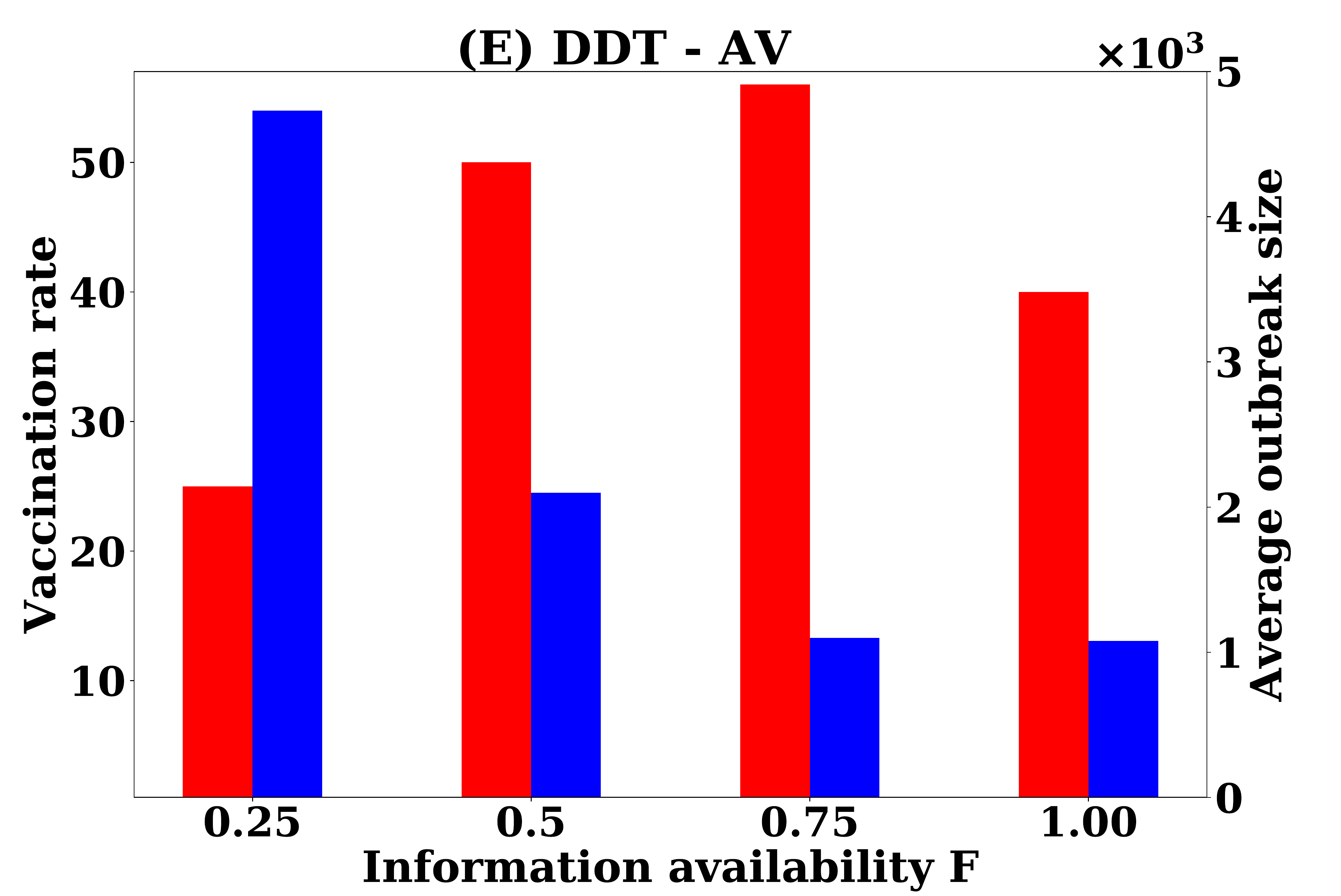}
\includegraphics[width=0.495\linewidth, height=5.0 cm]{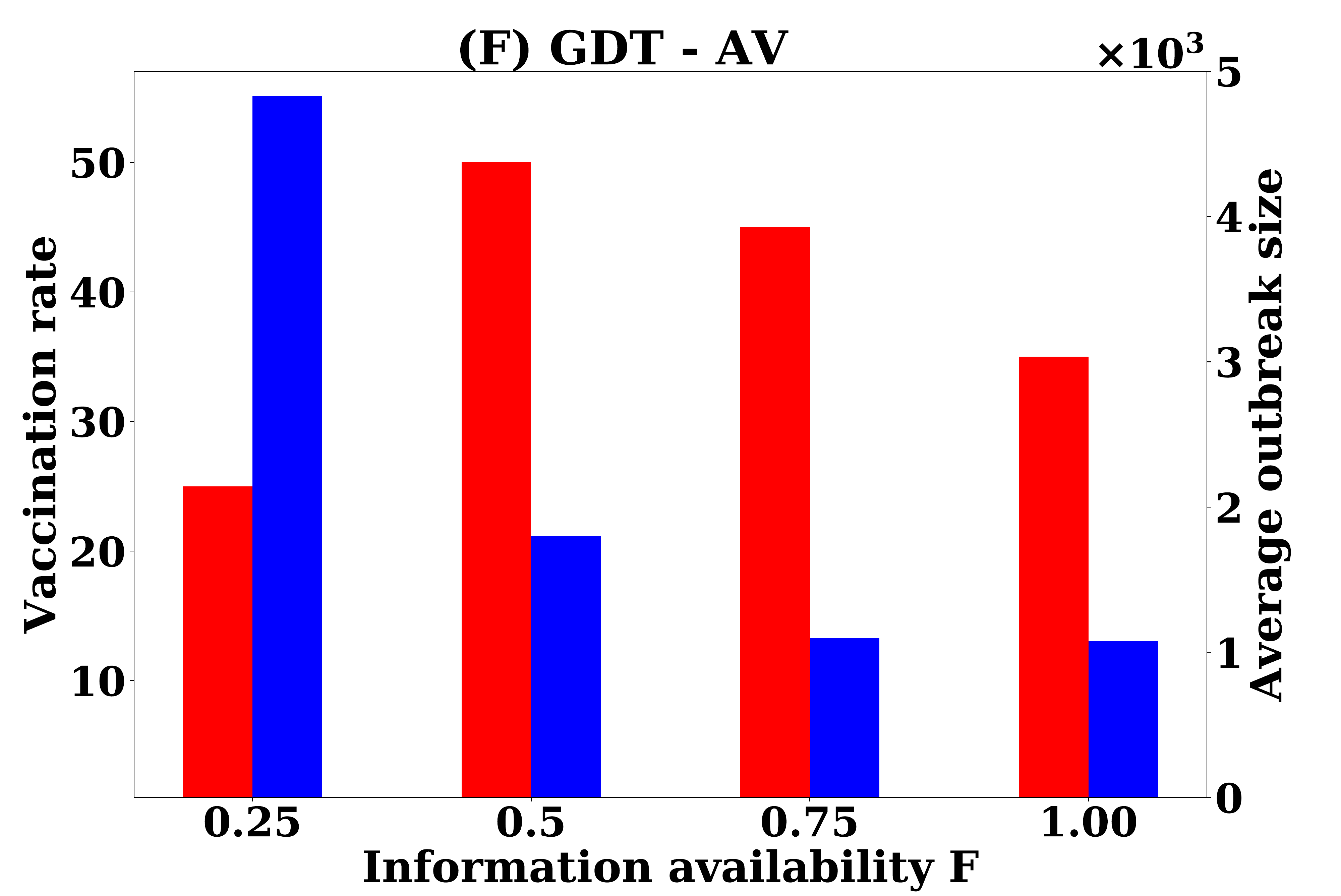}

\caption{Trade-off among information collection cost, vaccination cost and infection cost: (A, B) proposed IMV strategy, (C, D) degree based strategy (DV), and (E,F) acquaintance vaccination (AV) strategy }
\vspace{-1.0em}
\label{fig:fpvacf}
\end{figure}

Simulations are now conducted to understand the effectiveness of vaccination strategies with the scale of information availability regarding node's contact for post-outbreak scenarios. Simulations are carried out for the scenarios where the contact information of $F=\{0.25, 0.5, 0.75\}$ proportion of nodes are available for ranking procedure. At each value of $F$, the performance of the vaccination strategies is analysed for vaccination rates $P$ varying in the range [0,25]\% with a step of 5\%. Similar to the previous experiment, the disease starts with 500 seed nodes and continues for 42 days. The vaccination is implemented at the 7th day of simulations. The final outbreak sizes are obtained on both the DDT and GDT networks. The average outbreak sizes are presented in Figure~\ref{fig:fpvac}. The simulations are also run for all strategies until the average outbreak sizes are reduced by 90\% (1K infections) at a certain value of $P$. The results obtained show the trade-off among information collection cost, vaccination cost and infection cost for strategies and are presented in Figure~\ref{fig:fpvacf}.

The performance of the applied strategy is strongly affected by the scale of information availability regarding node contact. The AV strategy slowly reduces the average outbreak sizes within the studied vaccination rates in [0, 25]\%. They do not reduce outbreak sizes below 6K infections at $F=0.25$ in both the DDT and GDT networks for any value of $P$.
Therefore, it is not possible to contain the outbreak size by vaccinating nodes if only 0.25 proportion of nodes contact information is available. By increasing $F$ to 0.50, both IMV and DV strategies are capable of reducing the outbreak sizes by 80\% with vaccinating 50\% of nodes in both networks, i.e. all nodes picked up for ranking procedure are vaccinated. The AV strategy has the same average outbreak size at $F=0.50$ to that of IMV and DV strategy (Fig.~\ref{fig:fpvacf}). All these outbreak sizes are similar to that of random vaccination with 50\% vaccination rate. The differences at $F=0.5$ are that IMV and DV strategies have more efficiency at lower $P$ comparing to AV (Fig.~\ref{fig:fpvac}). Increasing $F$ to 75\% reduces the requirement of vaccination rate significantly in both IMV and DV. Now, IMV strategy requires 25\% of the nodes to be vaccinated and DV strategy requires 20\% in both networks. At this value of $F$, acquaintance vaccination (AV) shows the ability to contain outbreak sizes below 1K infections (outbreak sizes reduced by 90\%) with 55\% vaccination in the DDT network and 45\% vaccination in the GDT network. Up to $F=0.5$, the maximum performance can be achieved by the strategies that are same to random vaccination with 50\%. However, IMV and DV strategies for $F> 0.5$ shows that the strong protection is achievable with the lower value of $P$. There is a strong trade-off between information collection cost, vaccination cost and infection cost in applying a vaccination strategy (Fig.~\ref{fig:fpvacf}). The proposed IMV strategy still achieves better efficiency than AV strategy with the constrained of information availability. 

\subsubsection*{Node level vaccination}
\begin{figure}[h!]
\centering
\begin{tikzpicture}
    \begin{customlegend}[legend columns=2,legend style={at={(0.42,1.00)},draw=none,column sep=3ex ,line width=6pt,font=\small}, legend entries={ outbreak size, nodes vaccinated}]
    \addlegendimage{solid, color=red}
    \addlegendimage{color=blue}
    \end{customlegend}
 \end{tikzpicture}\\  \vspace{2ex}
\includegraphics[width=0.495\linewidth, height=5.0 cm]{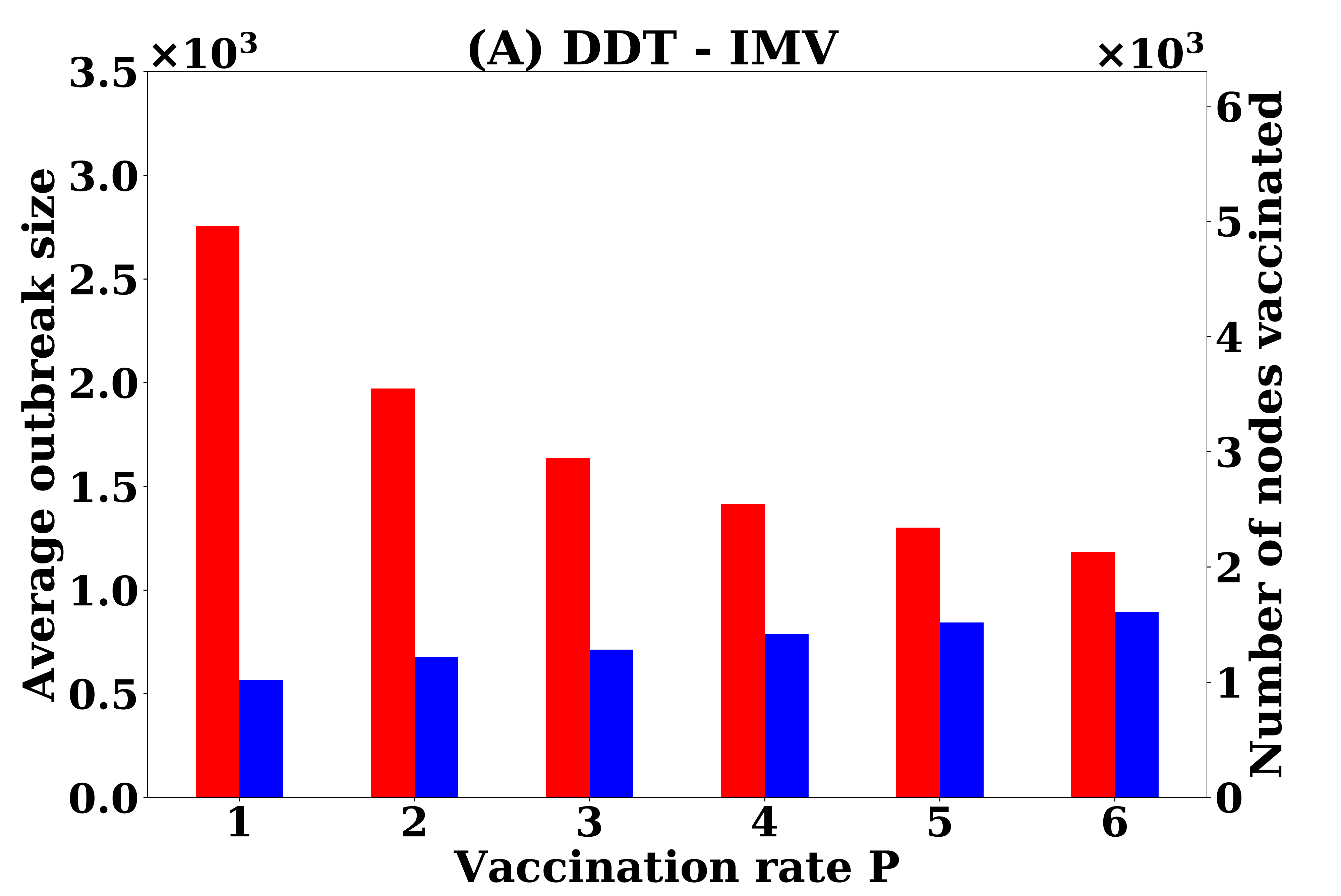}
\includegraphics[width=0.495\linewidth, height=5.0 cm]{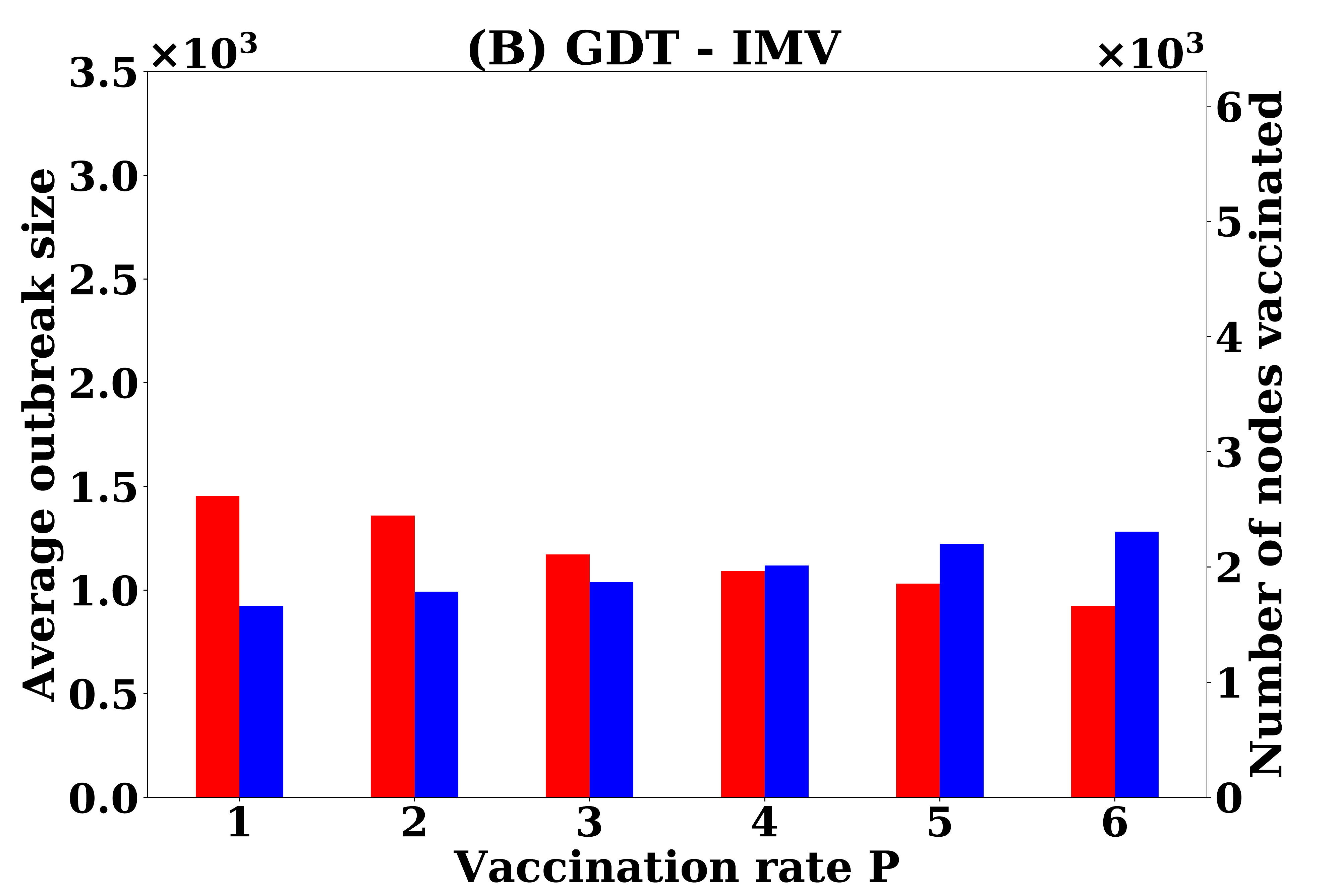}\\ [2ex]
\includegraphics[width=0.495\linewidth, height=5.0 cm]{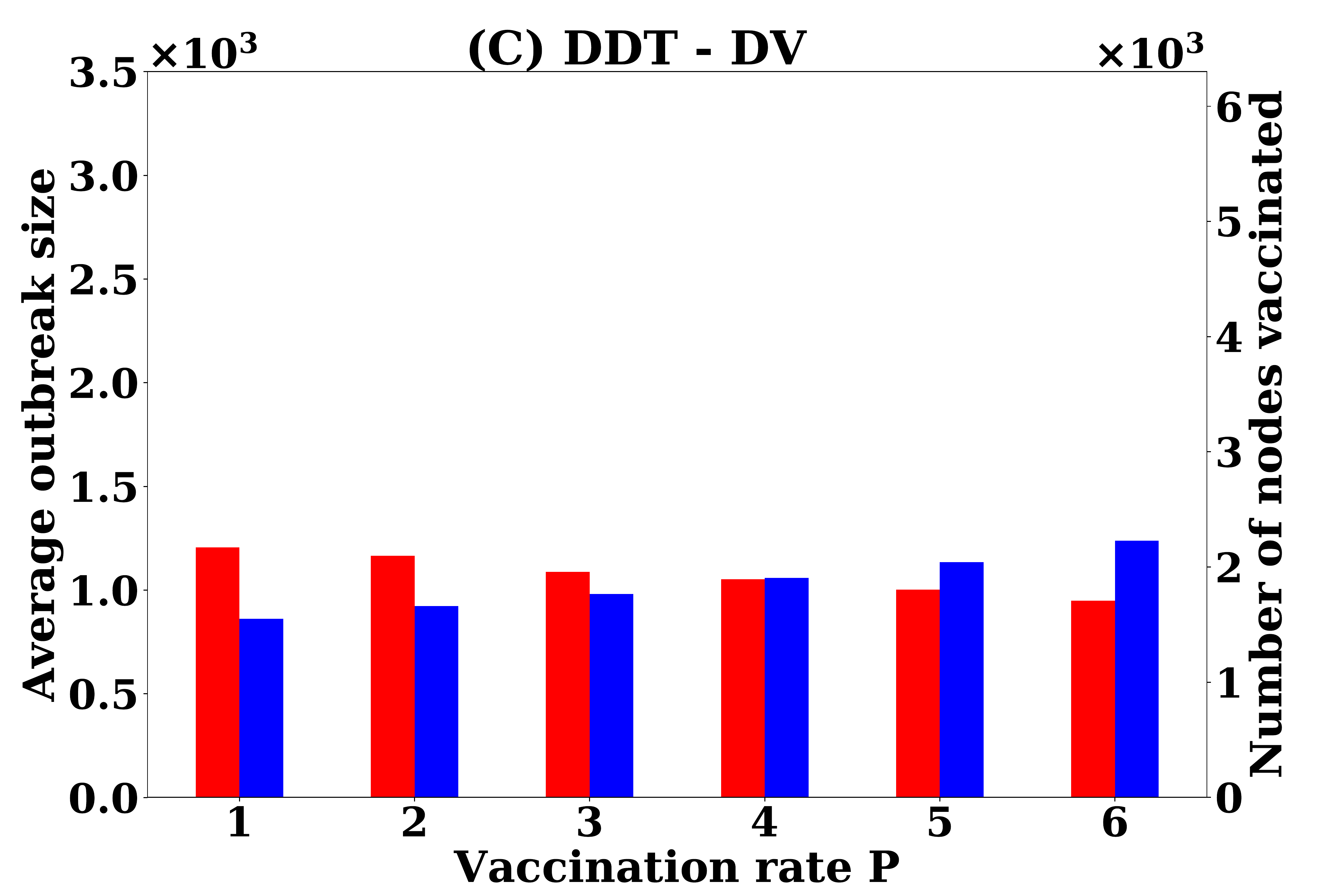}
\includegraphics[width=0.495\linewidth, height=5.0 cm]{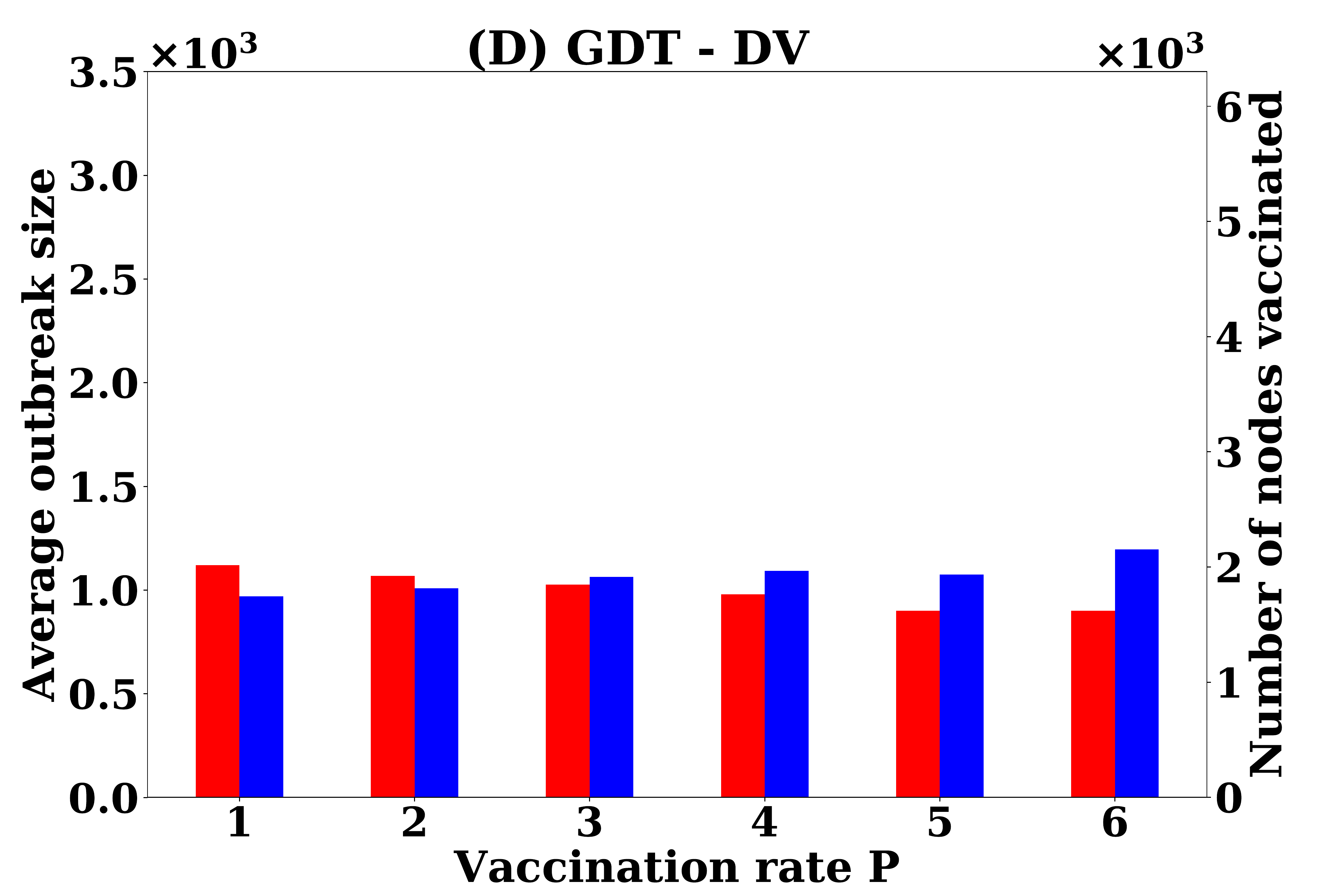}\\ [2ex]
\includegraphics[width=0.495\linewidth, height=5.0 cm]{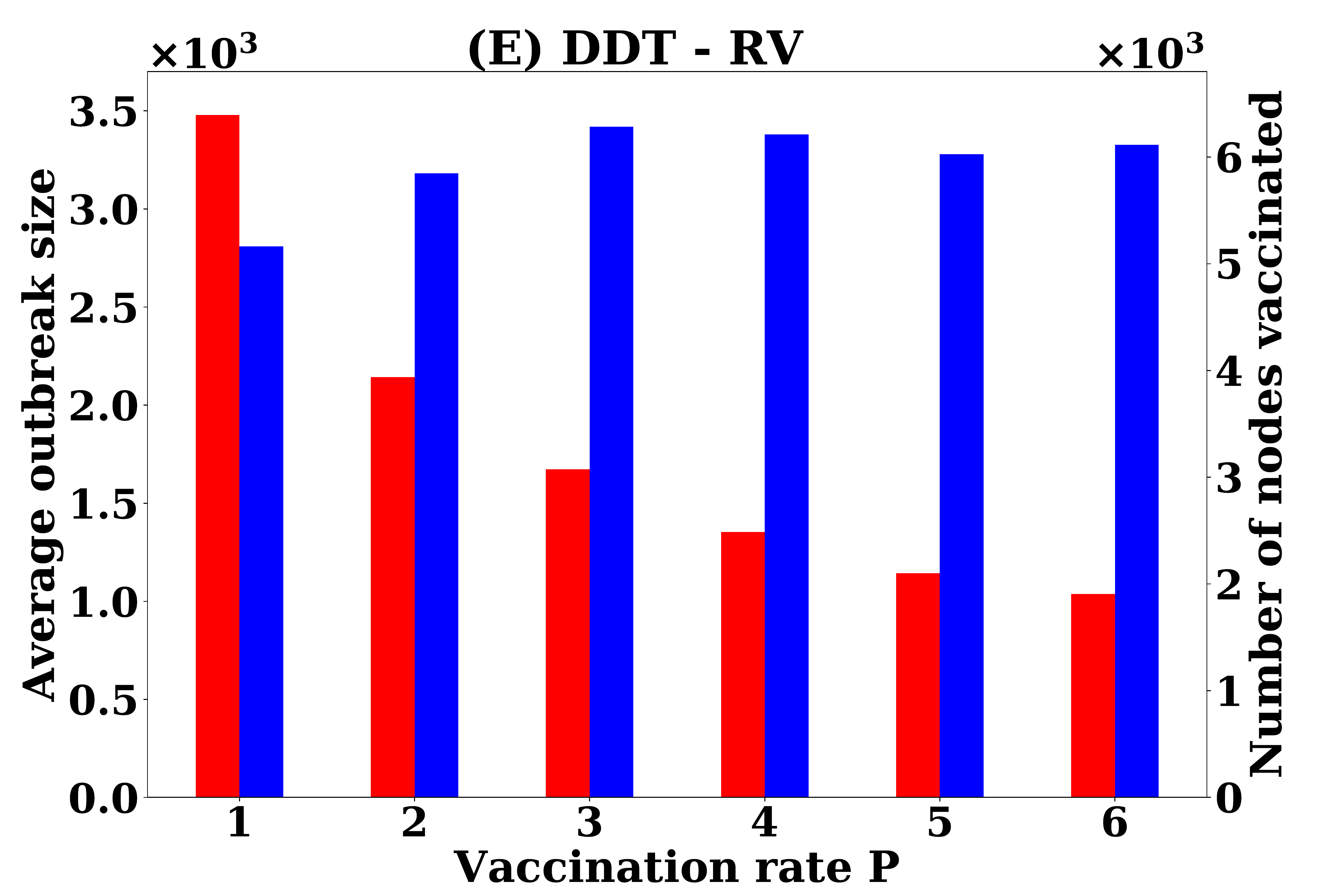}
\includegraphics[width=0.495\linewidth, height=5.0 cm]{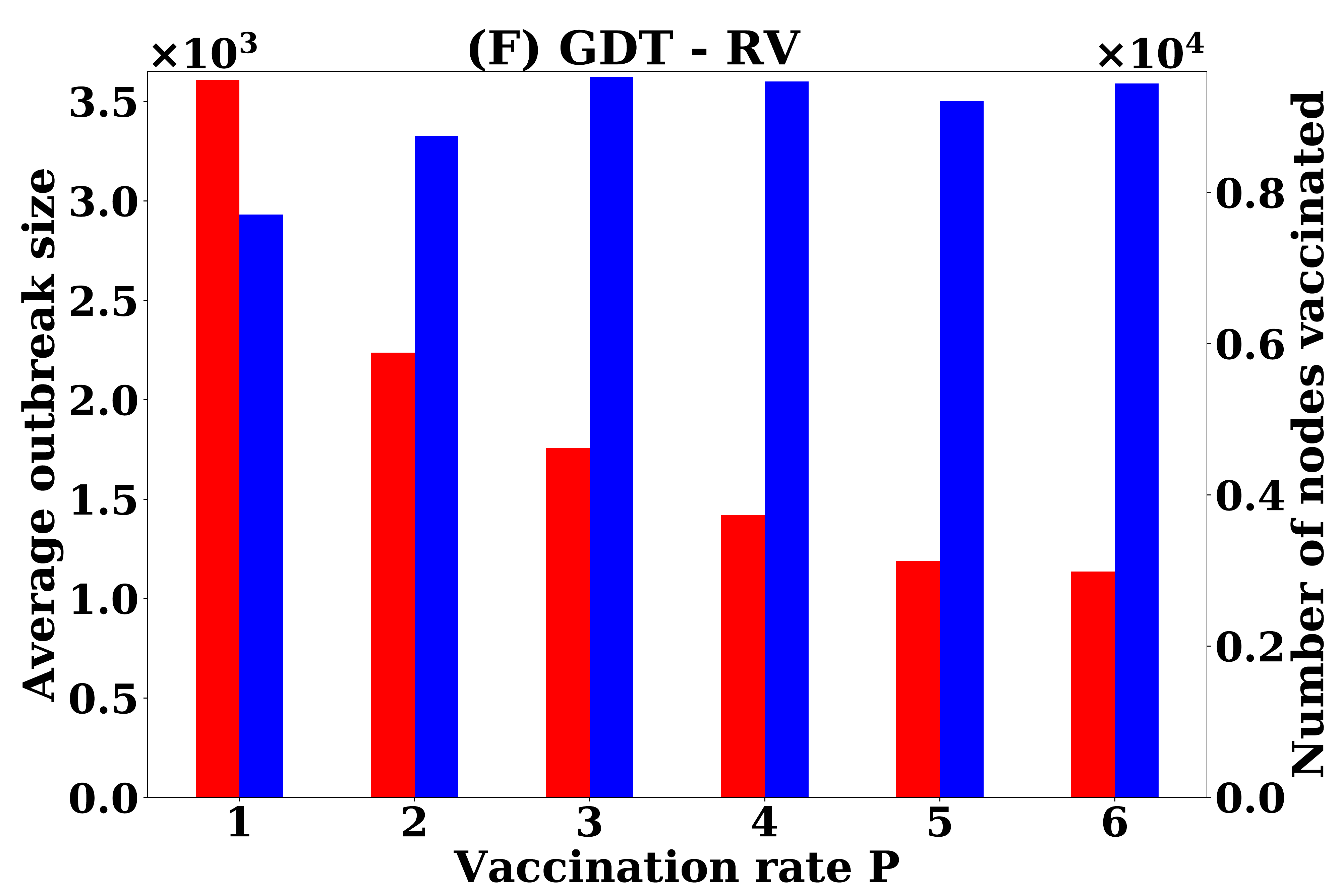}

\caption{Performance of node level vaccination with various strategies: (A, B) proposed IMV strategy, (C, D) DV strategy, and (E,F) RV strategy}
\label{fig:nvaca}
\vspace{-1.0em}
\end{figure}

Population level vaccination requires a high vaccination rate to contain the disease spreading and requirement increases with the constrained of information collection. Alternatively, node level vaccination is often applied through ring vaccination. In this approach, a proportion of infected node's neighbour is vaccinated. In this experiment, neighbour nodes to be vaccinated are selected using three methods, namely randomly, degree-based ranks and IMV based ranks. Similar to the previous experiments, the disease starts with 500 seed nodes and continues for 42 days. Initially, nodes are vaccinated at the 7th day of simulation and after that neighbour nodes are vaccinated when a new node is infected. It is assumed in the first experiment that all infected nodes are identified and their neighbour nodes are vaccinated based on applied strategy. Then, a proportion $F$ of all infected nodes is identified and their neighbour nodes are vaccinated based on the applied strategy. In these simulations, a threshold value of the ranking score is set to select neighbour nodes following DV and IMV strategies. If a neighbour node has scored more than a threshold, it will be vaccinated. The threshold value is a ranking score above which a proportion $P$ of total nodes have scored. This ensures that a proportion $P$ of neighbour nodes will be vaccinated through a threshold value. In random vaccination (RV), a proportion $P$ of neighbour nodes are randomly chosen for vaccination. The simulations are run for $P$ in the range [1,6]\% for each strategy and are repeated 100 times for each value of $P$. The main focus in this experiment is to understand the reduction in outbreak sizes against the number of nodes of vaccinated through the above neighbour selection process. The results are presented in Figure~\ref{fig:nvaca}. 

Node level vaccination is more efficient than the population level vaccination. In IMV strategy, average outbreak size at $P=1$\% is reduced by 75\% in the DDT network and these are slightly lower in the GDT network. If $P$ are increased to 6\%, the outbreak sizes is reduced by 90\% in both networks with the vaccination of 2000 nodes. To achieve this performance in population level vaccination, IMV strategy requires vaccination of about 4\% (14,400) nodes. The DV strategy in DDT network shows about 80\% reduction in the average outbreak sizes with vaccination of 1,700 nodes at $P=1$\% and average outbreak size is reduced by 95\% with vaccinating 2,000 nodes at $P=4$\%. The GDT network with DV strategy shows slightly higher outbreak size at $P=1$\% but is reduced by 94\% with increasing $P=6$\% and vaccination of 2,600 nodes. However, the RV strategy shows very poor performance at a lower value of $P$ with the vaccination of about 5,000 nodes in both networks. Increasing $P$ reduces outbreak sizes quickly with the increasing of the number of nodes to be vaccinated. It is found that the number of nodes vaccinated is stabilised at about 6,000 nodes in the DDT network and 10,000 nodes in the GDT network regardless of $P$ in RV strategy. Further, increasing of $P$ shows that RV strategy reduces outbreak sizes to about 1K infections (90\% reduction) at $P=6$\%. For node level vaccination, coarse-grained information based IMV strategy achieves the performance of DV strategy and better than RV strategy.

\begin{figure}[h!]
\centering
\begin{tikzpicture}
    \begin{customlegend}[legend columns=3,legend style={at={(0.32,1.00)},draw=none,column sep=3ex ,line width=6pt,font=\small}, legend entries={ F=0.25,F=0.5, F=0.75}]
    \addlegendimage{solid, color=red}
    \addlegendimage{color=blue}
        \addlegendimage{color=green}
    \end{customlegend}
 \end{tikzpicture}\\  \vspace{2ex}

\includegraphics[width=0.495\linewidth, height=5.0 cm]{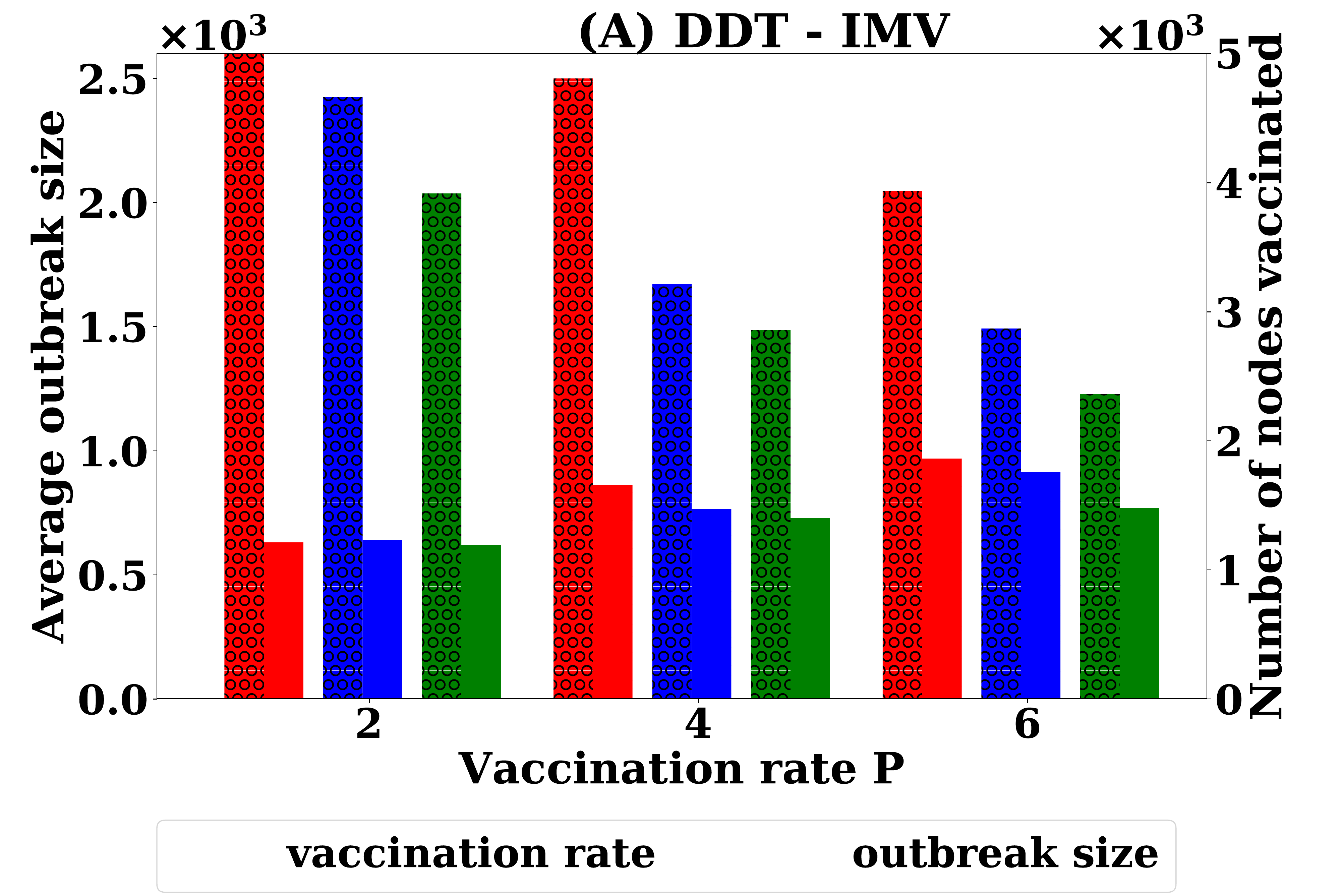}
\includegraphics[width=0.495\linewidth, height=5.0 cm]{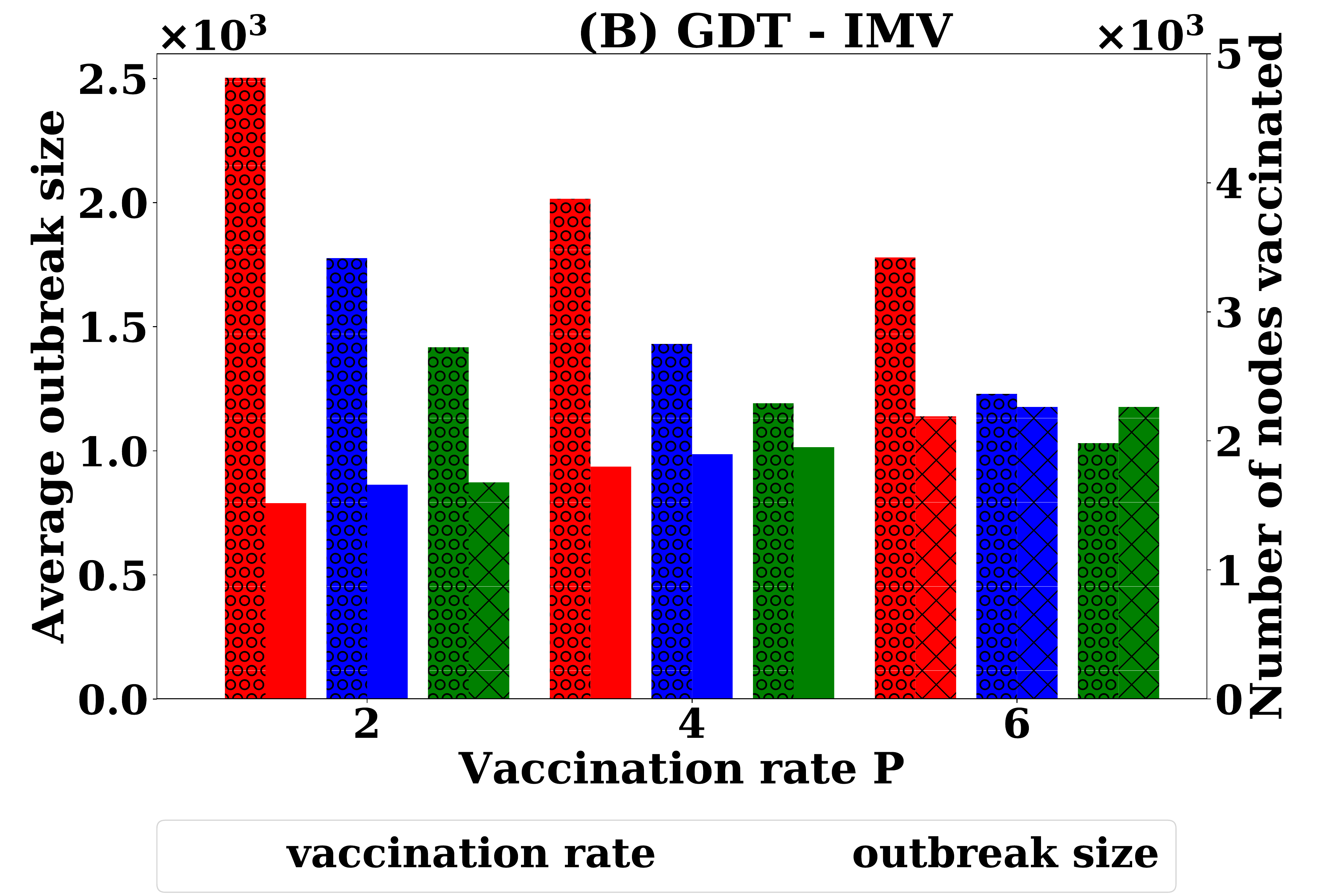}\\ [3ex]
\includegraphics[width=0.495\linewidth, height=5.0 cm]{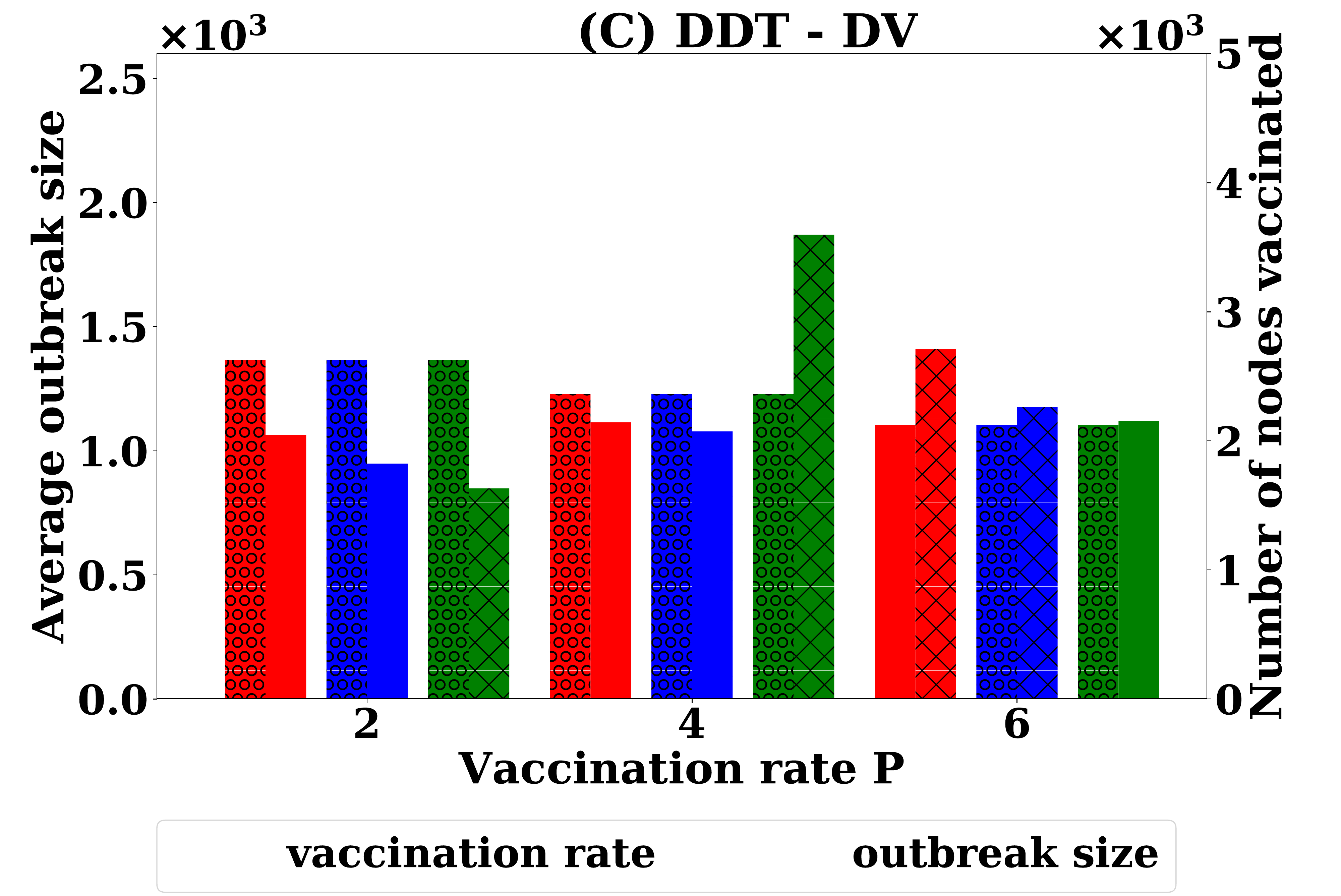}
\includegraphics[width=0.495\linewidth, height=5.0 cm]{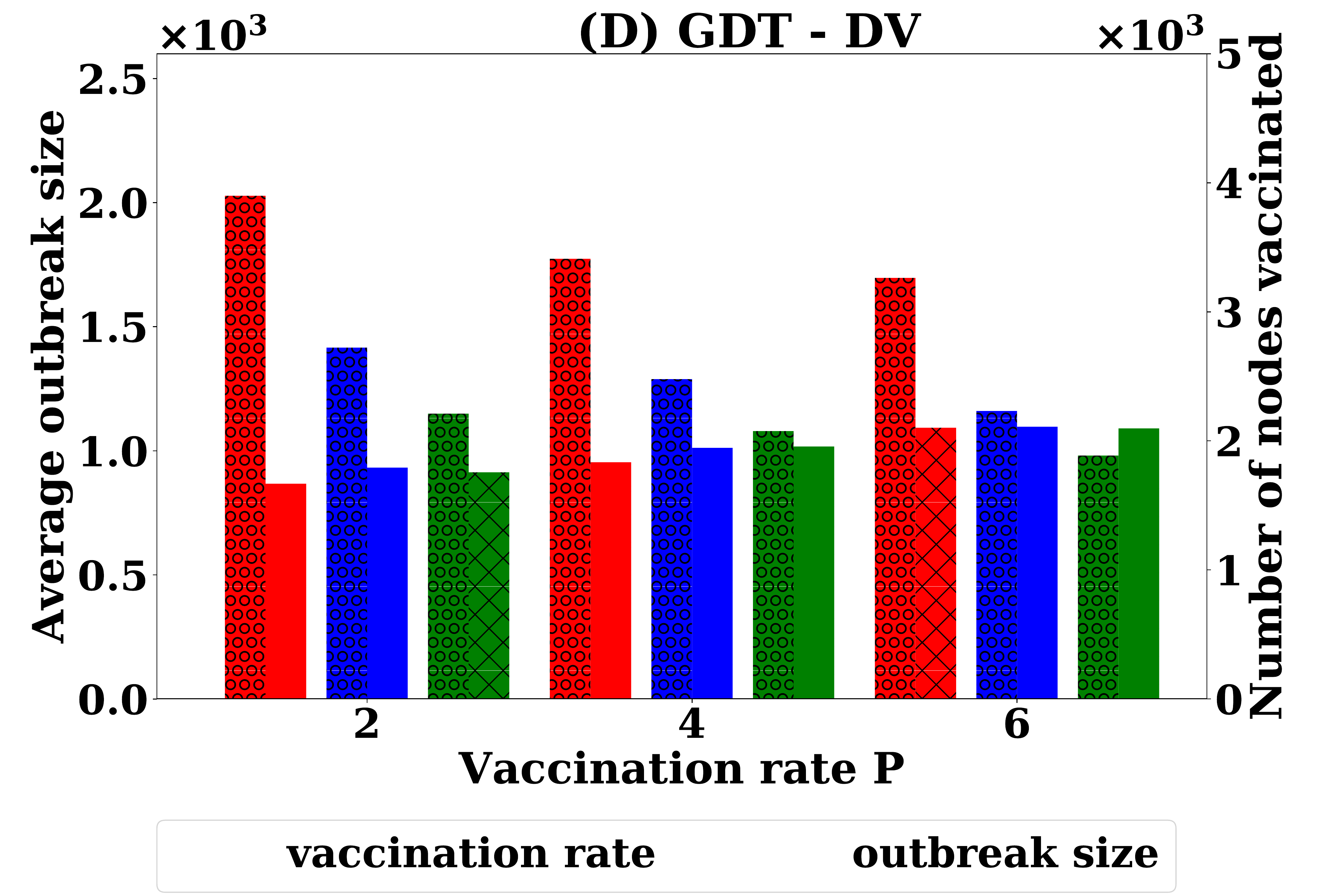}\\ [3ex]
\includegraphics[width=0.495\linewidth, height=5.0 cm]{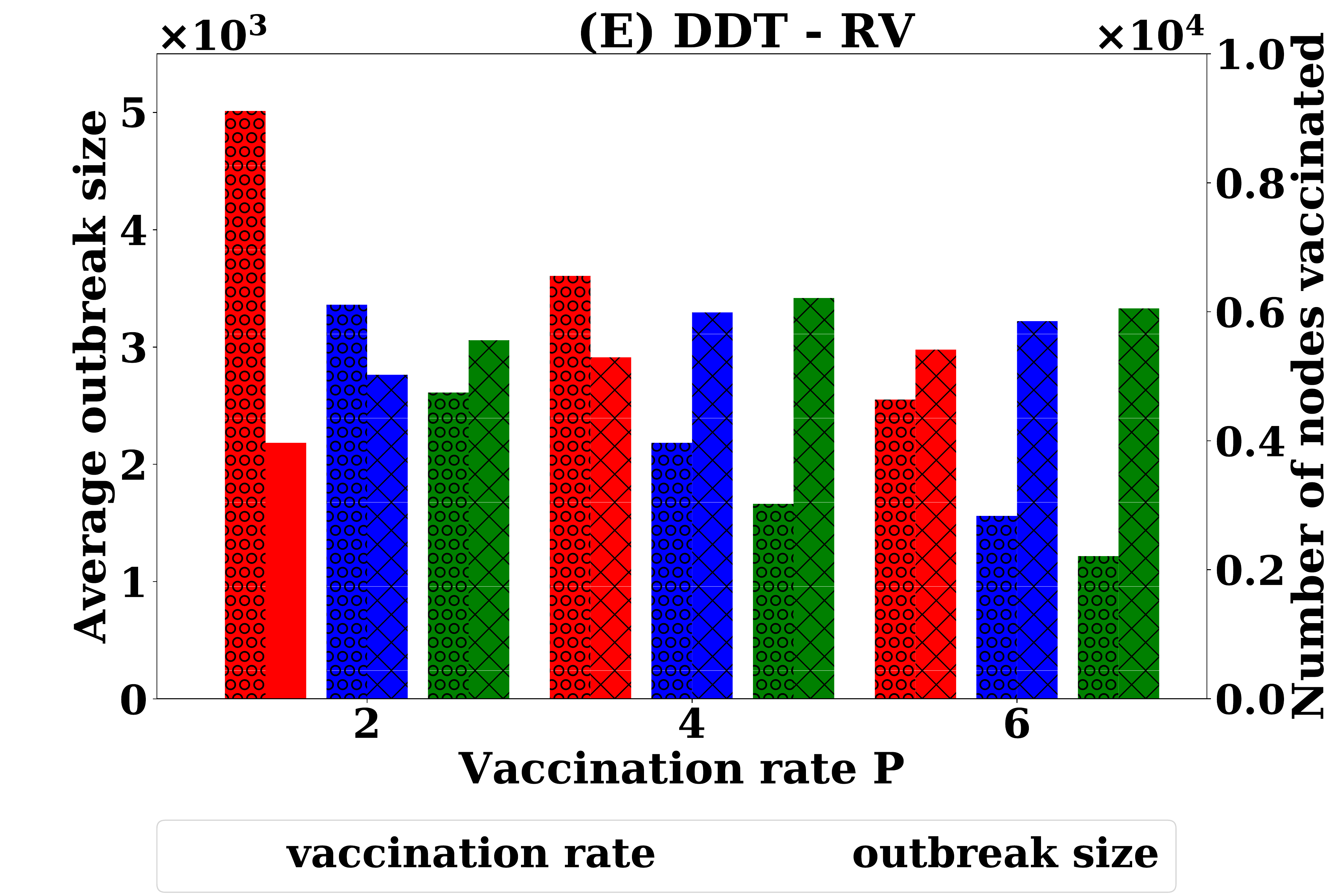}
\includegraphics[width=0.495\linewidth, height=5.0 cm]{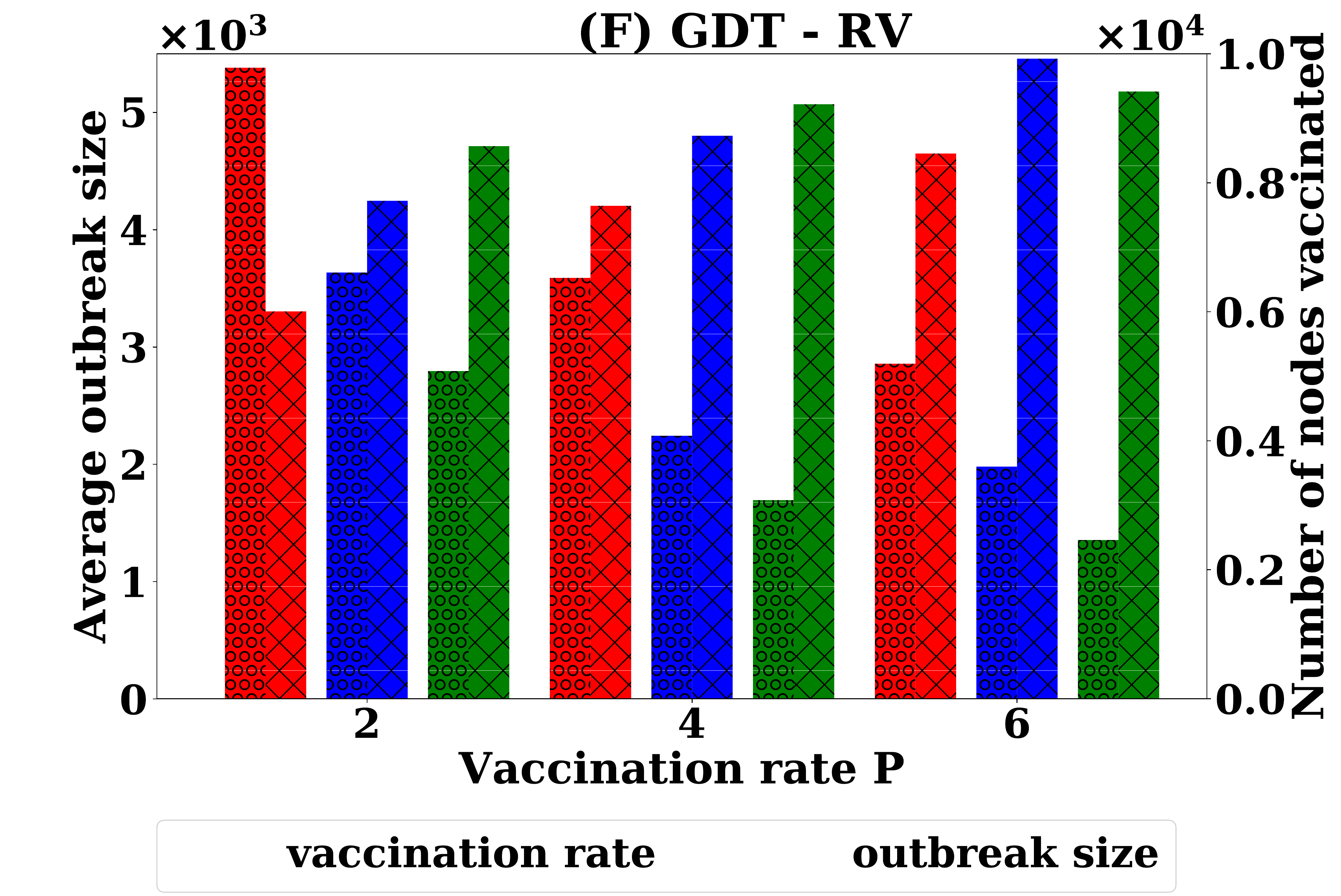}

\caption{Performances of node level vaccination with F proportion of infected nodes identified: (A, B) proposed vaccination strategy, (C, D) degree based vaccination, and (E, F) acquaintance based vaccination}
\label{fig:nvacb}
\vspace{-1.0em}
\end{figure} 

\begin{figure}[h!]
\centering
\begin{tikzpicture}
    \begin{customlegend}[legend columns=3,legend style={at={(0.32,1.00)},draw=none,column sep=3ex ,line width=6 pt,font=\small}, legend entries={ F=0.25,F=0.5, F=0.75}]
    \addlegendimage{solid, color=red}
    \addlegendimage{color=blue}
        \addlegendimage{color=green}
    \end{customlegend}
 \end{tikzpicture}\\  \vspace{2ex}
\includegraphics[width=0.46\linewidth, height=5.0 cm]{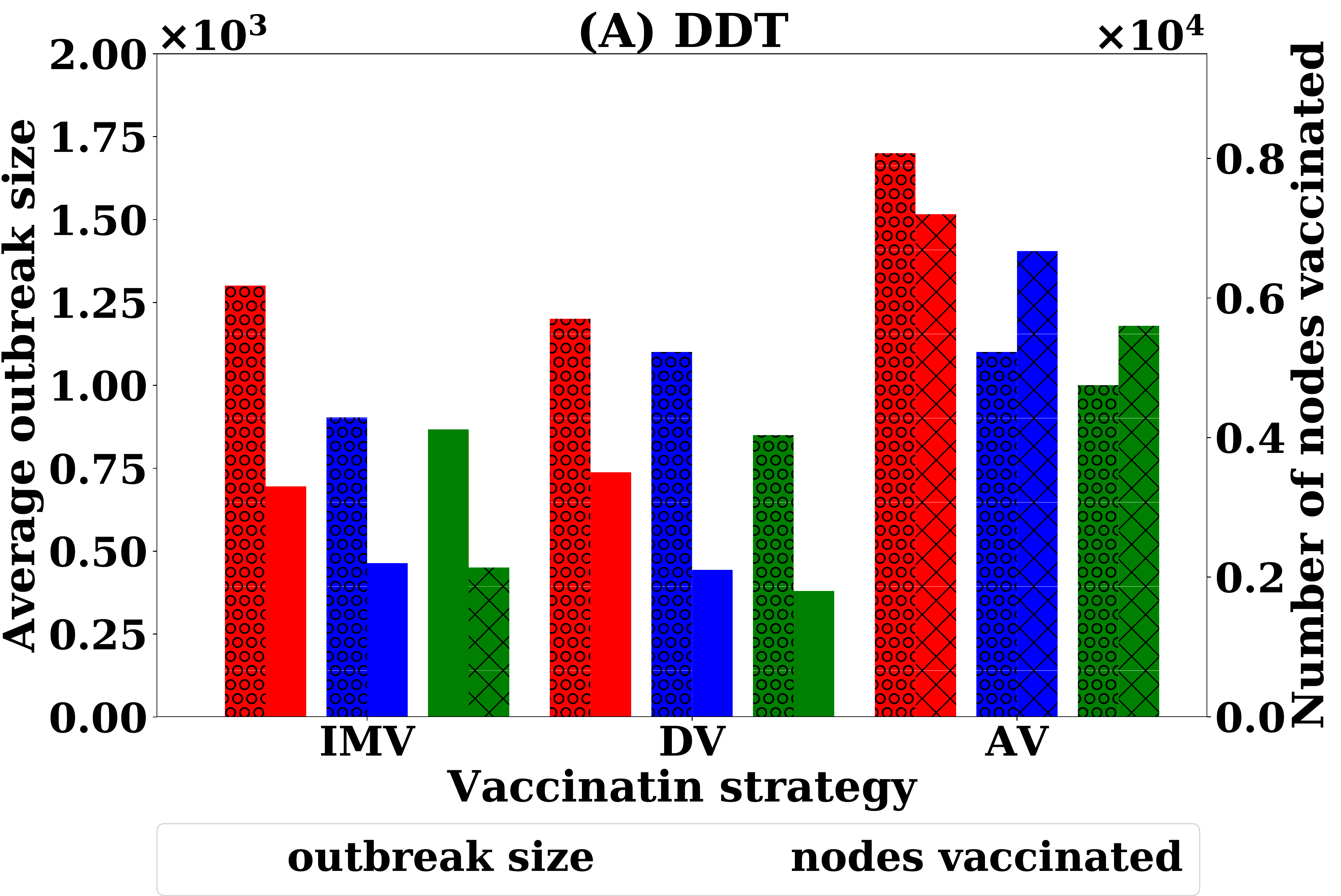}\quad
\includegraphics[width=0.46\linewidth, height=5.0 cm]{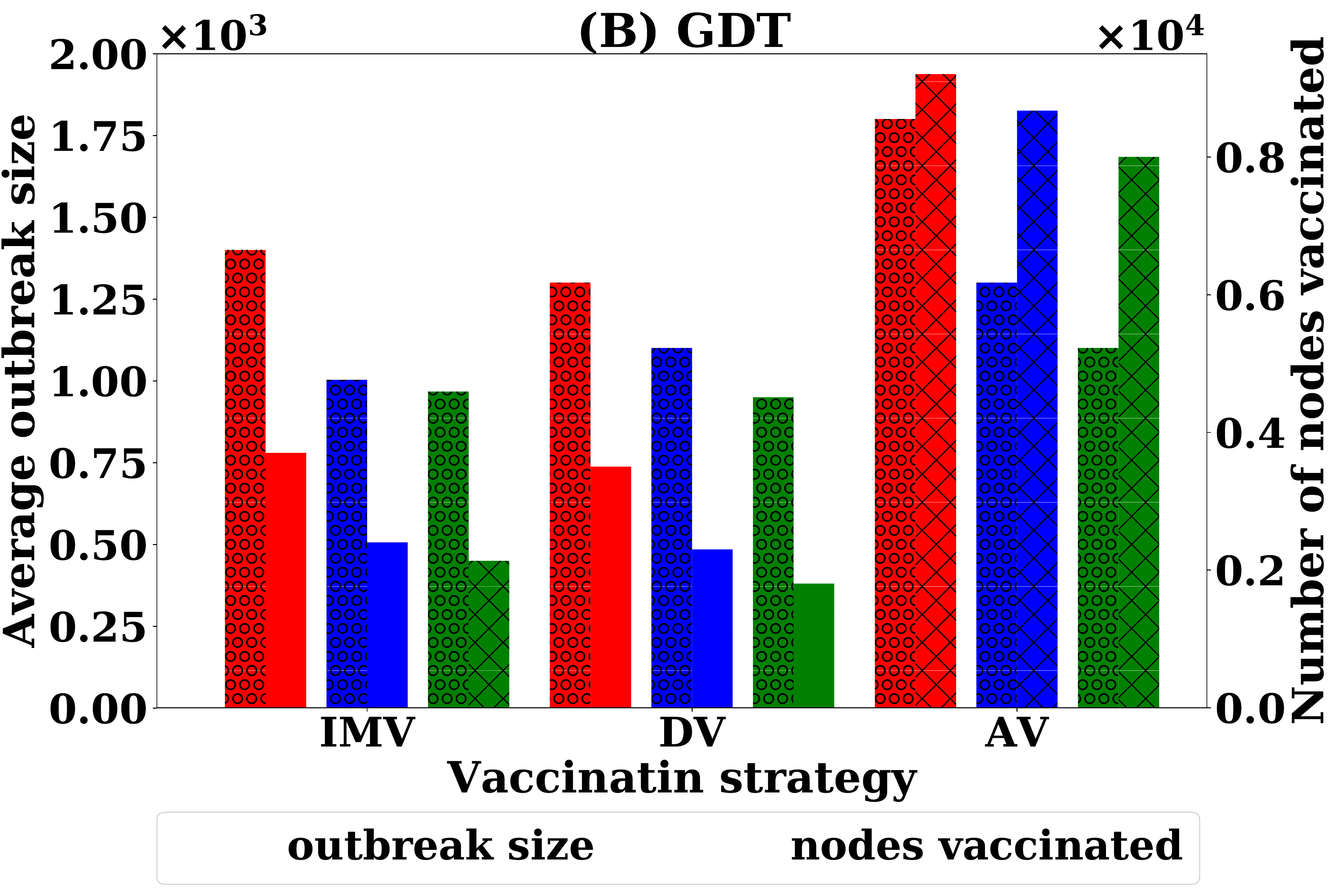}

\caption{Required number nodes to be vaccinated in node level vaccination for achieving outbreak sizes below 1K infections with F proportion of infected nodes identified with different strategies}
\vspace{-1.0em}
\label{fig:nvacc}
\end{figure}

The performance is now studied varying $F$, the proportion of infected nodes can be identified, for each vaccination strategy. In these simulations, a proportion $F$ of infected nodes are selected and then the above threshold based procedures of vaccinating nodes are used according to the applied vaccination strategy. Similarly, disease spreading is simulated for 42 days with the initial set of 500 seed nodes. The simulations are conducted for $P$ in the range [2,6]\% for each strategy and are repeated 1,000 times for each value of $P$. The average outbreak sizes and the corresponding number of nodes vaccinated at different $F$ for each vaccination strategy are presented in Figure~\ref{fig:nvacb}. Then, simulations are run for each strategy at different $F$ until the average outbreak sizes become below 1K infections. The results are presented in Figure~\ref{fig:nvacc}. In the IMV strategy, the outbreak sizes is reduced by 80\% at F=0.25 in both networks at the highest $P=0.6$ with vaccinating about 2K nodes (Fig.~\ref{fig:nvacb}A and Fig.~\ref{fig:nvacb}B). For other values of F, the outbreak sizes reduce and the number of nodes vaccinated reduces in the DDT network at $P=0.6$ while the number of nodes vaccinated is same to 2K nodes in GDT network with reducing outbreak sizes. For the IMV strategy, it is not possible to reduce outbreak sizes more than 90\% at F=0.25 in both networks while for other F it is possible by vaccinating 2K nodes. For $F>0.25$, the required number of vaccination to achieve outbreak sizes about 1K infections are same as the 2K nodes (Fig~\ref{fig:nvacc}). The similar trend is found for the DV strategy. However, the RV strategy requires a large number of nodes to be vaccinated to reduce outbreak size below 1K infections. Similar to the other strategy, the outbreak sizes do not reduce to 1K infections in RV strategy even after vaccinating 3\% nodes. Under the constrained of information collection, IMV strategy still performs well compared to RV strategy.

\section*{Conclusion}
This paper has investigated vaccination strategies on dynamic contact networks having indirect transmission links. We have developed a simple vaccination strategy called individual movement based vaccination (IMV). This strategy can work with the coarse-grained contact information instead of exact contact information. The IMV strategy is examined for both preventive and post-outbreak scenarios. The performance is compared with other three vaccination strategies namely random vaccination (RV), acquaintance vaccination (AV) and degree based vaccination (DV). For preventive vaccination scenarios, the proposed IMV strategy shows the performance of the DV strategy where only 6\% of nodes to be vaccinated so that no seed nodes have outbreaks of more than 100 infections. However, the RV strategy requires vaccination of 70\% nodes to achieve the same efficiency and the AV requires vaccination of 40\% nodes. The sensitivity of IMV strategy is tested against applying exact contact information and temporal information. There is no significant improvement in applying that information. The vaccination strategies are also studied against the scale of information availability $F$ on nodes. It is found that all strategies have the maximum performance of RV strategy if 50\% of nodes provide contact information for the vaccination procedure. However, IMV and DV strategies show better performance with the higher infection cost for $F<0.5$. Then, the performance of IMV strategy is examined in the post-outbreak scenarios. Post-outbreak vaccination is implemented in two ways population-level vaccination and node-level vaccination. In this case, the IMV strategy shows a similar performance than the DV strategy and requires vaccination of 4\% nodes to contain the outbreak sizes below 1K infections. On the other hand, RV and AV strategies require a higher number of nodes to be vaccinated with 70\% and 40\% respectively. Under the constrained of information availability, the IMV strategy shows better performance in the population level vaccination as well. In the node-level vaccination, the number of nodes to be vaccinated for containing outbreaks about 1K infections with IMV strategy reduces significantly and it is about 2K nodes (0.75\%). This is similar to that of DV strategy. Under the constrained of information availability, the number of nodes to be vaccinated for containing outbreak sizes below 1k infections does not change for $F>0.25$. On the other hand, the RV strategy requires to up 10K (about 3\% of total nodes) nodes at $F=1$ and it is increased if $F$ is reduced. The proposed strategy IMV performs better than RV and AV strategies in both preventive and post-outbreak scenarios.

\section*{Acknowledgments}

This study was supported by the Australian Research Council Discovery Grant (ARC - DP170102794) and Commonwealth Scientific and Industrial Research Organisation (CSIRO). Md.S. was partially supported by the CSIRO and ARC- DP170102794. B.M. was partially supported by the ARC-DP170102794. R.J. and F.d.H. were supported by the CSIRO. The authors gratefully acknowledge the Distributed Sensing System Group, Data61, CSIRO for providing research facilities for this research. 



%
%
%
\bibliography{references}

\end{document}